\documentclass[a4paper,11pt]{article}

\newcommand{\Rmnum}[1]{\expandafter\@slowromancap\romannumeral #1@}
\usepackage{jheppub} 

\usepackage{epsfig}
\usepackage{natbib}
\usepackage{epstopdf}
\usepackage{mathrsfs}
\usepackage{slashed}
\usepackage{amsmath}
\usepackage{verbatim}
\usepackage{graphicx}
\usepackage{amssymb}
\usepackage{psfrag}
\usepackage{array}
\usepackage{lipsum}
\usepackage{float}
\usepackage[dvipsnames]{xcolor}
\usepackage[normalem]{ulem}

\newcommand{\bs}{\boldsymbol }

\newcommand{\rmd}{{\rm d}}

\def\pTo{{p_{\rm T,1}}}
\def\pTt{{p_{\rm T,2}}}

\newcommand{\be}{\begin{equation}}
\newcommand{\ee}{\end{equation}}
\def\u{{\bf u}}
\def\v{{\bf v}}

\def\aSC{{\kappa_{\rm sc}}}

\newcommand{\kt}[1]{{\bf  \textcolor{magenta}{(KT: #1)}}}

\title{Deep learning jet modifications in heavy-ion collisions}

\author{Yi-Lun Du,}
\emailAdd{yilun.du@uib.no}
\author{Daniel Pablos and}
\emailAdd{daniel.pablos@uib.no}
\author{Konrad Tywoniuk}
\emailAdd{konrad.tywoniuk@uib.no}
\affiliation{Department of Physics and Technology, University of Bergen, Postboks 7803, 5020 Bergen, Norway}

\abstract{
Jet interactions in a hot QCD medium created in heavy-ion collisions are conventionally assessed by measuring the modification of the distributions of jet observables with respect to the proton-proton baseline. However, the steeply falling production spectrum introduces a strong bias toward small energy losses that obfuscates a direct interpretation of the impact of medium effects in the measured jet ensemble. Modern machine learning techniques offer the potential to tackle this issue on a jet-by-jet basis. In this paper, we employ a convolutional neural network (CNN) to diagnose such modifications from jet images where the training and validation is performed using the hybrid strong/weak coupling model. By analyzing measured jets in heavy-ion collisions, we extract the original jet transverse momentum, i.e., the transverse momentum of an identical jet that did not pass through a medium, in terms of an energy loss ratio. Despite many sources of fluctuations, we achieve good performance and put emphasis on the interpretability of our results. We observe that the angular distribution of soft particles in the jet cone and their relative contribution to the total jet energy contain significant discriminating power, which can be exploited to tailor observables that provide a good estimate of the energy loss ratio. With a well-predicted energy loss ratio, we study a set of jet observables to estimate their sensitivity to bias effects and reveal their medium modifications when compared to a more equivalent jet population, i.e., a set of jets with similar {\emph {initial}} energy. Finally, we also show the potential of deep learning techniques in the analysis of the geometrical aspects of jet quenching such as the in-medium traversed length or the position of the hard scattering in the transverse plane, opening up new possibilities for tomographic studies. 
}

\begin{document}

\maketitle

\section{Introduction}
\label{sec:intro}

Jets are collimated sprays of hadrons and other particles which originate from the fragmentation of high energy quarks, gluons or highly boosted bosons. Due to color confinement, only colorless states can be observed in experiments. Thus, jets manifest fundamental and important signatures of the underlying quantum physics probed in collider experiments. In relativistic heavy-ion collisions, jets are modified by elastic and inelastic processes that take place during their passage through the hot and dense debris of the collision \cite{dEnterria:2009xfs,Majumder:2010qh,Mehtar-Tani:2013pia,Qin:2015srf}.

Historically, the jet quenching phenomenon has been primarily attributed to the observed strong suppression of intermediate-$p_T$ hadrons at the Relativistic Heavy-Ion Collider (RHIC)~\cite{adcox2001suppression,adler2002centrality} 
and years later via the dijet asymmetry and the suppression of high energy reconstructed jets at the Large Hadron Collider (LHC)~\cite{aad2010observation,chatrchyan2011observation,Abelev:2013kqa,Adam:2015ewa,aad2015measurements,Aaboud:2017eww,aaboud2019measurement,Acharya:2019jyg}. More recently, considerable efforts have been devoted to measuring the modifications of the internal properties of jets, generally referred to as jet substructure measurements~\cite{Chatrchyan:2013kwa,Chatrchyan:2014ava,Khachatryan:2015lha,Aad:2015bsa,Adam:2015doa,Khachatryan:2016tfj,Aaboud:2017bzv,Acharya:2017goa,Sirunyan:2017bsd,Aaboud:2018hpb,Sirunyan:2018jqr,Sirunyan:2018gct,Acharya:2018uvf,Sirunyan:2018ncy,Sirunyan:2018qec,Aad:2019igg,Sirunyan:2019dow,Aaboud:2019oac,Acharya:2019ssy,Acharya:2019djg,ALICE-PUBLIC-2020-006,CMS:2020hvy}. Such observables provide new constraints for the theory and modelling assumptions. Jet quenching has arguably come to serve as one of the most powerful experimentally accessible probes of the properties of the hot, deconfined QCD medium produced in heavy-ion collisions~\cite{PhysRevC.90.014909}. The ability to properly correlate the level of medium modifications to intrinsic properties of individually reconstructed jets will help enhance the potential of these probes to accurately diagnose properties of hot QCD medium, provided that the mechanisms by which jets interact with the medium are under good theoretical control.

When studying modifications of jet observables in both in proton-proton and heavy-ion collisions, it is typical to select jets within a certain $p_T$ range or at least above a minimum $p_T$. Most frequently, one presents the ratio of the distributions of such observables measured in the two colliding systems. However, such an imposed $p_T$ cut introduces a selection bias due to the steeply falling jet production spectrum, typically ${\rm d} \sigma_{\rm jet}/{\rm d} p_T \propto p_T^{-n}$, where $n\gtrsim 5$ at the LHC and even larger at lower collisions energies. It is indeed unlikely to observe jets above the $p_T$ cut which have lost a significant amount of energy,  simply because such more quenched jets had to be produced at a higher $p_T$, where the spectrum is increasingly suppressed. One strategy to mitigate part of these bias effects is to match the cumulative jet cross-sections in pp and AA collisions  \cite{Brewer:2018dfs,Takacs:2020Que,Brewer:2020chg}.

It is generally believed, both from theoretical arguments \cite{Kurkela:2014tla,Mehtar-Tani:2017web} and in most Monte Carlo implementations, that the momentum scales related to high-$p_T$ jet production are typically much larger than the local medium scales. This implies that hard radiation, occurring on short time-scales inside the jet, takes place independently of the medium. These initial emissions play an important role for the subsequent evolution of the jet, both considering the final jet properties and the amount of medium modifications that it can experience. One therefore expects that final, measurable jet properties, such as the jet width and fragmentation functions, correlate with the amount of energy that was lost in the medium. Considering at the same time that these properties could be modified during the passage through the medium, it becomes clear that the existence of the selection bias obscures the actual impact of medium effects in final observables and complicates the task of extracting robust information about medium properties.

In order to minimize selection biases and get a better handle on the medium effects, it would be desirable to discern whether a given jet has actually suffered energy loss, or even estimate the amount of energy loss it has experienced. This is a highly nontrivial task on a jet-by-jet basis. Even in vacuum, a jet population of a certain radius within a given momentum range can vary considerably due to the random nature of jet fragmentation. In the medium, randomly distributed path lengths through the medium introduce additional fluctuations. Both sources of fluctuations, adding on top of other kinds related, for example, to the process of hadronization, hinder our ability to identify the degree of medium modification of a specific jet in a heavy-ion collision. Unfolding these fluctuations to access information about the magnitude of jet-medium interactions for each individual reconstructed jet looks a priori as a daunting task.

Machine learning techniques have been widely applied in jet physics, such as QCD/W jet tagging~\cite{cogan2015jet,baldi2016jet,de2016jet,dreyer2018lund,louppe2019qcd}, top tagging~\cite{almeida2015playing,pearkes2017jet,kasieczka2017deep}, quark/gluon jet discrimination~\cite{komiske2017deep,cheng2018recursive} and heavy flavor classification~\cite{guest2016jet}. Various neural networks have been employed in tackling these issues, including convolutional neural networks (CNNs) with jet image as input~\cite{baldi2016jet,de2016jet,komiske2017deep,kasieczka2017deep} -- often supplemented by deep neural networks analyzing jet substructure observables for the purpose of interpretability \cite{kasieczka2017deep,Faucett:2020vbu} --, recurrent neural networks (RNNs) with the Primary Lund plane as input~\cite{dreyer2018lund}, recursive neural networks (RecNNs) with declustering history tree as input~\cite{louppe2019qcd,cheng2018recursive} and point cloud networks with an unordered set of jet constituent particles as input~\cite{qu2019particlenet}. First attempts of using machine learning techniques for jet physics in heavy-ion collisions include distinguishing between quark and gluon energy loss \cite{Chien:2018dfn}. 
These techniques match or outperform conventional physically-motivated features in the tasks above. It is therefore tempting to apply these machine learning techniques for diagnosing the jet quenching phenomenon on a jet-by-jet basis to help identify the features most sensitive to the process of energy loss. 

In this work we explore the power of deep learning techniques and study its feasibility to extract the energy loss information that individual inclusive jets experience from final jet measurable properties. The problem is formulated as a regression task with the objective to find the hadronic energy loss ratio, i.e., the ratio of the transverse momentum of a hadronic jet after traversing a hot QCD medium over the transverse momentum of its ``vacuum'' equivalent jet. This is, the energy of an identical jet that has not propagated through the medium. With the predicted energy loss ratio, we classify jets from an in-medium sample in a range of categories spanning from strongly quenched to almost completely unmodified. We analyze a set of jet observables, including groomed observables, e.g., the groomed momentum sharing fraction $z_g$, the groomed jet radius $R_g$ and Soft Drop multiplicity $n_{SD}$, and also ungroomed ones, such as the jet shape and jet fragmentation functions. We show how some of the properties of the jets from the unquenched class do not necessarily converge to the ones from the jets in a vacuum sample. This is due to the effect of the selection bias, which affects more strongly those observables that more intimately relate to the amount of energy loss experienced. Finally, an exploratory discussion on jet tomography assisted by deep learning is discussed in the end. 

Throughout this work, we put strong emphasis on interpreting our results by comparing various setups for extracting the energy loss ratio. This includes varying the inputs and network architectures. These additional attempts aid our understanding of what features of the jet allow the network to successfully extract key information. It turns out that the combination of hard, small-angle and soft, large-angle structures of the jet are crucial in obtaining the best performance. Furthermore, a main goal of our work is to examine the output of the machine learning in terms of human understandable variables. This goal is shared with the main body of classification studies in proton-proton collisions, which compare the performance between human defined observables \cite{baldi2016jet,Faucett:2020vbu}, and a neural network. In heavy-ion collisions, in lieu of a theoretically well motivated observable, or set of observables, to gauge the effects of energy loss, we instead attempt to construct such observables by studying the sensitivity of the network. First, we define an energy fraction of hard particles in a jet and, second, a non-linear combination of inputs from the jet shape, which are explored in Sec.~\ref{sec:interpret}. Neither construct resulted in a sensitivity on the same level as that of the full neural network. Nevertheless, this constitutes a first attempt at providing physical intuition on the problem of jet energy loss from machine learning.

The rest of the paper is organized as follows: in Section~\ref{sec:event-generation} we first explain the energy loss model used, matching procedure between vacuum and medium jets and introduce definitions of the relevant physical quantities and jet observables. Then we describe the regression task and detail the network architectures. Last, we generate the jet samples and assign sample weights for unbiased training. Section~\ref{sec:ML} contains information about the pre-processing steps undertaken to use the jet image as an input to the deep neural network architecture. The correlations between jet energy loss and jet observables are briefly presented. Next, we present the prediction performance of the CNN and compare the results from various scenarios, with different inputs and different networks, to discuss the interpretability of the CNN's efficiency. Some first applications of our procedure are proposed in Section~\ref{sec:applications}, where we study the response of jet observables to the amount of energy loss using two different jet selections, 
and also present the geometrical information that can be extracted due to its correlations with energy loss. Finally, Section~\ref{sec:Conclusions} summarizes the results and discusses the path ahead.

\section{General setup and main variables}
\label{sec:event-generation}

This section introduces three main concepts. Firstly, we describe the particular Monte Carlo event generator that was used to generate the analyzed jet images. Secondly, we discuss the main physical observables that will be used for the analysis. Finally, we describe the machine learning frameworks used in this work.

\subsection{Modeling energy loss using the hybrid model}
\label{sec:elossmodel}

Energetic partons produced in hard scatterings are created with a high initial virtuality, $Q\sim p_T$. The high virtuality is relaxed through successive splittings, as dictated by the DGLAP evolution equations. In the presence of a deconfined QCD medium, as the one created in heavy-ion collisions, these jets will interact with the degrees of freedom of the plasma, whose scale is typically characterised by the local temperature $T$. Given the wide scale separation between the two relevant scales of the jet-plasma system, i.e., $Q \gg T$, one can to a good approximation factorise the high virtuality showering process from the interaction with the plasma at lower energies. This assumption has been used in the development of jet quenching Monte Carlos where the interaction with the medium is described at weak coupling, such as MARTINI \cite{Young:2011ug}, LBT \cite{He:2015pra} (both available within the JETSCAPE framework~\cite{Putschke:2019yrg}), PYQUEN~\cite{Lokhtin:2005px} and Saclay model \cite{Caucal:2019uvr},\footnote{JEWEL \cite{Zapp:2008gi,Zapp:2012ak} also belongs to this category to a certain extent.} and also in those where the interaction with the medium is strongly coupled, such as the hybrid strong/weak coupling model \cite{casalderrey2015erratum}. In the present work we will analyse data from the hybrid strong/weak coupling model, leaving the extension to more models for future work. 

In the hybrid model, parton showering is described using the event generator PYTHIA \textcolor{black}{8.244} \cite{Sjostrand:2007gs}, supplemented with the nuclear PDF modifications from EPS09 \cite{Eskola:2009uj}. The spacetime picture of the parton shower is based on a formation time argument \cite{CasalderreySolana:2011gx}, such that each parton propagates through the QGP for a distance $t_{\rm f}\equiv 2E/Q^2$, with $E$ the energy of the parton and $Q$ its virtuality. The shower is then embedded in a heavy-ion environment. First, one selects the initial position of the hard scattering in the transverse plane through an optical Glauber sampling. Local properties of the QGP necessary to describe the interaction, such as the temperature $T(x)$ and fluid velocity $\u(x)$, \textcolor{black}{where $x$ is the four-vector that describes the space-time position of the travelling parton,} are read from hydrodynamic profiles that describe the expansion and cooling down of the liquid QGP droplet \cite{Shen:2014vra}. The strongly coupled interaction is modelled using an energy loss rate obtained within gauge/gravity duality for $\mathcal{N}=4$ supersymmetric Yang-Mills theory at large $N_c$ \cite{Chesler:2014jva,Chesler:2015nqz},
\be
\label{eq:CR_rate}
\left. \frac{\rmd E}{\rmd x}\right|_{\rm strongly~coupled}= - \frac{4}{\pi} E_{\rm in} \frac{x^2}{x_{\rm therm}^2} \frac{1}{\sqrt{x_{\rm therm}^2-x^2}}\quad , \quad \quad x_{\rm therm}= \frac{1}{2\aSC}\frac{E_{\rm in}^{1/3}}{T^{4/3}} \, ,
\ee
where $E_{\rm in}$ is the initial energy of the parton and $T$ is the local temperature of the plasma. The quantity $\aSC$, which depends on the 't Hooft coupling but whose precise expression varies depending on how the energetic parton is prepared in the holographic calculation, is taken as a  free parameter that is fit to hadron and jet suppression data \cite{Casalderrey-Solana:2018wrw}.
The results in Eq.~\eqref{eq:CR_rate} are derived in the local fluid rest frame. In order to take into account the effect from the flowing medium, we need to replace $E_{\rm in}$ and $x$ by their corresponding values in the local fluid rest frame, $E_{\rm in}^F$ and $x_F$, which one can express in terms of the quantities in the laboratory frame as \cite{Casalderrey-Solana:2015vaa}     
\begin{align}
E^F_{\rm in} &= E_{\rm in} \,\gamma_F \left(1-\v \cdot \u\right) \, , \label{eq:EF} \\
x_F(t)&=\int_{t_0}^{t} \rmd t'\, \sqrt{\v^2 + \gamma_F^2\big(\u^2 - 2 \u\cdot \v + (\u \cdot \v)^2\big)} \, , \label{eq:xF}
\end{align}
where $\v \equiv{\bs p}/E$ is the parton velocity, $\u$ and $\gamma_F$ are the fluid velocity and Lorentz factor, $t_0$ the time the parton 
was produced
and $t$ is the observation time. By following the branching history of a given parton $j$, we can compute the total length traversed through the plasma as 
\be
\label{eq:lengtheq}
L^j = \sum_{i \in H_j} x_F^i\big( \textrm{min}(t_{\rm f}^i, t_c^i \big)) \, ,
\ee
where the sum runs over the parent history $H_j$ of the given parton $j$, while $t_c^i$ is the time, since it was created, at which the parton $i$ exits the QGP phase by encountering a temperature below the pseudo-critical temperature $T_c =145$ MeV.
\footnote{In principle, there could also be quenching in the hadron gas phase, below $T_c$. This has so far been ignored by jet quenching models based on the general argument that hadrons take too long to form~\cite{Dokshitzer:1991wu}. Nevertheless, there are studies that point to the importance of this phase 
in a variety of observables, specially for low $p_T$ particles (see, e.g., \cite{Cassing:2003sb,Werner:2012sv,Bierlich:2018xfw,Dorau:2019ozd}), which are precisely the kind of hadrons that form the fastest and the ones more affected by further rescattering. The inclusion of these effects, whose implementation within the current state-of-the-art quenching models is still ongoing work, is left for the future.}

The amount of energy and momentum lost by the energetic parton, as described by Eq.~\eqref{eq:CR_rate}, exactly corresponds to the amount of energy and momentum flowing into the QGP hydrodynamic modes \cite{Chesler:2015nqz}. This will generate a wake that is correlated with the direction of the jet \cite{Chesler:2007an}, whose contribution to the experimentally observable jet properties has to be taken into account. The hybrid model provides an estimate of the wake contribution to the final hadron spectrum by performing an expansion of the Cooper-Frye formula at the perturbed freeze-out hypersurface, which yields \cite{Casalderrey-Solana:2016jvj}
\begin{equation}
\label{eq:onebody}
\begin{split}
E\frac{\rmd\Delta N}{\rmd^3p}=&\frac{1}{32 \pi} \, \frac{m_T}{T^5} \, \textrm{cosh}(y-y_j)  \exp\left[-\frac{m_T}{T}\, \textrm{cosh}(y-y_j)\right] \\
 &\times \Bigg\{ p_T \Delta P_T \cos (\phi-\phi_j) +\frac{1}{3}m_T \, \Delta M_T \, \textrm{cosh}(y-y_j) \Bigg\} \, ,
\end{split}
\end{equation}
where $p_T$, $m_T$, $\phi$ and $y$ are the transverse momentum, transverse mass, azimuthal angle and rapidity of the emitted thermal particles and where $\Delta P_T$ and $\Delta M_T=\Delta E/\cosh y_j$ are the transverse momentum and transverse mass transferred from the jet, with azimuthal angle and rapidity $\phi_j$ and $y_j$, respectively.
The distribution in Eq.~\eqref{eq:onebody} has been obtained by considering that the background behaves as a Bjorken flow, which only has \textcolor{black}{a} longitudinal expansion. Generalizing it to the case in which there is transverse expansion can modify such distribution, depending on the orientation of the jet with respect to the background radial flow components~\cite{Yan:2017rku,Tachibana:2020mtb,Casalderrey-Solana:2020rsj}. The consequences of these observations will be explored in the near future.

The partons that do not completely hydrodynamize are hadronized using the Lund string model included in PYTHIA \textcolor{black}{8.244}. The contributions from the hadrons of the wake, together with the fragmented hadrons, ensure event-by-event energy-momentum conservation.\footnote{The distribution in Eq.~\eqref{eq:onebody} can become negative, most notably in the direction opposite to the jet in the transverse plane. This reflects the absence of soft particles in such region of phase space compared to an unperturbed QGP background as a result to the boost experienced by the fluid cell due to the injection of momentum from the jet. In the present work we will ignore such negative contributions, since they would show up as negative energy pixels in the jet images used in Section~\ref{sec:jet-image} (one would need to devise a procedure to cancel out such negative contributions using particles from a real background which are close in momentum and configuration space, such as in~\cite{Casalderrey-Solana:2016jvj}, which we leave for future work).
It has been shown that their contribution to jet observables with relatively small jet radius, such as the one used in the present work, $R=0.4$, is almost negligible \cite{Pablos:2019ngg}, which guarantees that none of our conclusions will be affected by the omission of such contribution. A study of jets with a larger radius will be done in future publications.}

\subsection{Jet energy loss ratio \texorpdfstring{$\chi_{jh}$}{Xjh} and traversed path-length \texorpdfstring{$L$}{L}}
\label{sec:def-chiE}

The main goal of this work is to determine, on a jet-by-jet basis, the amount of energy loss, quantified through the variable
\be
\label{eq:chi-definition}
\chi_{jh} \equiv \frac{E_f^h}{E_i^h} \,,
\ee
suffered by jets due to the propagation through a hot and dense QCD medium. Here, the subscript ``$jh$'' refers to the energy of the \emph{jet} measured at \emph{hadronic} level. These jets are reconstructed with FastJet \textcolor{black}{3.3.1}\cite{Cacciari:2011ma} using the anti-$k_T$ algorithm \cite{Cacciari:2008gp} with reconstruction parameter $R=0.4$. In this definition, $E_f^h$ is the $p_T$ of a given jet in the presence of a medium, and $E_i^h$ is the $p_T$ of the \emph{same} jet had there been no medium.
This relies on a carefully devised matching procedure that is explained below. The variable $\chi_{jh}$ was carefully chosen for several reasons. On the one hand, it is well suited to gauge the energy shift on the level of observable particles, since $E_f^h$ is, in principle, directly measured in experiment. This helps mitigating the event generator bias mentioned earlier. On the other hand, we have also found that $\chi_{jh}$ is quite well approximated by the neural network compared to other possible quantities, as will be discussed extensively below. All needed information, including $E_i^{h}$, is readily available in the hybrid model, where each unmodified event is stored together with its medium-modified version.

Other jet quenching models, in which the vacuum evolution is factorized from the interactions with the medium, should also allow such a jet-by-jet correspondence. In this case, $E_i^h$ should be thought of as a measure of the $p_T$ of an \emph{equivalent} jet in vacuum, e.g., a jet with a similar $p_T$ in the cone before the stage where medium interactions are applied to the jet. 

In this work, we also consider the amount of QGP traversed by a specific jet. While it is not a quantity directly extracted by the neural network from the provided images, it is readily available from the numerical model used to generate the data. This quantity provides meaningful information that should be strongly correlated to the modifications and energy loss experienced by a jet. Given that the quantity in Eq.~\eqref{eq:lengtheq} refers to the length traversed by a single parton $i$, we construct the length traversed by a parton jet, $L$, from the $p_T$ weighted sum of the individual lengths of the jet constituents on partonic level $L_i$, as
\be
\label{eq:lengthjet}
L=\frac{\sum\limits_{i \in \rm{jet}} p_{T,i} L_{i}}{\sum\limits_{i \in \rm{jet}} p_{T,i}} \, .
\ee
This biases the extracted jet in-medium length to the one of the leading particle.

\subsection{Matching procedure}
\label{sec:matching-proc}
Given a quenched jet of energy 
\begin{equation}
p_T^{\rm jet}\big\vert_{{\rm PbPb}} \equiv E_f^h\,,
\end{equation}
(\textcolor{black}{at the} hadronic level), in order to find its vacuum partner we perform the following procedure:
\begin{enumerate}
    \item Extract the vacuum jets by clustering the list of vacuum hadrons, i.e., the hadrons one would obtain if there was no medium modifications. 
    \item Extract the medium jets by clustering the list of medium hadrons, which include the hadrons fragmented from the quenched parton shower as well as the hadrons from the wake.
    \item For each medium jet, get its vacuum partner by selecting the highest $p_T$ vacuum jet whose axis is within $\Delta R<0.4$ from the medium jet axis, where $\Delta R\equiv \sqrt{\Delta \phi^2+\Delta y^2}$.
\end{enumerate}
Finally, the $p_T^{\rm jet}$ of the corresponding vacuum jet, that has not experienced any medium modifications or energy loss, is identified as the initial jet energy, i.e.,
\begin{equation}
p_T^{\rm jet} \big\vert_{{\rm equivalent\, pp}} \equiv E_i^h \,,
\end{equation}
before quenching.

In order to identify the medium partonic jet that produced the medium hadronic jet under consideration, we need to carry out a matching procedure analogous to the one outlined above. We now need to match a given hadronic jet with one of the medium partonic jets, 
which are the jets reconstructed using the quenched partons that were not completely absorbed in the medium. 

\subsection{Observables}
\label{defobservables}
The set of jet observables used at different stages of this work is presented here. They are classified into groomed and un-groomed observables. Starting out with the un-groomed class of observables, these include:
\begin{itemize}
    \item Jet mass $M$, defined as $M^2\equiv \big(\sum_{i \in {\rm jet}} p_i^{\mu}\big)^2$.
    \item Differential jet shape $\rho(r)$ (JS), i.e., the transverse energy distribution as a function of the distance $r$ to the jet axis in the $\lbrace \eta,\phi \rbrace$ plane with $\eta$ the pseudorapidity and $\phi$ the azimuthal angles, defined as \cite{Chatrchyan:2013kwa,Sirunyan:2018ncy}
    \begin{equation}
    \label{eq:diff-jet-shape}
      \rho(r)\equiv \frac{1}{N_{\textrm{jets}}}\frac{1}{\delta r}\sum\limits_{\textrm{jets}} \frac{\sum\limits_{i \, \in \, r\pm\delta r/2}
	  p_T^{i,\textrm{track}}}{p_T^{\textrm{jet}}}  \, ,
    \end{equation} 
\textcolor{black}{where $N_\textrm{jets}$ is the number of jets samples and $\delta r$ is the specified interval in $r$.}
    \item Jet fragmentation function \textcolor{black}{$D(z)$} (JFF), i.e., the distribution of hadrons with an energy fraction $z$ with respect to the jet energy, defined as \cite{Aaboud:2017bzv,Aaboud:2018hpb} $z\equiv p_T^{\textrm{track}}/p_T^{\textrm{jet}}\, \textrm{cos}\, \Delta R$, with $\Delta R$ the distance to the jet axis in the $\lbrace y,\phi \rbrace$ plane. \textcolor{black}{It's defined as}
        \begin{equation}
    \label{eq:diff-jet-ff}
       \textcolor{black}{D(z)\equiv \frac{1}{N_{\textrm{jets}}}\sum\limits_{\textrm{jets}}\frac{\mathrm{d}N}{\mathrm{d}z}.}
    \end{equation} 
\end{itemize}

Groomed jet observables are those obtained from a jet after applying a so-called grooming procedure that removes soft and/or wide-angle particles from the jet. Grooming can be achieved through a set of techniques developed to reduce the sensitivity to the soft, non-perturbative components of a jet, with the intention of gaining access to the partonic, perturbative aspects of jet substructure,
see \cite{Marzani:2019hun} for extensive reviews. Focusing on the so-called Soft Drop (SD) procedure \textcolor{black}{\cite{Larkoski:2014wba}} in the following, this procedure consists in looking into the clustering history of a jet following the hardest branch of the clustered pair of sub-jets, analyzing any number of steps that satisfy the SD condition, defined as
\be
\label{eq:Soft DropCondition}
z_g \equiv \frac{{\rm min}  \left(\pTo,\, \pTt\right)}{\pTo+\pTt} > z_{\rm cut} \left(\frac{R_g}{R}\right)^\beta \, ,
\ee
where $R$ is the jet radius, $\pTo$ and $\pTt$ are the momentum of the two subjet branches under consideration, $z_{\rm cut}$ and $\beta$ are grooming parameters, while $z_g$ and $R_g$ are the groomed momentum sharing fraction and groomed angular separation in $\lbrace y,\phi \rbrace$ plane  between the branches, respectively. The most common setup explored in the literature corresponds to the reclustering of a given anti-$k_T$ jet with the angular ordered Cambridge/Aachen (C/A) algorithm \cite{Dokshitzer:1997in,Wobisch:1998wt}, and setting $\beta=0$ and $z_{\rm cut}=0.1$. This will also be the setup adopted in this work, leaving the analysis of other choices for the grooming parameters or the use of alternative grooming scenarios, such as the recently developed dynamical grooming \cite{Mehtar-Tani:2019rrk}, for future publications.

The groomed observables we will use throughout the paper are:
\begin{itemize}
    \item The groomed momentum sharing fraction $z_g$, as defined in \eqref{eq:Soft DropCondition}. It refers to the first declustering step that satisfies the SD condition, unless explicitly stated otherwise.  
    \item The groomed jet radius $R_g$, i.e., the angle between the two sub-jets that satisfy \eqref{eq:Soft DropCondition}. Referring to the first declustering step that satisfies the SD condition, unless explicitly stated otherwise. 
    \item The SD multiplicity $n_{\rm SD}$, i.e., the number of times a given jet satisfies the condition \eqref{eq:Soft DropCondition} along the hardest branch during the declustering procedure.
    \item The groomed jet mass $M_g$, i.e., the 
    sum of the four-momenta of the first two sub-jets that pass the SD condition \eqref{eq:Soft DropCondition}, $M_g^2 \equiv (p_1^\mu + p_2^\mu)^2$.
\end{itemize}

We will present results for these observables in Section~\ref{sec:observables}, and will show its sensitivity to the amount of energy loss $\chi_{jh}$ 
as well as the physics behind their modifications with respect to the vacuum, or the absence of these. 

\subsection{Network architectures and task description}
\label{sec:neural-arch}

In this work we use two conventional machine learning architectures in order to extract the energy loss fraction $\chi_{jh}$ on a jet-by-jet basis. These are the fully-connected neural network (FCNN) and the convolutional neural network (CNN). The former architecture takes as arguments a set physically motivated observables \cite{baldi2016jet} and will mainly be employed as a check on the results obtained via the latter method. We will come back to describe its details in Sec.~\ref{sec:interpret}. Since most of our results will rely on the image recognition capabilities of the CNN, we will describe it in more detail below.

\begin{figure*}[t!]
\centering
\includegraphics[width=0.9\textwidth]{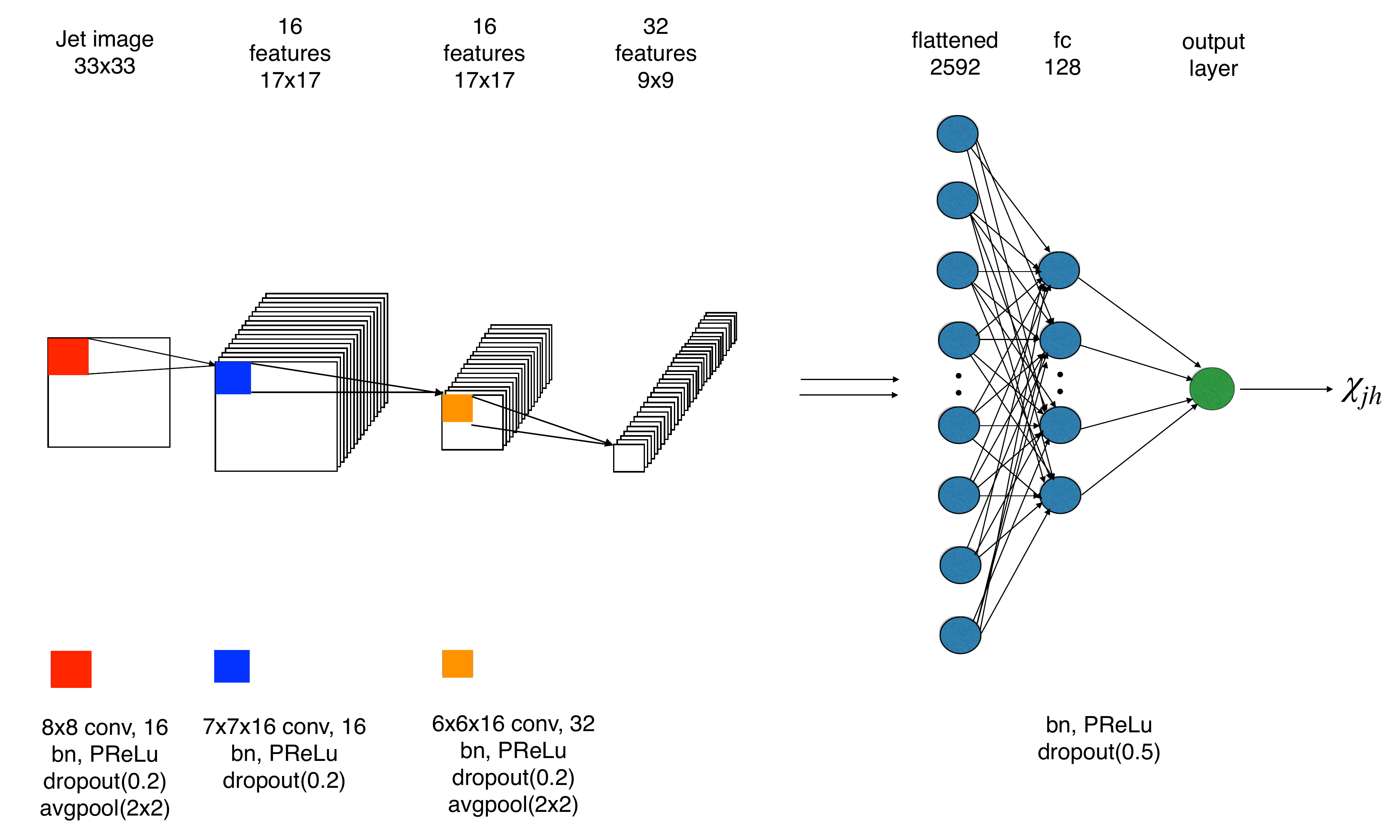} 
\caption{The architecture of our convolution neural network (CNN) for predicting the energy loss ratio $\chi_{jh}$ from pre-processed jet image with 33 pseudorapidity $\eta$ bins and 33 azimuthal angle $\phi$ bins.} 
\label{fig:cnn}
\end{figure*}

The CNN architecture used in this work is similar to that discussed in Ref.~\cite{du2020identifying,pang2018equation}. We refer to these papers for technical details. Fig.~\ref{fig:cnn} shows the neural network architecture. We use three convolutional layers and one subsequent fully-connected layer. All the convolutional layers and the fully-connected one are followed by a batch normalization~\cite{ioffe2015batch}, PReLu activation~\cite{he2015delving}, dropout~\cite{srivastava2014dropout} (with a rate of 0.2 and 0.5, respectively) and average pooling (of pool size $2\times2$, following first and third convolutional layers) layer, one by one. There are 16, 16, 32 filters of size $8\times8$, $7\times7$ and $6\times6$, respectively, in these three convolutional layers, scanning through the input $J(\eta,\phi)$, or the previous layers, and creating 16, 16, 32 features of size $33\times33$, $17\times17$, $17\times17$, respectively. The weight and bias matrix of these convolutional kernels and dense layers are initialized with ``He normal" initializer~\cite{he2015delving}, i.e., truncated normal distribution with zero mean and standard deviation $\sqrt{2 /N_\mathrm{in}}$ with $N_\mathrm{in}$ the number of input units in the weight tensor. They are constrained with L2 regularization~\cite{ng2004feature} in the loss function. Each neuron in a convolutional layer does connect only locally to a small chunk of neurons in the previous layer by a convolution operation. This is a key reason for the success of the CNN architecture. Dropout, batch normalization, PReLU layers and L2 regularization, all work together to prevent overfitting, which would generate model-parameter-dependent features from the training dataset and thus hinder the generalizability of the method. The resulting 32 features of size $9\times9$ from the last average pooling layer are flattened and connected to a 128-neuron fully-connected layer. The output layer is another fully-connected layer with one special neuron indicating the energy loss ratio $\chi_{jh}$. There are overall 395789 trainable and 134 non-trainable parameters in the present neural network.

The supervised learning is performed in tackling this regression task. The difference between the true label and the predicted label from the single output neuron is quantified by the Log-Cosh loss function, log(cosh($x$)), which is approximately equal to $x^2$/2 for small $x$ and to abs($x$) - log(2) for large $x$. The loss is function of the trainable parameters $\theta$ of the neural network. The training minimizes the loss function $l(\theta)$ by updating $\theta\to\theta-\delta\theta$. Here $\delta\theta=\alpha\partial l(\theta)/\partial\theta$, where $\alpha$ is the learning rate, with initial value 0.0001, which is adaptively changed by the AdaMax method~\cite{kingma2014adam}.

The architecture is built by Keras \textcolor{black}{2.3.1} with a TensorFlow backend and the training is performed with Google Colab GPUs. The training datasets are fed into the network in batches with an empirically selected size of 1024. One traversal of all the batches in the training datasets is called one epoch. The training datasets are reshuffled before each epoch to speed-up the convergence. The neural network is trained with 400 epochs. The model parameters are saved to a new checkpoint whenever a smaller validation loss is encountered.

The fully-connected neural networks used in this work, despite inputs of different size, consist of two hidden dense layers of size 128 and 32, respectively, which are initialized with ``He normal" initializer and constrained with L2 regularization. Each dense layer is followed by a dropout (with a rate of 0.2) and PReLu activation layer. 

\subsection{Jet sample generation and re-weighting procedure}
\label{sec:re-weighting}

Inclusive jet samples are generated from \textcolor{black}{approximately 400,000 hybrid model events} using $\hat{p}_{T,{\rm min}}$ = 50 GeV at $\sqrt{s}=5.02$ ATeV with an oversampling factor of the hard cross section of $p_T^4$ to obtain sufficient statistics at high $p_T$. The heavy-ion samples correspond to PbPb collisions in the 0-5\% centrality bin, with an average temperature of $T \simeq 250$ MeV. 
Reconstructed jets with anti-$k_T$ and $R=0.4$ are required to be within $|\eta|<2$ and to have momentum $100\,\, \mathrm{GeV}< p_T^{\mathrm{jet}} < 2000\,\, \mathrm{GeV}$.  \textcolor{black}{Within the aforementioned cuts, we get approximately 250,000 jets}. 80\% of these will serve as training samples and the rest 20\% will serve as validation samples which will not be fitted by the network in the training process.

\begin{figure}[t]
\centering
\includegraphics[width=0.48\textwidth]{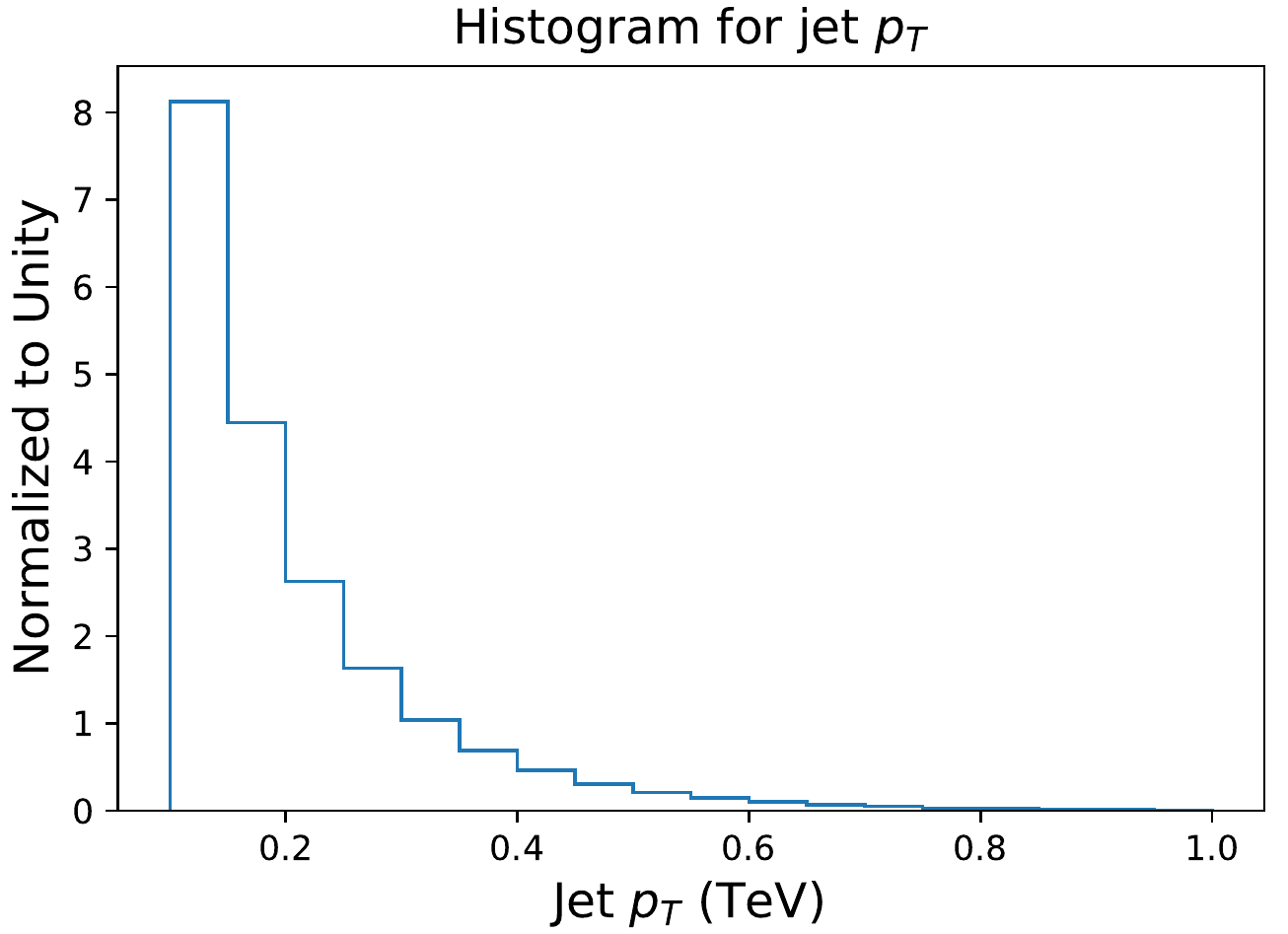}
\includegraphics[width=0.48\textwidth]{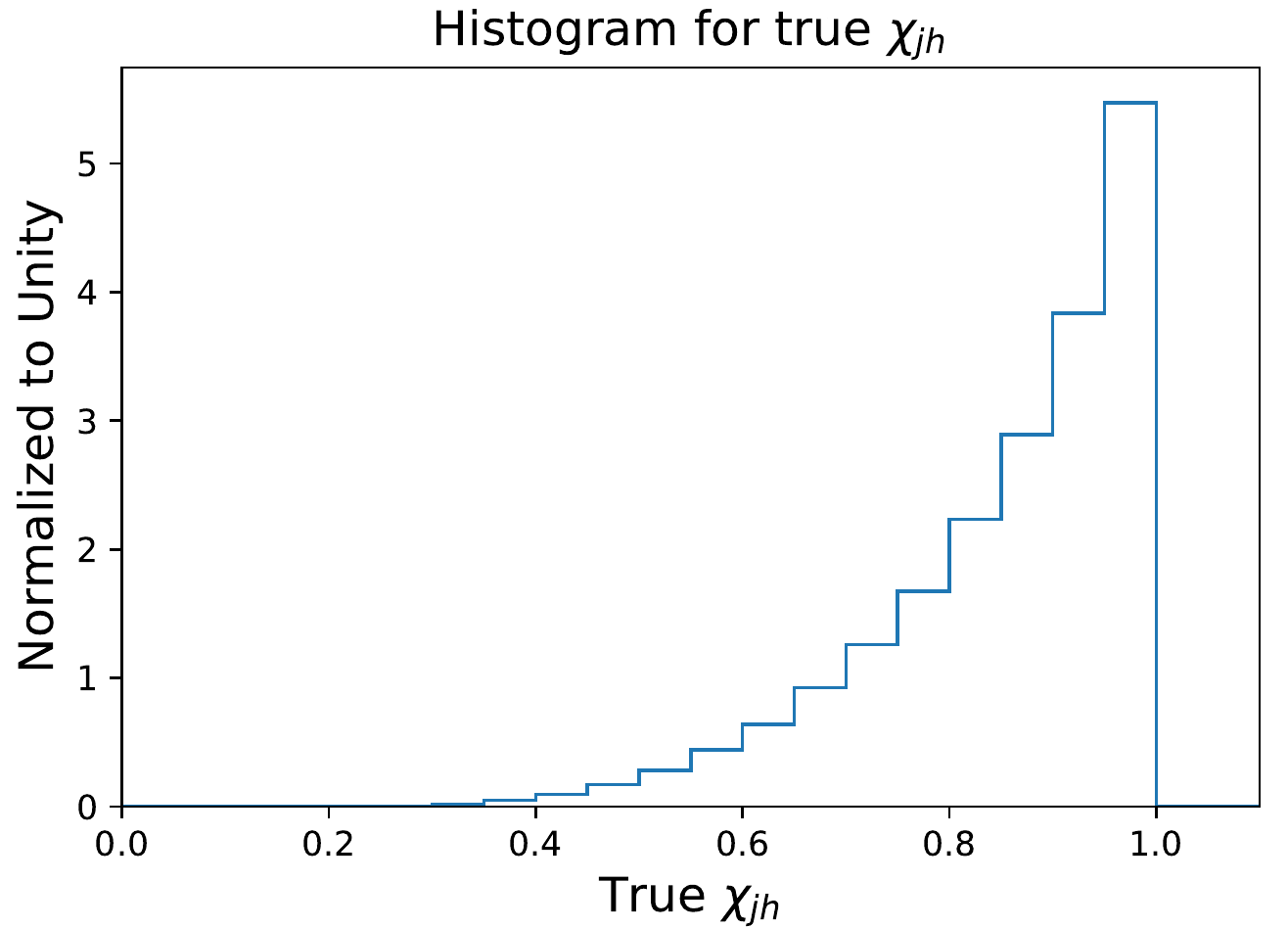}
\caption{The normalized histograms for jet $p_T$ after oversampling (left) and $\chi_{jh}$ samples (right) from the generated MC samples.} 
\label{fig:steeply-falling}
\end{figure}
Given the wide range of jet momenta studied in this work, it is important to ensure that the most common features of the events, such as the shape of the jet spectrum or the typical energy loss fraction, do not introduce any bias in our results. From the point of view of training the neural networks, it is desirable to deal with flat distributions. Since the jet spectrum is steeply falling, in order to obtain enough statistics at higher $p_T$ we use the oversampling method, consisting in multiplying the hard cross section by a power of the $p_T$ involved in the hard process, and weighting the event accordingly at the moment of analysis.
It is hard, however, to obtain flat distributions merely with this procedure, as one can see in the top left panel of Fig.~\ref{fig:steeply-falling} (left). Moreover, the energy loss ratio $\chi_{jh}$, the main object of interest for us, also presents a very non-flat distribution, which could again lead to biases in our results, cf. Fig.~\ref{fig:steeply-falling} (right). 

\begin{figure}[t]
\centering
\includegraphics[width=0.49\textwidth]{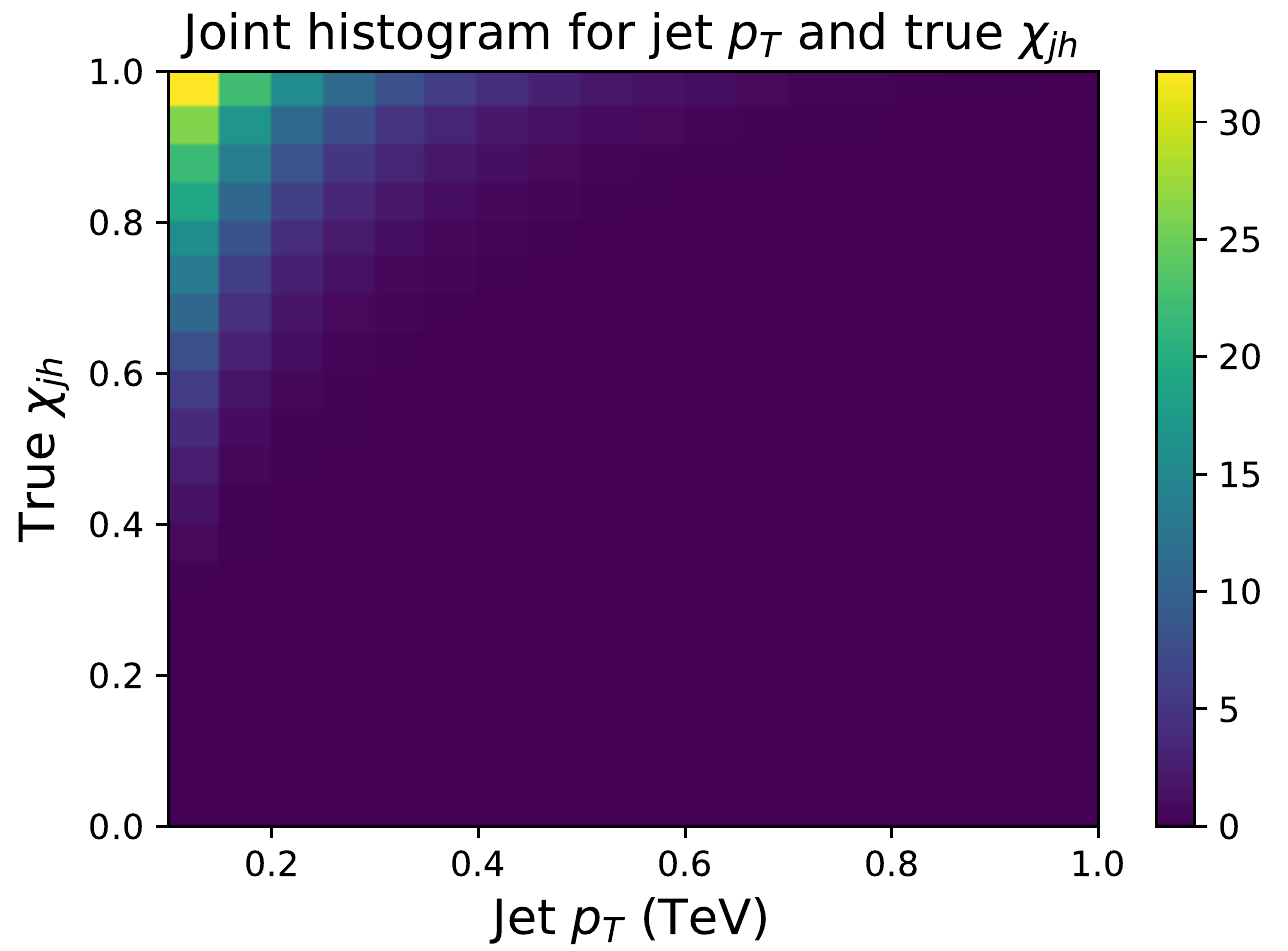}
\includegraphics[width=0.49\textwidth]{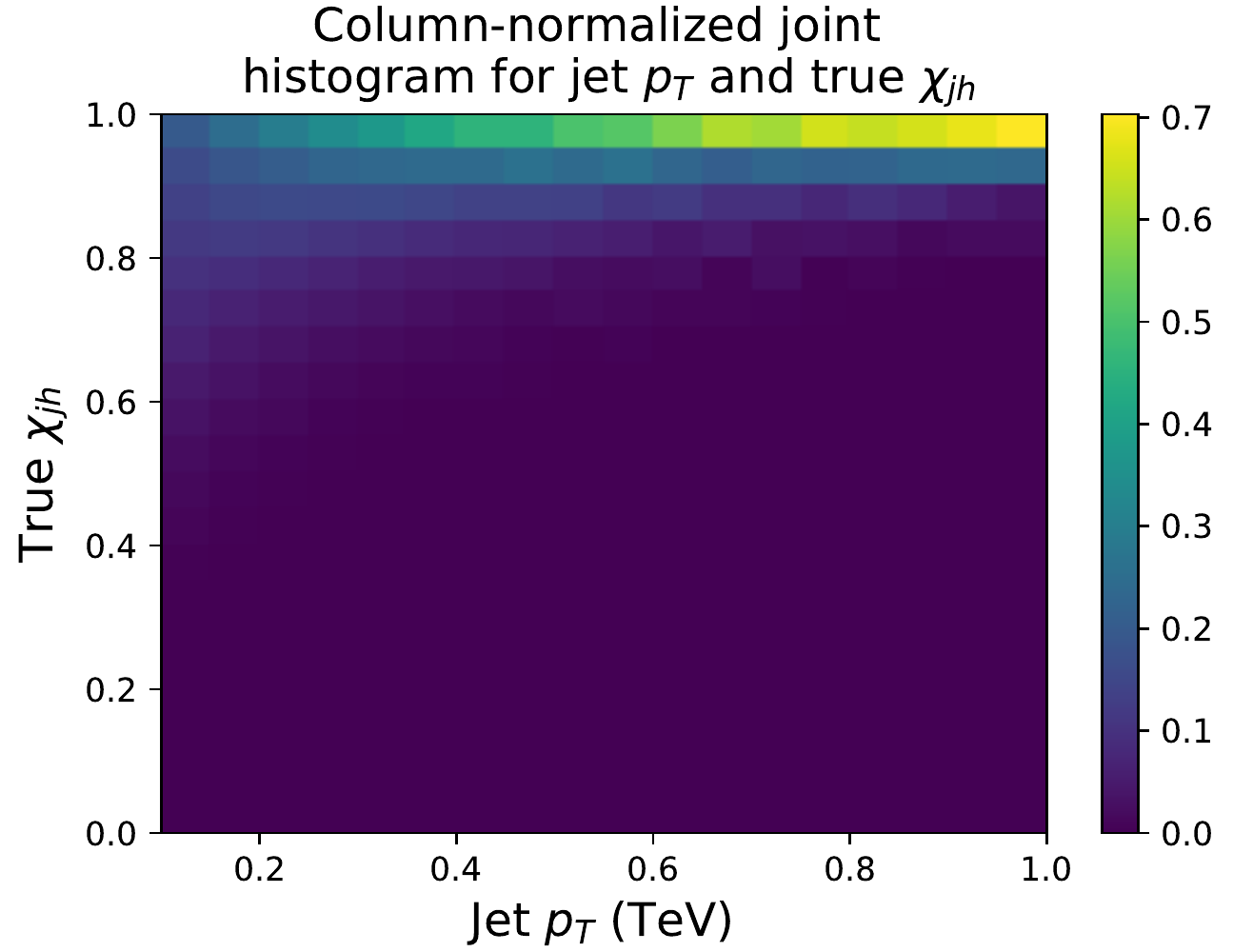}
\caption{The normalized joint histogram (left) and column-normalized joint histogram (right) for jet $p_T$ and energy loss fraction $\chi_{jh}$.}
\label{fig:joint-histogram-pt-chi}
\end{figure}
The normalized joint $(p_T,\chi_{jh})$ distribution, shown in Fig.~\ref{fig:joint-histogram-pt-chi} (left), clearly visualizes that most of the analyzed jets are at low $p_T$ and have lost little energy, $\chi_{jh} \approx 1$. Normalizing the joint distribution in a column-wise fashion, cf. Fig.~\ref{fig:joint-histogram-pt-chi} (right) reveals rather that the typical $\chi_{jh}$ for a given jet $p_T$ gets increasingly more peaked at high $p_T$. This means that high-$p_T$ jets tend to lose less energy compared to their jet $p_T$. Such behavior is expected whenever the amount of energy loss is weakly dependent on the initial jet energy \cite{Baier:2001yt} and will be discussed in detail below. Hence, providing the bare, joint distribution to the network could bias the performance of jet samples with low $p_T$ and large $\chi_{jh}$.

\begin{figure}[t]
\centering
\includegraphics[width=0.48\textwidth]{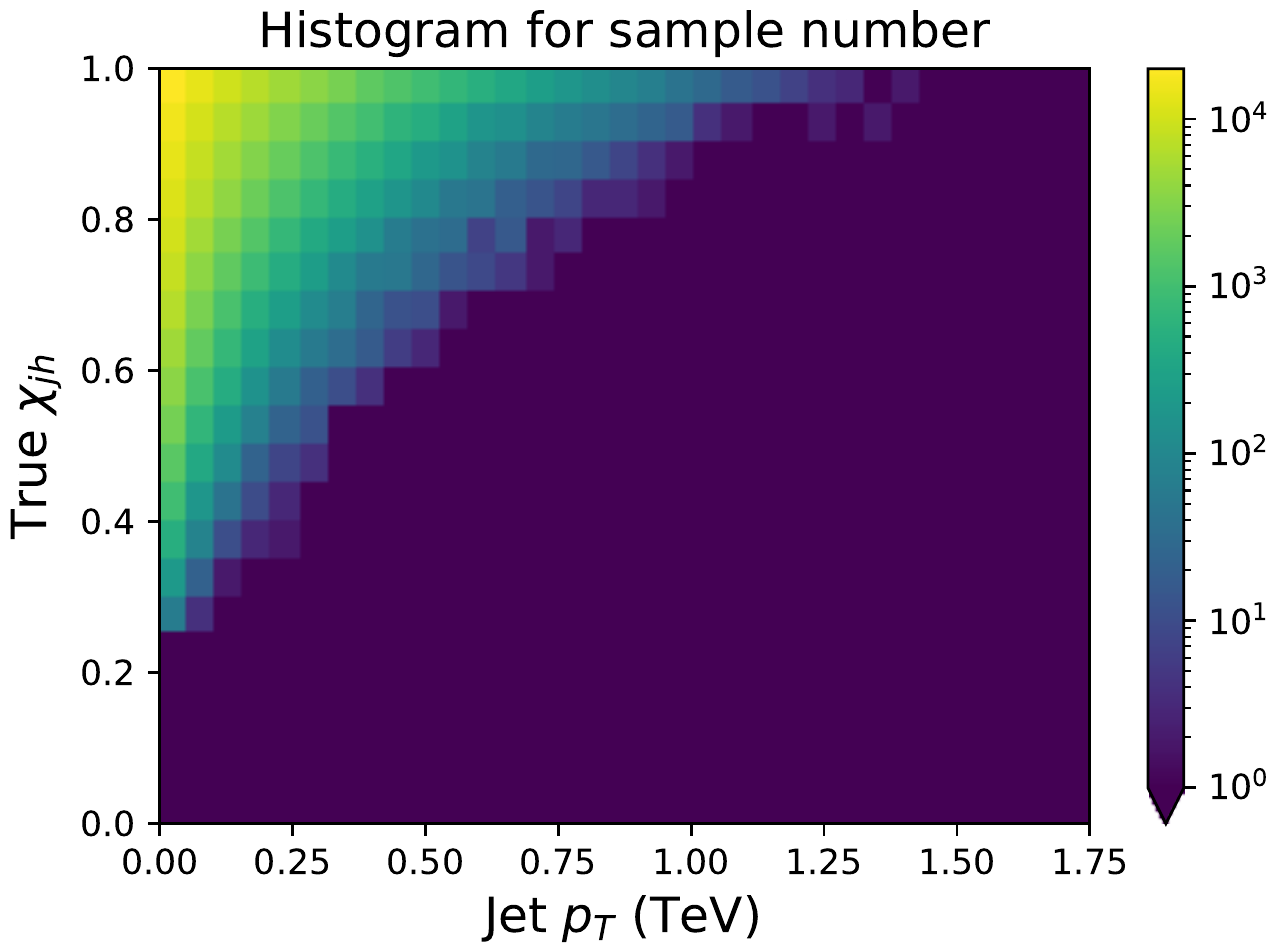}
\includegraphics[width=0.48\textwidth]{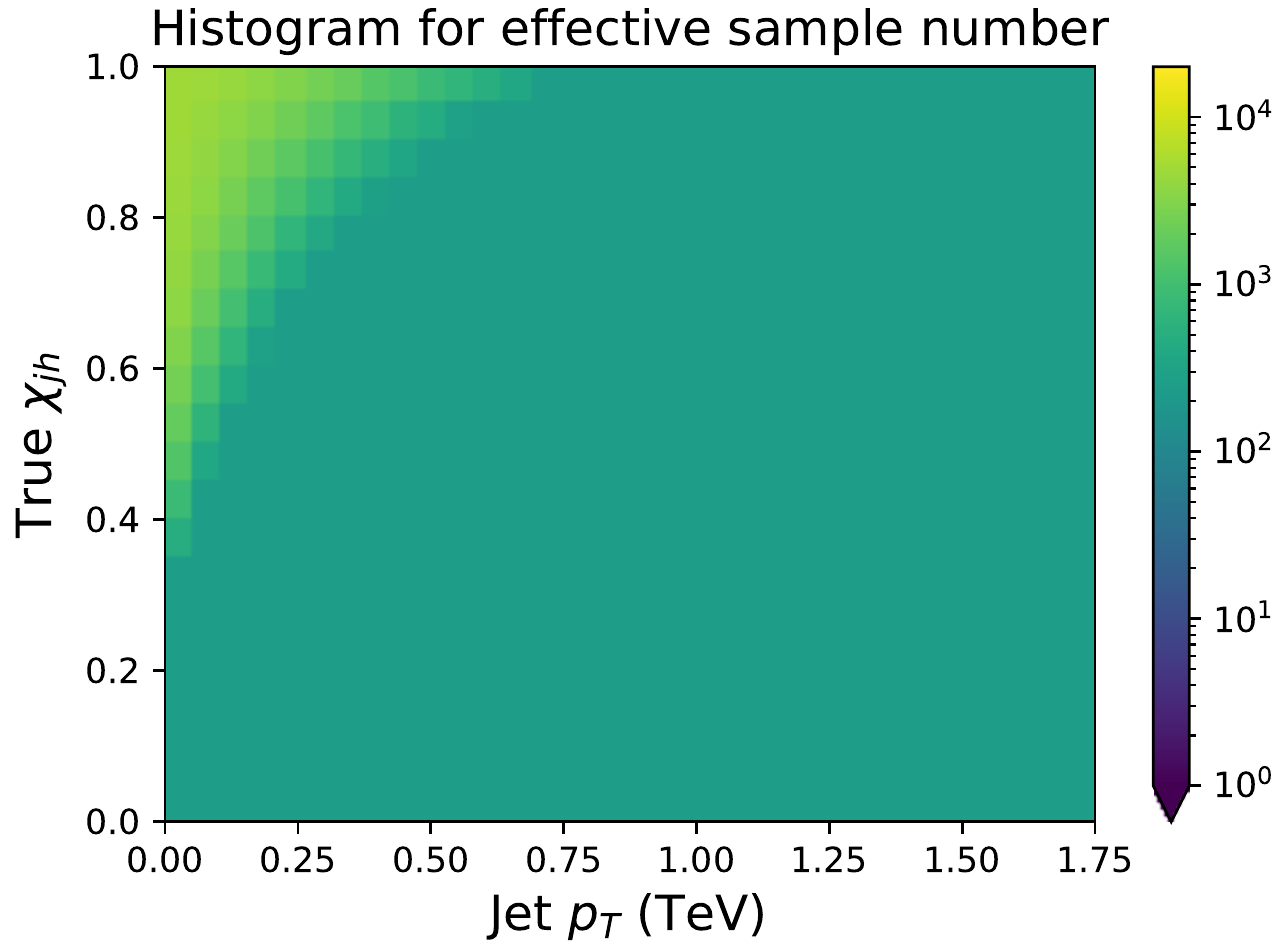}
\caption{Sample number (left) and effective sample number (right) in 2-D joint histogram of $p_T$ and $\chi_h$.} 
\label{fig:Effective sample number}
\end{figure}

\textcolor{black}{We will address this problem by assigning each sample a weight in the loss function in the training and validation~\cite{cui2019class}. The aim is to obtain a relatively flat 2-D ($p_T$, $\chi_{jh}$) joint distribution after re-weighting.
This is done in the following way. The weight of each sample is inversely proportional to the effective sample number $N_\mathrm{eff}$~\cite{cui2019class} of the ($p_T$, $\chi_{jh}$) bin which that particular sample belongs to. 
The effective sample number in a certain bin is $N_\mathrm{eff}= (1-\beta^N)/(1-\beta)$, where $N$ is the total sample number in that bin and $\beta$ is the probability that a new sample in that bin is independent of the previous samples. 
For example, the $j$th sample in that bin contributes $\beta^j$ to the effective sample number. In this work we set $\beta=0.9998$ and, therefore, the effective number of samples in each bin is limited to maximally 5000. This choice of the value of $\beta$ results in a high resolution of the sample number between bins.}

\textcolor{black}{On the other hand, in order to restrict the variation of the magnitude of sample weights in the loss function and to avoid biasing the training by the dominant samples in a training batch of $\mathcal{O}(10^3)$ samples, the sample weight is restricted to be smaller than 20 times of the smallest sample weight. In other words, the effective sample number is restricted to be bigger than 1/20 of the biggest effective sample number, leading to a minimum effective sample number of 250. We see the effect of the re-weighting procedure by looking at the joint $(p_T,\chi_{jh})$ histogram for the (effective) sample number in Fig.~\ref{fig:Effective sample number} where the left (right) plot shows the histogram before (after) the re-weighting (note the log-scale on the z-axis in these histograms).}

\section{Jet image analysis}
\label{sec:ML}

This Section first introduces the input to the network, i.e., jet image, and the pre-processing procedure applied on it. We present the average of pre-processed jet image for different $\chi_{jh}$ ranges, respectively, and the correlations between $\chi_{jh}$ and  each pixel of jet images as well as jet observables to hint at the possibility of extracting, jet-by-jet, the amount of suffered energy loss from measurable jet properties. Next, we present our main result on the prediction performance of $\chi_{jh}$ and explore the robustness of the performance and interpret the success of the prediction made by the algorithm.

\subsection{Jet image and pre-processing}
\label{sec:jet-image}

The input to the neural network is a so-called jet image $J(\eta,\phi)$. It represents the total $p_T$ of jet constituents deposited in the pixel of $( \eta, \phi)$ space with 33 $\eta$-bins and 33 $\phi$-bins. \textcolor{black}{Since we use a fixed jet radius $R=0.4$, it's natural to have pseudorapidity $|\eta|\leq 0.4$ and $|\phi|\leq 0.4$ in our jet images.}

In general, training algorithms may benefit from pre-processing of the datasets. The input here, jet image $J(\eta,\phi)$, is a $33\times33$ matrix. Following Refs.~\cite{de2016jet,komiske2017deep}, the jet image is first pre-processed by translation: the hardest groomed subjet is at $(\eta, \phi) = (0, 0)$. Then the rotation of the jet image around the center is applied so that the second hardest groomed subjet is at $-\pi/2$. If the second hardest groomed subjet does not exist, then the jet image is rotated by aligning the first principal component of pixel intensity distribution of jet image along the vertical axis.
The final step is a parity flip such that the right side of jet image has a larger pixel intensity sum.

\begin{figure}[t!]
\centering
\includegraphics[width=1\textwidth]{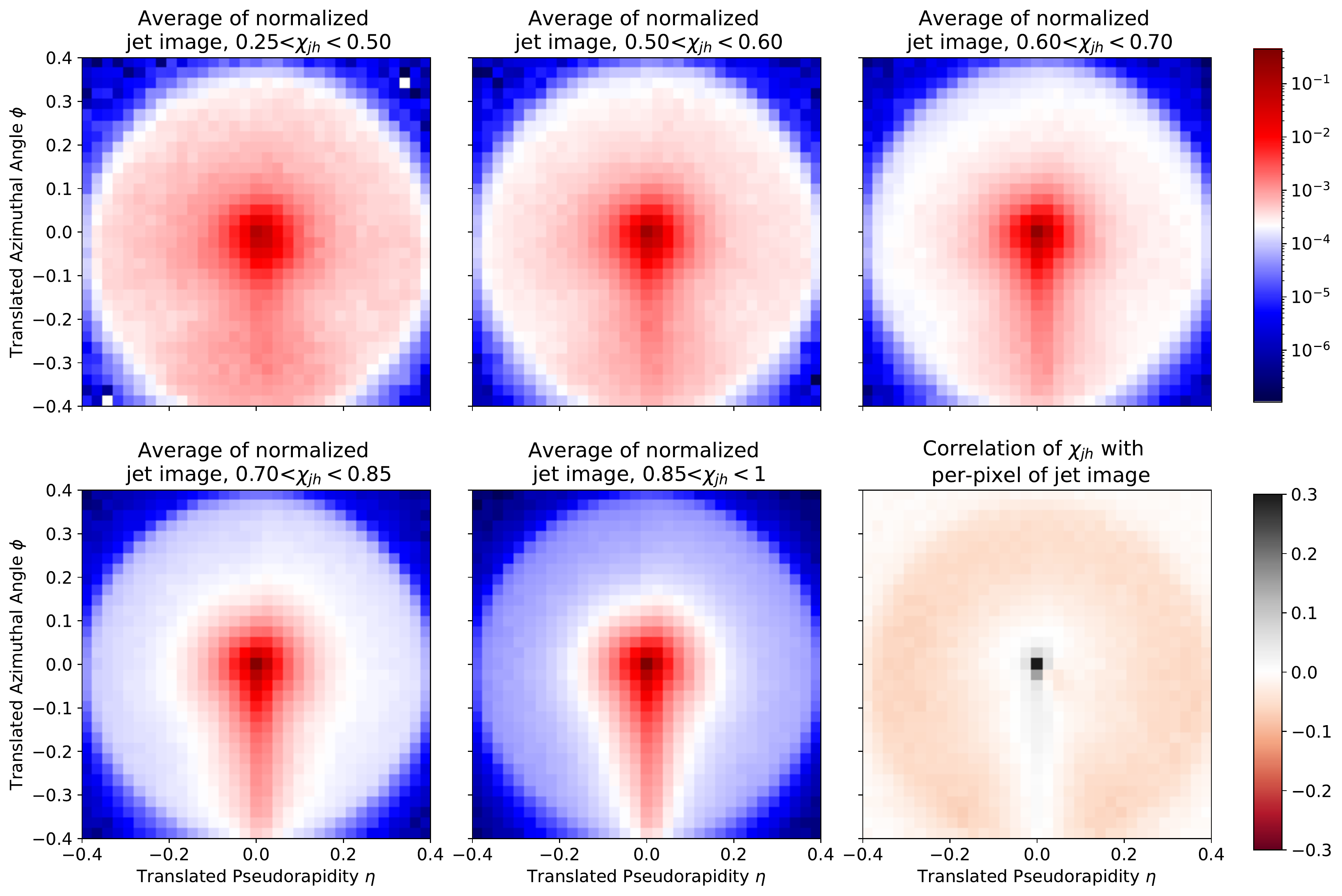}
\caption{The average of jet image normalized by jet $p_T$ within \textcolor{black}{5} different $\chi_{jh}$ cut bins \textcolor{black}{(3 in the top row and 2 in the bottom row, respectively)} and Pearson correlation coefficient of $\chi_{jh}$ with per-pixel of the \textcolor{black}{unnormalized} jet image \textcolor{black}{(rightmost panel in the bottom row)}.}
\label{Jet image}
\end{figure}

In Fig.~\ref{Jet image}, we show the average jet image normalized by jet $p_T$ for different ranges of values of $\chi_{jh}$. We can easily recognize the general features of jet quenching phenomenology, namely that quenched objects (top left, most quenched) present a larger number of softer particles at larger angles than unquenched ones (bottom center, least quenched). The amount of soft radiation also smears the hard prong structure, which appears vertically below the core pixel, corresponding to the next-to-hardest subjet in the jet. For the most quenched sample, see Fig.~\ref{Jet image} top left, the jets are rotationally invariant. The bottom rightmost plot in Fig.~\ref{Jet image} visualizes the correlation of each pixel of the jet image (not the normalized one) with the jet energy loss ratio $\chi_{jh}$ and will be explained in next Subsection.

Besides the above pre-processing steps of jet image, some widely used pre-processing methods in computer vision, standardization of image, could be applied. We refer to each pixel of the jet image as one ``feature" and each jet image as one ``sample". The pre-processing of the jet image could be in a feature-wise (per pixel) or sample-wise (per image) manner.

In the feature-wise standardization, the jet images $J(\eta,\phi)$ of all the training samples are pre-processed in a sample-interdependent manner. Each feature is subtracted with the mean over all training samples and is divided by their standard deviation. In this way, all features are centered around zero and have variances of the same order. Thus it is prevented that one feature with larger variance dominates the objective function over other features. The transformation is saved and then will be applied to the testing samples. We refer to such transformed jet images as ``feature-wise standardized jet image" in the following discussion.

In the sample-wise standardization, or min-max normalization, the jet images $J(\eta,\phi)$ are pre-processed in a sample-independent manner. The pixels of each jet image are rescaled either to have a zero mean and a unit variance, or to a specific range, such as $[0, 1]$. In this work we use the jet image normalized by the jet $p_T$ as an example of the sample-wise pre-processing method.

In this work, we will use the raw jet image, the pre-processed jet image with only translation, rotation and flipping (referred to as ``jet image" in the following unless explicitly stated otherwise, which are mostly used), the feature-wise standardized jet image and the jet image normalized by jet $p_T$, respectively, as inputs to the neural network to probe the impact of these pre-processing methods.

\subsection{A first look at correlations}
\label{sec:correlations}

\begin{figure}[t]
\centering
\includegraphics[width=0.95\textwidth]{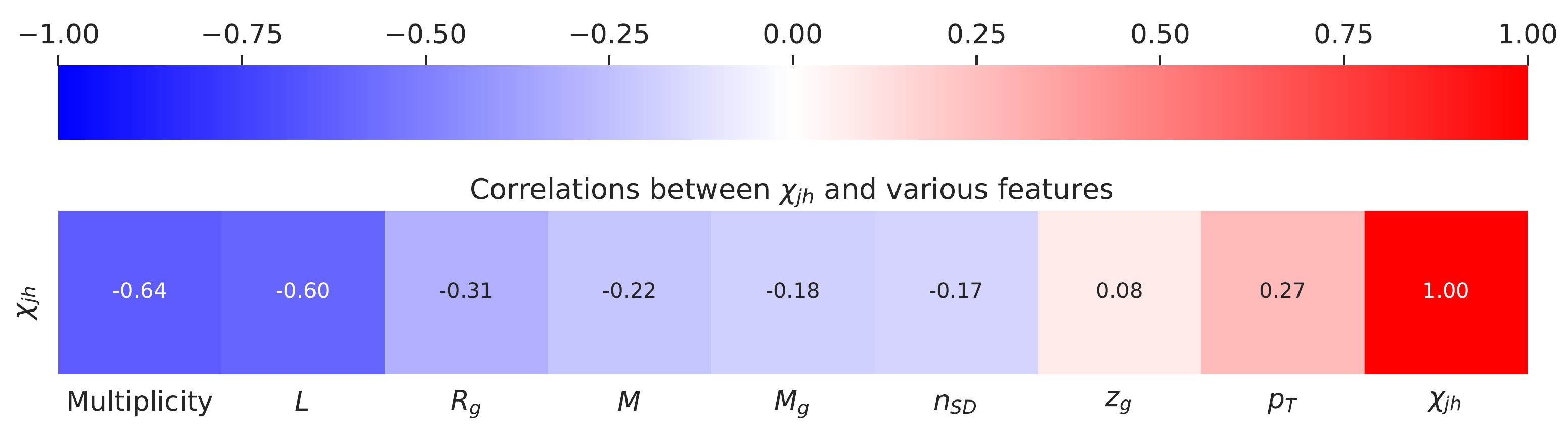}
\caption{Pearson correlation coefficients between $\chi_{jh}$ and jet observables}
\label{fig:correlations}
\end{figure}

In the bottom rightmost panel of Fig.~\ref{Jet image}, we show the Pearson correlation coefficient between $\chi_{jh}$ and each pixel of the jet image (not the normalized one). The Pearson correlation coefficient between samples $x$ and $y$ with $n$ population is defined as 
\be
r_{xy} =\frac{\sum ^n _{i=1}(x_i - \bar{x})(y_i - \bar{y})}{\sqrt{\sum ^n _{i=1}(x_i - \bar{x})^2} \sqrt{\sum ^n _{i=1}(y_i - \bar{y})^2}}
\ee
The value of the coefficient $r$ varies in the range $r\in [-1,1]$. A value of 1 means total positive linear correlation, 0 means no linear correlation, and -1 means full  linear anti-correlation. Indeed, larger values of $\chi_{jh}$ are characteristic of those jets that retain most of its energy in the hard structures at the main subjets. The anti-correlation between $\chi_{jh}$ and the soft, large angle particles in the jet cone illustrates that the energy is taken away from the leading particle(s) and spread to large angles within the jet.

In Fig.~\ref{fig:correlations}, the Pearson correlation coefficients between $\chi_{jh}$ and the set of chosen of observables as well as the physically immeasurable quantity $L$ are shown in ascending order. 
As expected, one can see that $\chi_{jh}$ is strongly anti-correlated with the jet traversed length $L$ in the QGP. The energy loss ratio is also strongly anti-correlated with the jet multiplicity, while it is slightly anti-correlated with jet mass $M$ and the groomed substructure observables $R_g$, $M_g$ and $n_{SD}$, in decreasing order. Its correlation with $z_g$ is very slight. 

We observe that $\chi_{jh}$ is slightly correlated with the (final, quenched) jet $p_T$, which is also demonstrated in Fig.~\ref{fig:joint-histogram-pt-chi} (right). Such correlation arises mainly from the fact that $\chi_{jh}$ is a relative energy loss, so that at high $p_T$ the value of $\chi_{jh}$ increases for a fixed value of absolute energy loss $\Delta E$. Another reason is that for higher (final, quenched) $p_T$ it becomes unlikely to produce low values of $\chi_{jh}$; such a jet should have started with a very large $p_T$, close to the kinematical limit, where the spectrum dies off. 

The presence of the correlations briefly discussed here clearly hint at the possibility of extracting, jet-by-jet, the amount of suffered energy loss $\chi_{jh}$ from measurable jet properties. 

%

\subsection{Prediction performance}
\label{sec:robustness}

\begin{figure}[t]
\centering
\includegraphics[width=0.48\textwidth]{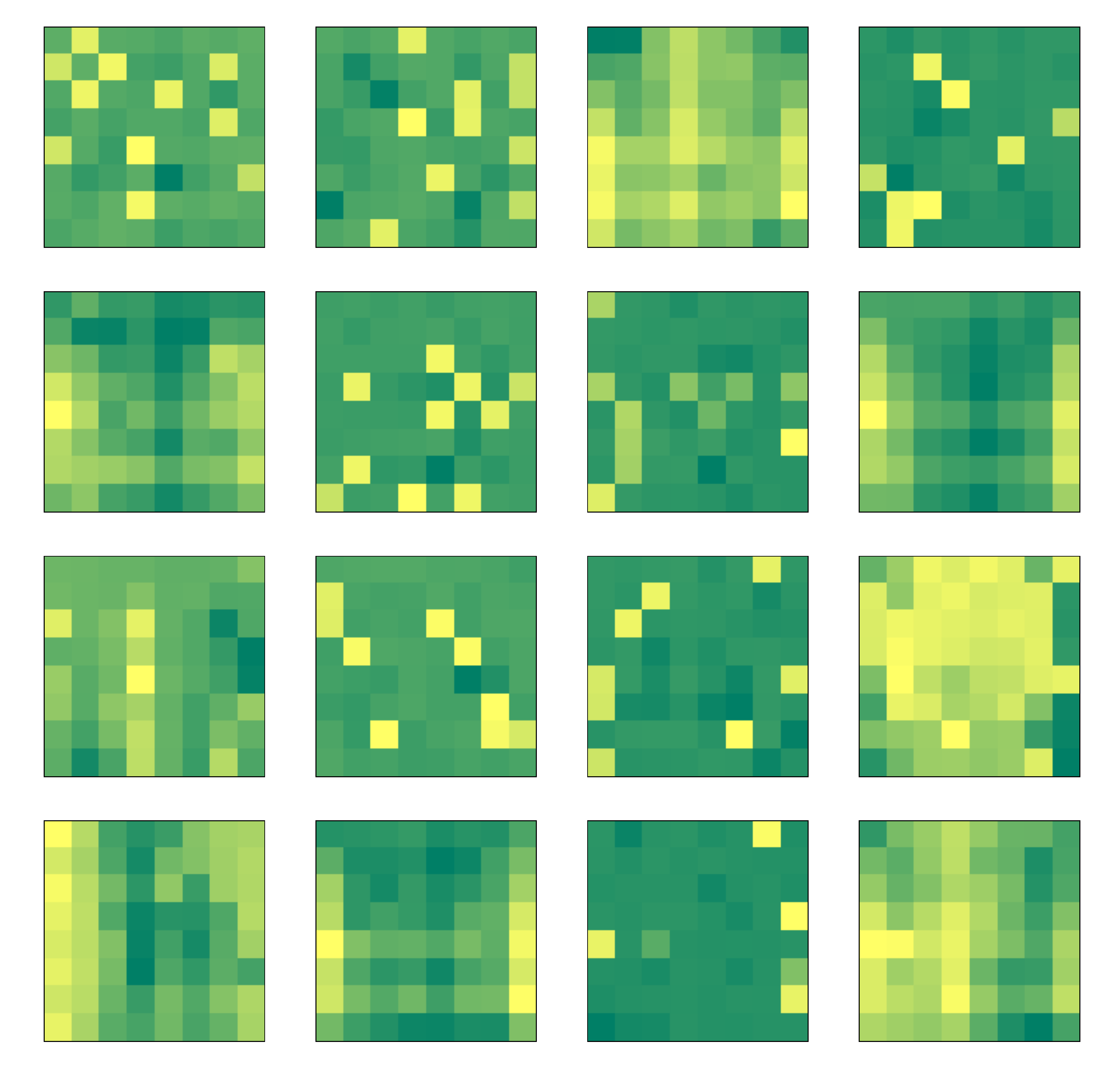}
\includegraphics[width=0.48\textwidth]{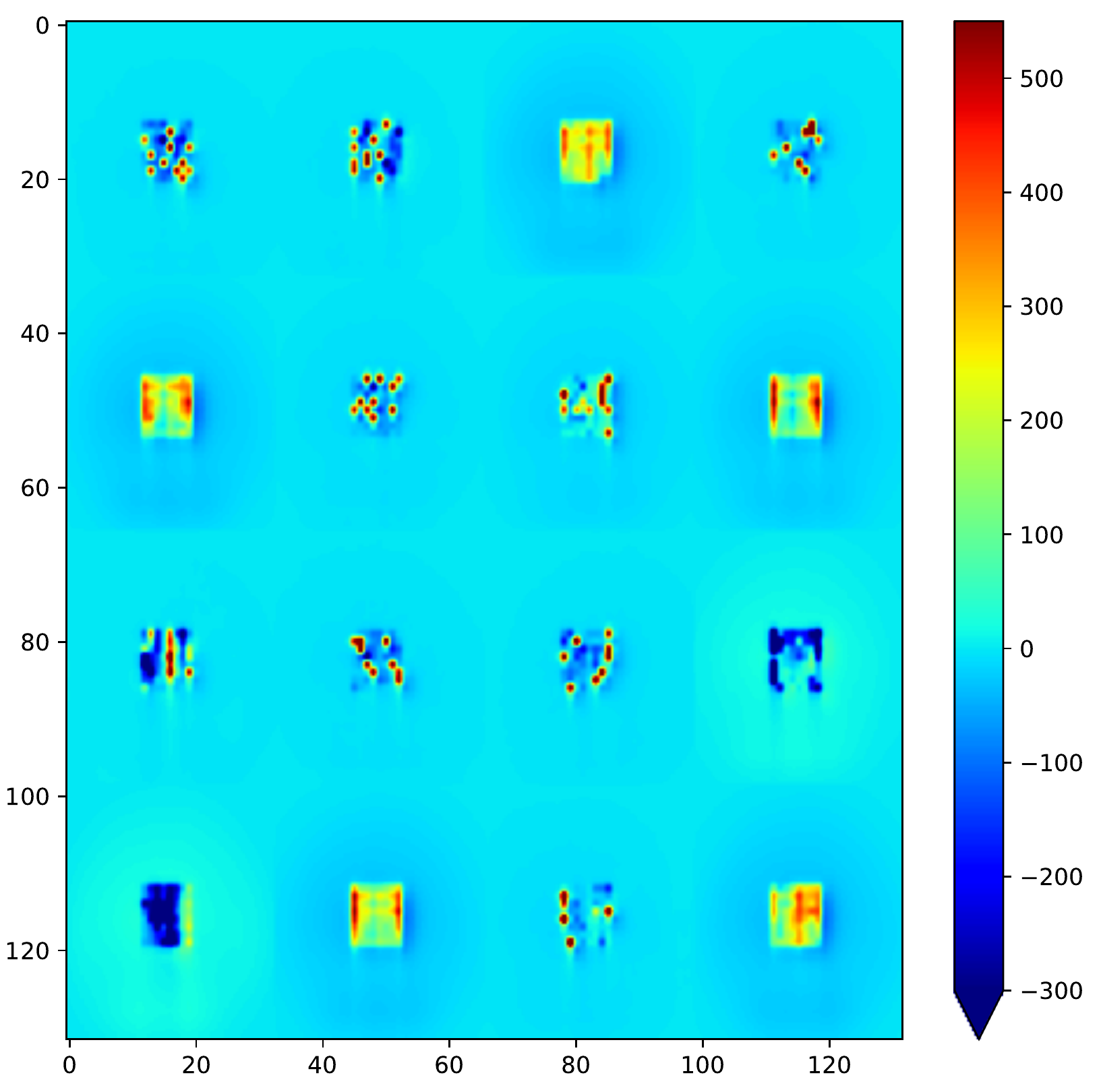}
\caption{16 convolutional filters ($8\times8$) from the first layer (left) and their activation difference of averaged unquenched and quenched normalized jet image by leading particle $p_T$ (right). These filters convolve with the center of jet image most so that strongly activated square area form with size of $\sim 8\times 8$ in the center of the activation maps.}
\label{Filters and activations}
\end{figure}
We train and validate the neural network with the above setup. Before examining the performance of the network, we try to understand what has been learned by the CNN by opening and visualizing it. In Fig.~\ref{Filters and activations}, we show the 16 filters of the first convolutional layer of the CNN by the learned weights (left panel) and the corresponding activation difference of the averaged unquenched ($0.25<\chi_{jh}<0.5$) and quenched ($0.85<\chi_{jh}<1$, see Fig.~\ref{Jet image}) normalized jet image by jet $p_T$ (right panel). One can see that these 16 filters are quite different which indicates they tend to extract different features. Some filters tend to be activated by the quenched jet images while others by the unquenched one. Features including the hardest and second hardest subjet, the distance between them and the pattern of soft particles could be captured by these filters. The jet-by-jet internal structure of soft particles are smeared in these averaged jet image so their activation is not directly visible here.

\begin{figure}[t]
\centering
\includegraphics[width=0.75\textwidth]{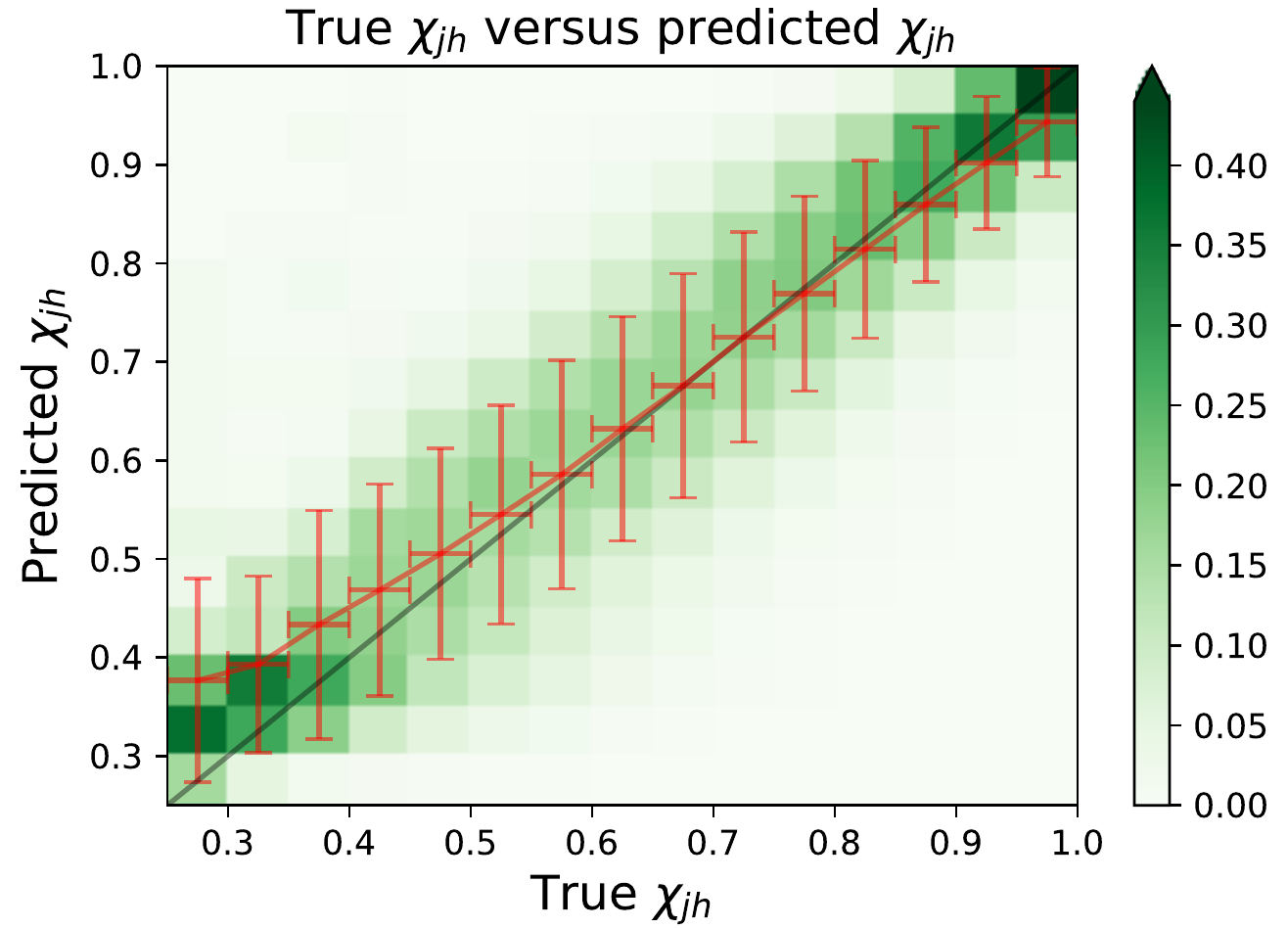}
\caption{Prediction performance. The green color represents the probability of predicted $\chi_{jh}$ along y-axis given true $\chi_{jh}$ in 2-D histogram. Each column is normalized here. The red line with error bar quantifies the average and standard deviation of the predicted $\chi_{jh}$ within the given true $\chi_{jh}$ bin.}
\label{Prediction Performance}
\end{figure}
Fig.~\ref{Prediction Performance} shows the $\chi_{jh}$ prediction performance of CNN from the pre-processed jet image \textcolor{black}{with sample re-weighting applied in the training and validation}. The green column-normalized joint distribution  represents the probability of predicted $\chi_{jh}$ within the given true $\chi_{jh}$ bin. The red line with error bar quantifies the average and standard deviation of the predicted $\chi_{jh}$ within the given true $\chi_{jh}$ bin. The error bar decreases with $\chi_{jh}$.
Overall, we can see that the CNN can predict $\chi_{jh}$ successfully over the whole range. 
As we check, by applying the sample weights in the training and validation, the prediction performance has a slighter dependence on true $\chi_{jh}$ \textcolor{black}{than that without sample weights applied}. Meanwhile, the prediction performance still decreases with decreasing $p_T$ obviously, which shows the re-weighting procedure cannot eliminate this trend and there has to be intrinsic physical reason, as explained above. In Appendix.~\ref{app:Prediction Performance}, we show the prediction performance against various jet observables in detail. In particular, those against $\chi_{jh}$ and jet $p_T$ are shown in Figs.~\ref{Performance chi}-\ref{Performance pT}.

We also checked the individual performance of quark or gluon initiated jets, as can be assigned by following a matching procedure analogous to that explained in Section~\ref{sec:def-chiE}. Even though there are small differences around extreme values of $\chi_{jh}$, we observe no notable bias on the jet species in the overall performance. 

\begin{table*}[b]
\small
\centering
\begin{tabular}{|c|c|c|c|c|}
\hline
Input & Output & Network  & Loss  \\
\hline
Raw jet image & $\chi_{jh}$  & CNN & 0.0028 \\
\hline
Pre-processed jet image & $\chi_{jh}$  & CNN &  0.0029 \\
\hline
Feature-wise standardized jet image & $\chi_{jh}$ & CNN & 0.0031\\
\hline
Jet image normalized by jet $p_T$ & $\chi_{jh}$ & CNN & 0.0036\\
\hline
\end{tabular}
\caption{Predictive performance with different inputs.}  
\label{Performance List A}
\end{table*}
\normalsize

In Tab.~\ref{Performance List A}, we present the prediction performance from different jet images by CNN in terms of validation loss by measuring the difference between true and predicted $\chi_{jh}$. One can see that the prediction performance from pre-processed jet image is very close to that from raw one, which means that the pre-processing is not obviously beneficial to this regression task and CNN can get the rotation-invariant quantity $\chi_{jh}$ automatically. We find that the feature-wise standardized jet image and normalized jet image as aforementioned in Section.~\ref{sec:jet-image} could only give well-matched or worse performance, which is due to the fact that the feature-wise standardization may distort the internal structure of \textcolor{black}{the} jet image and jet $p_T$ is an important feature in this task from above analysis, respectively. 

An important test of the consistency of our procedure consists in making sure that the CNN assigns values $\chi_{jh}\simeq 1$ to vacuum jets created in proton-proton collisions. Having trained the network using medium jets only, we indeed predict that the energy loss ratio distribution for vacuum jets has an average value of $\chi_{jh}$ with standard deviation as 0.98(3).


\subsection{Sensitivity to soft and large-angle radiation}
\label{sec:interpret}

One of the most interesting outcomes from a machine learning task such as the one we performed consists in learning ourselves which are the features and correlations that the algorithm deems as most relevant to carry out a successful prediction. By feeding the network with different combinations of jet properties and observing the change in the performance, in this Section we demonstrate our ability to discern which are the most relevant features of the jet image. Moreover, with such information, we can construct human understandable observables which are not too far from the machine's level of performance.

\begin{figure}[t]
\centering
\includegraphics[width=1\textwidth]{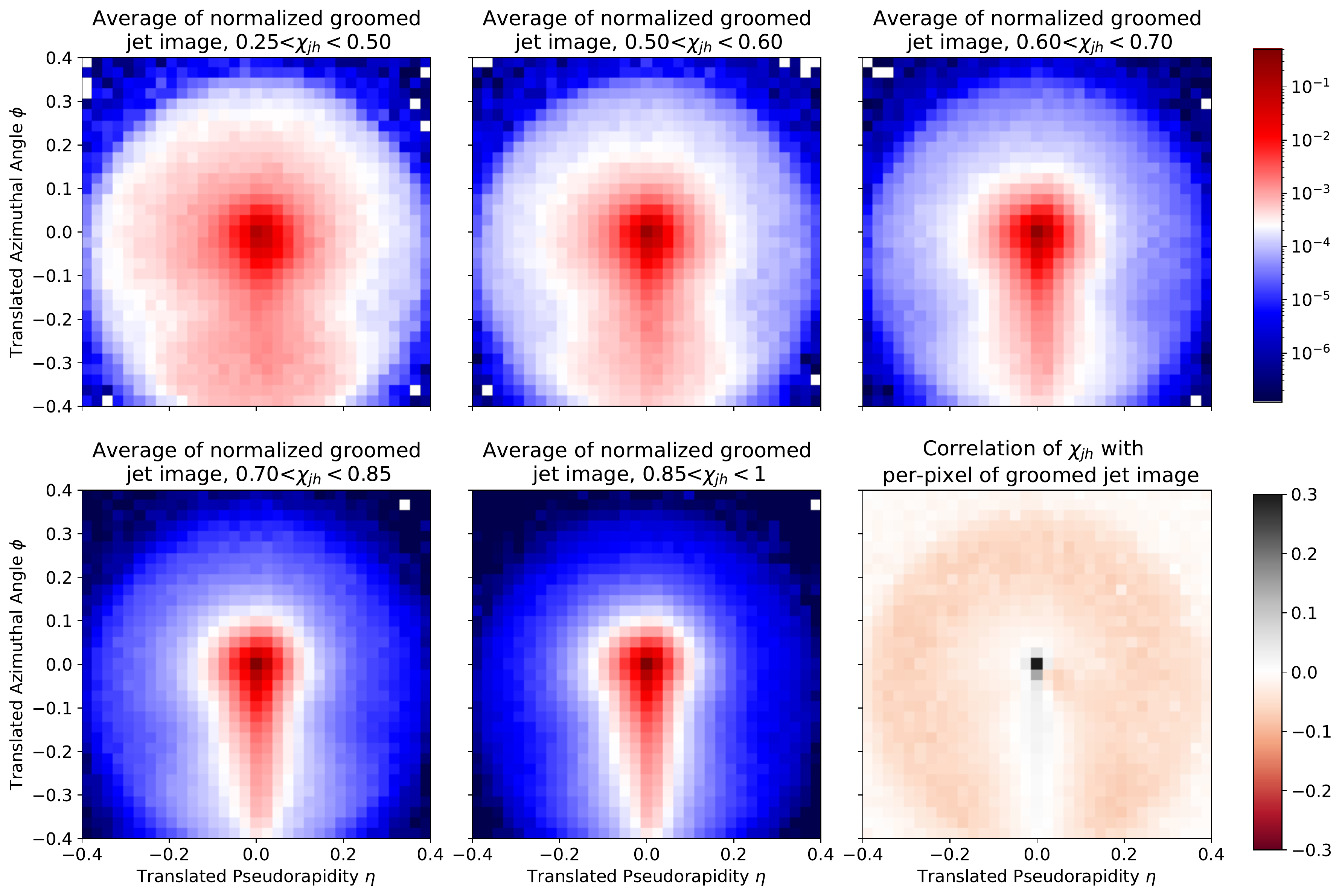}
\caption{The average of groomed jet image normalized by jet $p_T$ within \textcolor{black}{5} different $\chi_{jh}$ cut bins \textcolor{black}{(3 in the top row and 2 in the bottom row, respectively)} and Pearson correlation coefficient of $\chi_{jh}$ with per-pixel of the \textcolor{black}{unnormalized} jet image \textcolor{black}{(rightmost panel in the bottom row)}.}
\label{Groomed jet image}
\end{figure}

As a first check, we use a groomed jet image, which is the jet image after all the branches prior to the one satisfying the SD condition for the first time are discarded, as input to CNN. In Fig.~\ref{Groomed jet image}, we show the average of groomed jet images normalized by jet $p_T$ sliced by different $\chi_{jh}$ bins as Fig.~\ref{Jet image}. One can see some soft and large angle particles are removed by the jet grooming procedure, but the correlation of groomed jet image with $\chi_{jh}$ survives to a large extent. We notice that the performance of using groomed jet image is reduced compared to using the full jet image as shown in Tab.~\ref{Performance List C}. Even though grooming certainly reduces contamination from soft, less well-known processes such as hadronization, the lowered performance hints that it is precisely in the soft, large angle particles where important imprints of the energy loss effects lie. It's worth mentioning that the SD parameters could be tuned to improve the performance.

\begin{table*}[t]
\small
\centering
\begin{tabular}{|c|c|c|c|c|}
\hline
Input & Output & Network  & Loss   \\
\hline
Groomed jet image & $\chi_{jh}$ & CNN & 0.0065 \\ 
\hline
Jet image above 1 GeV & $\chi_{jh}$ & CNN & 0.0042\\
\hline
Jet image above 2 GeV & $\chi_{jh}$ & CNN & 0.0066 \\
\hline
\end{tabular}
\caption{Predictive performance with different inputs. Jet image is pre-processed by default.}  
\label{Performance List C}
\end{table*}
\normalsize

In a more crude approach, we can also use the jet image where we remove soft particles ($p_T<1$ GeV and $p_T<2$ GeV) as input to the CNN and compare with the full jet image. From Tab.~\ref{Performance List C} one can see that the soft particles (e.g., with $p_T<2$ GeV) contain considerable discriminating information, given the big associated loss in performance, consistent with the conclusions drawn from the study of the groomed jet image. The jet grooming and crude soft particles removing actually belong to hard attention mechanism where we focus our attention on the left particles in order to understand the decision-making of neural network, see further application in~\cite{Li:2020bvf}. This sensitivity is  an interesting feature of the problem, but presents at the same time a challenge for the detailed modeling of jet quenching in the presence of a full heavy-ion background. 

\begin{figure}[t]
\centering
\includegraphics[width=0.75\textwidth]{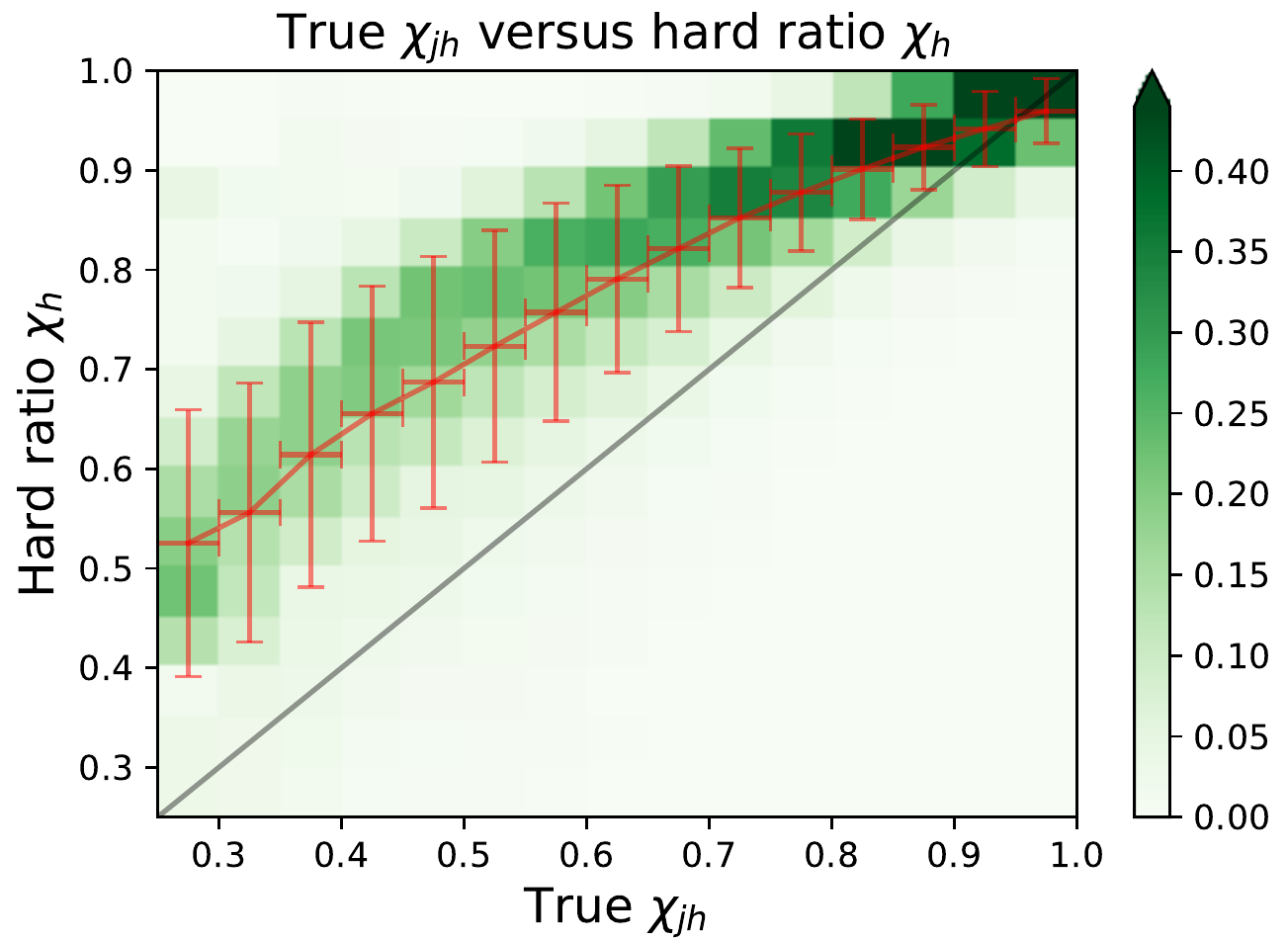}
\caption{$\chi_{jh}$ versus hard ratio $\chi_h$. The green 2-D histogram and red line have same explanation as Fig.~\ref{Prediction Performance}.}
\label{Hard ratio}
\end{figure}

These observations lead us to construct a simple quantity that we will call the hard ratio $\chi_h$, defined as
\begin{equation}
    \chi_h \equiv \frac{\sum_{i\in \textrm{jet}} p_T^i \, \Theta(p_T^i>2 \, \textrm{GeV})}{p_T^{\textrm{jet}}} \, ,
\end{equation}
which is, the percentage of the total jet transverse momentum carried by ``hard'' particles, where here by hard we mean particles with $p_T>2$ GeV. \textcolor{black}{This hard ratio $\chi_h$ serves as our first constructed observable to provide insight into the energy loss ratio $\chi_{jh}$.} It is found that $\chi_{h}$ is strongly correlated with $\chi_{jh}$ as shown in column-normalized joint distribution of $\chi_{jh}$ and $\chi_h$ in Fig.~\ref{Hard ratio}. Note that even \textcolor{black}{though} the fluctuation of $\chi_h$ is comparable with that of the predicted $\chi_{jh}$ by the CNN for the samples within a given \textcolor{black}{true} $\chi_{jh}$ bin, the hard ratio $\chi_h$ is not as discriminating as the network, since its non-linear correlation with \textcolor{black}{true} $\chi_{jh}$ will not give a direct correspondence. Specially for samples with large \textcolor{black}{true} $\chi_{jh}$, the hard ratio $\chi_h$ has a poor discriminating power due to the small slope. 

We also train fully-connected neural network to predict $\chi_{jh}$ with physics-motivated high-level features \cite{baldi2016jet} which include jet shape (JS), jet fragmentation function (JFF), jet $p_T$, $z_g$, $n_{SD}$, $R_g$, jet mass $M$, groomed jet mass $M_g$, jet multiplicity. Here, the JS and JFF are the projections of jet image along the radial and shared momentum $z$ direction, respectively. This is not necessarily a mutually orthogonal set of observables, but captures the physics information contained in the image, as we will see from the performance. We show the correlation matrix for this set of observables in Appendix~\ref{app: variables correlations}. In addition to the jet $p_T$, which is particularly important for our study, one could study instead an extended set of N-subjettiness variables \cite{Datta:2017rhs}, energy-flow polynomial \cite{Komiske:2017aww},
or counting observables, e.g., \cite{Frye:2017yrw}. For a more in-depth discussion on this point see, e.g., \cite{kasieczka2020towards,Faucett:2020vbu}. 

\begin{table*}[b]
\small
\centering
\resizebox{0.8\textwidth}{!}{
\begin{tabular}{|c|c|c|c|c|}
\hline
Input (size) & Output & Network  & Loss \\
\hline
JFF (10) & $\chi_{jh}$ & FCNN & 0.0058 \\
\hline
Jet shape (8)    & $\chi_{jh}$ & FCNN &  0.0033\\
\hline
JFF, jet shape (18)   & $\chi_{jh}$ & FCNN & 0.0032 \\
\hline
JFF, jet shape, features (25)& $\chi_{jh}$ & FCNN & 0.0028\\
\hline
Jet image \& JFF, jet shape, features (25) & $\chi_{jh}$ & API: CNN\&FCNN & 0.0028 \\
\hline
\end{tabular}}
\caption{Predictive performance with different inputs and neural networks. Jet image is pre-processed by default. Features include: jet $p_T$, $z_g$, $n_{SD}$, $R_g$, $M$, $M_g$, Multiplicity.} 
\label{Performance List B}
\end{table*}
\normalsize

\begin{figure}[tb]
\centering
\includegraphics[width=0.75\textwidth]{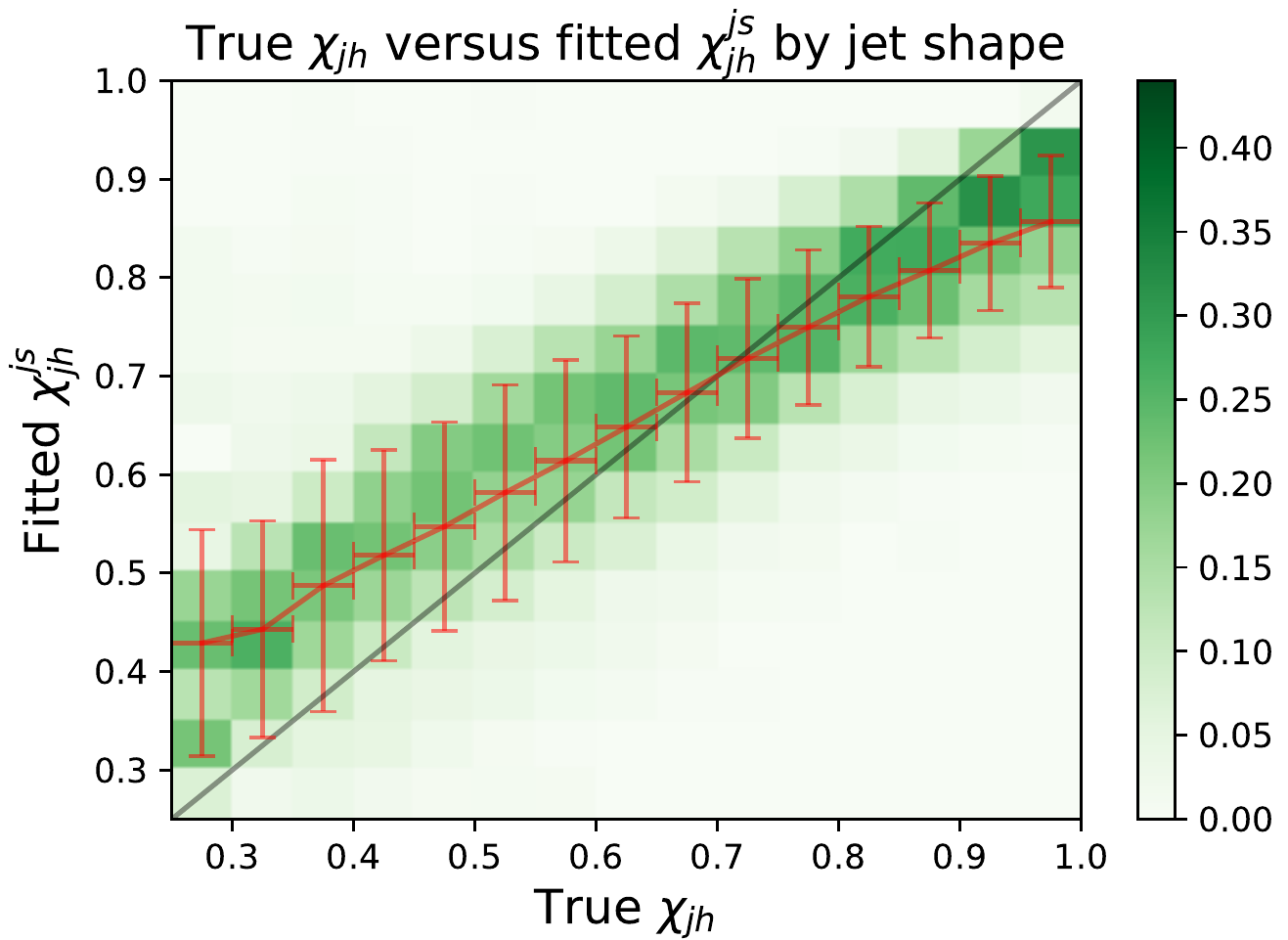}
\caption{$\chi_{jh}$ versus fitted $\chi_{jh}^{js}$ by jet shape. The green 2-D histogram and red line have same explanation as Fig.~\ref{Prediction Performance}.}
\label{JS_fitting}
\end{figure}

From Tab.~\ref{Performance List A} and~\ref{Performance List B}, one can see that the jet shape gives well-matched performance as the 
$p_T$-normalized jet image does and outperforms JFF. It's a fair comparison since jet shape also doesn't contain the jet $p_T$ information. It demonstrates that the energy distribution in the jet cone is more important than the energy sharing among these particles. It also indicates that the azimuthal distribution of particles in the jet cone contains little discriminating power, which can be attributed to the uniform splitting and radiation probability along the azimuthal angle in the jet cone. Since \textcolor{black}{the}  jet shape with 8 bin values contains \textcolor{black}{the} main information on \textcolor{black}{the} jet energy loss, we try to use it to fit the $\chi_{jh}$ with a simple parametrization:
\begin{equation}
    \chi_{jh}^{js} = \sum_i \left(\frac{p_{Ti}}{p_T} \right)^{\alpha_i} r_i^{\beta_i}+\gamma,
\end{equation}
where $p_{Ti}/p_T$ is the momentum fraction of jet $p_T$ at $r_i$ and $\{\alpha, \beta, \gamma\}$ is a 17-fitting-parameter set. Fig.~\ref{JS_fitting} shows the fitting performance which is though a bit worse than the network prediction but already much better than the hard ratio $\chi_h$.

The combination of jet shape, JFF and other physics-motivated features including jet $p_T$, $z_g$, $n_{SD}$, $R_g$, jet mass $M$, groomed jet mass $M_g$, jet multiplicity, results in a comparable performance to the jet images. If we take the combination of these physics-motivated features and jet image both as inputs to a more complex and flexible functional neural network model consisting of CNN and FCNN using the Keras Functional API, the performance will not improve with respect to each input alone, which indicates that this combination and jet image share almost the same discriminating information. To quantitatively estimate and rank the discriminating power of each jet observables and jet image, even in each pixel, layer-wise relevance propagation method could be applied in the future, as in Ref.~\cite{Agarwal:2020fpt}. 

\section{Applications}
\label{sec:applications}
Having proven our ability to directly access the degree of jet-by-jet modification due to energy loss within the QGP, expressed through the energy loss ratio $\chi_{jh}$, we now can start asking questions about the origins and fate of a reconstructed jet. Intuitively, a large energy loss, when $\chi_{jh}$ is significantly smaller than 1, implies significant jet-medium interaction. This could be due to the fact that the jet traversed a long distance in a hot QCD plasma or because the jet multiplicity in the medium was high, leading to a stronger quenching effect. While our current setup does not allow to disentangle these two effects, presently we introduce two novel applications we can now explore. The first deals with devising new jet observables to enhance the quenching effects. The second is a first step toward using jets truly as tomographic probes of the QGP.


\subsection{Sensitivity of observables to in-medium modification}
\label{sec:observables}

When we study a jet observable in heavy-ion collisions we are typically required to select jets within a certain momentum range, or at least with some minimum $p_T$. In order to quantify the modification of the observable due to the presence of the medium, we usually construct a ratio between the medium distribution for the observable and the one obtained in pp collisions, for the same kinematical selection. The interpretation of the resulting ratio is, alas, not unique. It is fair to pose the question to which extent the modification that we observe is a consequence of the modification of a given jet due to energy loss. Due to underlying steeply falling production spectrum, it is unlikely to measure a jet that has lost a lot of energy, as it would have had to start at a much higher energy where the spectrum is suppressed as a negative power of the $p_T$. This means that imposing a $p_T$ cut on the inclusive jet distribution naturally selects those kinds of jets that have \emph{on average} lost the least energy. This selection bias is an unavoidable fact about jet quenching physics that limits the interpretability of conventional observables.

A jet, reconstructed with a given radius $R$, is not only characterised by its energy, $p_T$. One key characteristic is the identity of the jet partonic ancestor, i.e., whether it is a quark or gluon jet. The latter has a larger color charge which both leads to a stronger interaction with the medium and a wider fragmentation pattern. In addition, within a given energy bin, jets can consist both of narrow, little fragmented structures, as well as wide, or more populated structures. Jets with more fragments will tend to lose more energy \cite{Milhano:2015mng,Casalderrey-Solana:2019ubu}, provided that the fragments are created in (and are resolved by) the medium \cite{Mehtar-Tani:2017web}, 
than those with \textcolor{black}{fewer} components. This is simply because each parton of the jet can in principle undergo processes of elastic scattering and medium-induced radiation, so that the total amount of energy lost by the components that build an extended object like the jet roughly scales like the number of partons resolved by the medium\footnote{Here, ``resolved'' means that two partons coming from a given splitting have had enough time to decohere, i.e., that they are separated enough to be resolved as two independent color currents interacting with the medium.
Otherwise, we say that the medium cannot resolve the dipole and it is quenched according to its combined charge only. The scales that govern the physics of resolution depend on the properties and extent of the medium and have been subject to a great amount of study~\cite{MehtarTani:2011tz,CasalderreySolana:2011rz,MehtarTani:2012cy}. These physics have been incorporated in the hybrid model \cite{Hulcher:2017cpt}, but for simplicity we have chosen to omit them in the present study and their effects will be explored in future work.}.

The different way in which jets with varying widths are quenched is now a key ingredient of our understanding of di-jet asymmetry~\cite{Milhano:2015mng}, which was for long regarded as stemming solely from the differences of in-medium path length traversed by the leading and the subleading jet. Also, the fact that wider jets lose more energy, combined with the steeply falling jet spectrum and the $p_T$-dependent quark-gluon jet fraction, allows us to understand the narrowing and hardening of the core of the inclusive jet samples measured in heavy-ion collisions compared to proton-proton collisions. Even though each individual jet could get wider as a result of the interaction with the medium, the fact that the wider jets tend to lose more energy can leave us with a narrower final jet ensemble if we impose a $p_T$ cut
~\cite{Rajagopal:2016uip,Casalderrey-Solana:2016jvj,Brewer:2017fqy}. Essentially, the steeply falling spectrum turns the medium into a very effective filter in which the jets we observe for a given $p_T$ range are those that lost the least energy. Therefore, losing less energy does not simply mean traversing less amount of QGP: there is a selection bias in terms of the properties of the jet, such that narrower jets, or those with less resolved color charge, are more likely to pass the selection criteria and enter the distributions of any given observable.

\begin{figure}[t!]
\centering
\includegraphics[width=0.49\textwidth]{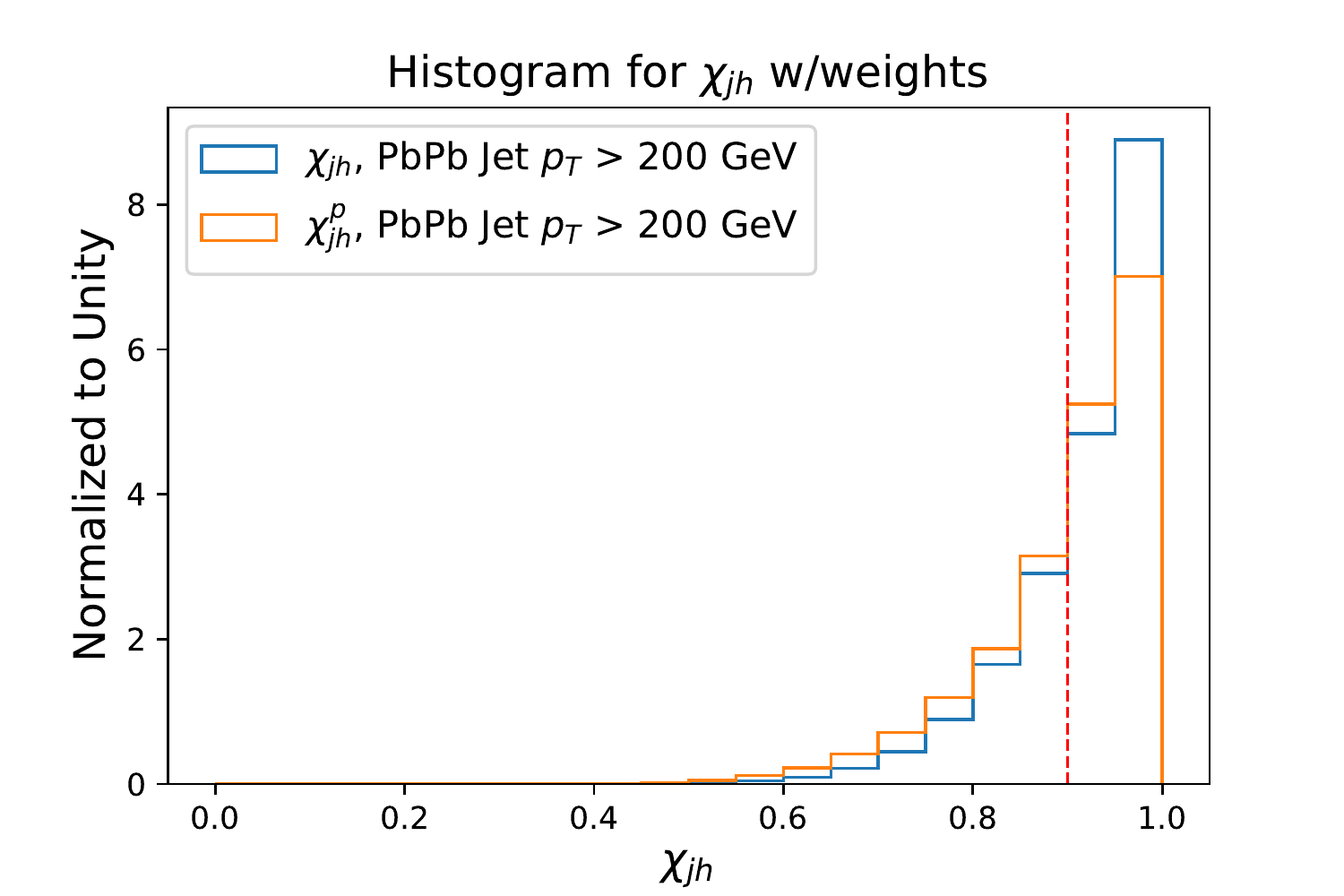}
\includegraphics[width=0.49\textwidth]{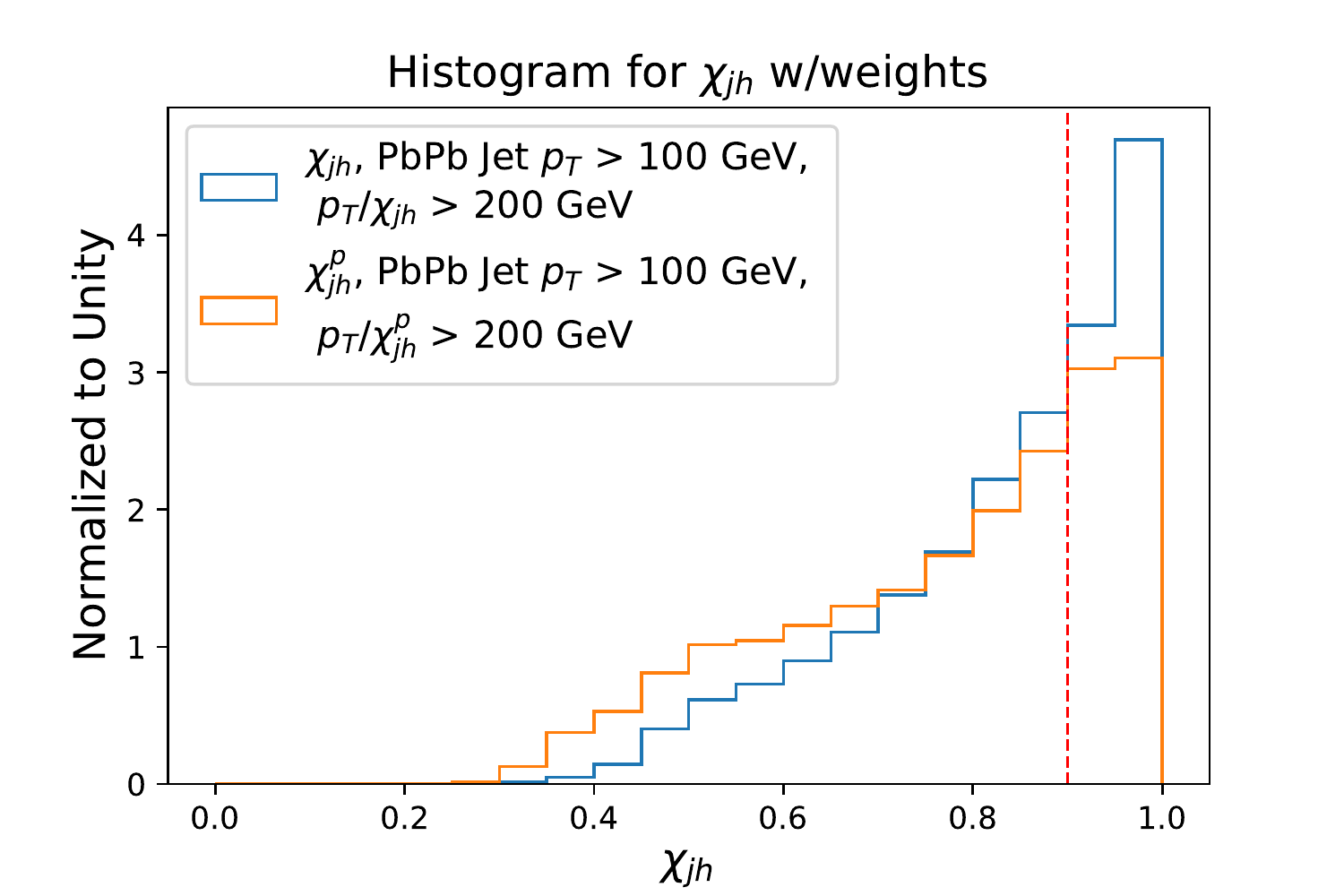}
\caption{The true (blue) and \textcolor{black}{CNN} predicted (orange) $\chi_{jh}$ distribution of jets with final energy selection, FES, (left panel) and initial energy selection, IES,  (right panel). The red vertical dashed line depicts the location of the $\chi_{jh}=0.9$ cut.}
\label{chijh histogram}
\end{figure}

In this Section, we employ our new technique to extract $\chi_{jh}$ to investigate the response of a set of selected observables to the amount of energy loss experienced by the jet. To do so, we divide the full jet population in two selections, the ``quenched'' class, with $\chi_{jh} < \chi_{jh}^{\rm cut}$, and the ``unquenched'' class, with $\chi_{jh}> \chi_{jh}^{\rm cut}$, and compare the resulting distributions of both selections. Within our regression setup, we can choose $\chi_{jh}^{\rm cut}$ freely, or even quantify the degree of quenching in finer detail using multiple classes. In our study we make a reasonable choice $\chi_{jh}^{\rm cut}=0.9$. The modification of observables for a finer set of $\chi_{jh}$ classes is discussed in more detail in Appendix~\ref{sec:observable-sliced-chijh}.

In order to understand the actual implications of performing  a cut in $\chi_{jh}$, we show the distribution of $\chi_{jh}$ in the left panel of Fig.~\ref{chijh histogram}. These results are obtained using the realistic jet spectrum for $R=0.4$ anti-$k_T$ jets produced at $\sqrt{s}=5.02$ ATeV, where the jet is required to have $p_T^{\textrm jet}>200$ GeV. We compare the results for the true label $\chi_{jh}$ versus the \textcolor{black}{CNN} predicted one $\chi_{jh}^p$. The vertical dashed line depicts the location of the chosen $\chi_{jh}=0.9$ cut. As argued in the discussion above, this distribution is biased towards jets that have lost little amount of energy as a result of the steeply falling jet spectrum. Therefore, the average values of $\chi_{jh}$ above and below the cut will be fairly similar, as will be the modifications of their observables with respect to vacuum. This is a consequence of the fact that the jet selection, as it is done in experiments, is applied on the \emph
{final}, measured jet energy, both for jets created in heavy-ion collisions and in proton proton collisions. 

However, we can now perform the jet selection on the \emph{initial} jet energy. Using our knowledge about the amount of energy loss $\chi_{jh}$ that a measured jet with momentum $p^{\rm jet}_T$ has experienced, we estimate this as $p_T/\chi_{jh}$, as shown in the right panel of Fig.~\ref{chijh histogram}. Hence, we can select jet samples according only to the energy they had before quenching, such that, in contrast to the previous choice, one introduces no implicit requirement about the necessity to 
``pass the cut'' or, in other words, to have lost relatively little energy.
More specifically, we choose jets 
whose final energy is at least $p^{\rm jet}_T > 100$ GeV, 
 and where the 
\emph{initial} jet $p_T$ was significantly higher, $p^{\rm jet}_T/\chi_{jh} > 200$ GeV.
This novel approach allows us for the first time to study jet observables in a way that before was only possible by looking under the hood of the energy loss model. 

We can see in the right panel of Fig.~\ref{chijh histogram} that the energy loss distribution has much more support at lower values of $\chi_{jh}$, so that by performing a simple cut at $\chi_{jh}=0.9$ we really are separating jets into two classes with very different amounts of medium modification. From now on, we will refer to the selection based on the final jet energy (left panel of Fig.~\ref{chijh histogram}) as Final Energy Selection (FES), while the selection based on the initial jet energy (right panel of Fig.~\ref{chijh histogram}) will be the Initial Energy Selection (IES).

In the following we will show results for medium over vacuum ratios of jet observables, defined in Section~\ref{defobservables}, in which we compare FES (left column of each figure) with IES (right column of each figure). We also confront the results obtained using our \textcolor{black}{CNN} algorithm to extract $\chi_{jh}^p$ (bottom row of each figure) against the true value of $\chi_{jh}$ as extracted from the Monte Carlo model (top row of each figure). All results using either the true or the predicted value of $\chi_{jh}$ share the same qualitative features, and in many cases the agreement reaches quantitative level, as we show in the Figures below.

For FES, jet selection is done with the following cuts:
\begin{itemize}
    \item Medium: $p_T^{\textrm{jet}}>200$ GeV.
    \item Vacuum: $p_T^{\textrm{jet}}>200$ GeV.
\end{itemize}

For IES, the jet cuts are:
\begin{itemize}
    \item Medium: $p_T^{\textrm{jet}}>100$ GeV, $p_T^{\textrm{jet}}/\chi_{jh}>200$ GeV.
    \item Vacuum: $p_T^{\textrm{jet}}>200$ GeV.
\end{itemize}
The reason why we have an extra lower cut of $p_T^{\textrm{jet}}>100$ GeV in medium jets for IES is because lower $p_T$ jets tend to be harder to reconstruct in experiments due to their lower signal to background ratio and because for the moment we have decided not to extend our machine learning analysis down to these lower momenta.

This way of presenting the sensitivity of our results to $\chi_{jh}$, based on the IES and FES setups, is akin to current phenomenology standards and emphasises the role of the selection bias and the difference in jet populations when treating inclusive samples -- but it is only one among several possible choices. As mentioned before, we also provide results for all the observables, finely sliced in the (true label) $\chi_{jh}$ for the FES setup in Appendix~\ref{sec:observable-sliced-chijh} which the reader can access for convenience.

\subsubsection{Groomed observables}
\label{sec:groomed-observables}

\begin{figure}[t!]
\centering
\includegraphics[width=0.8\textwidth]{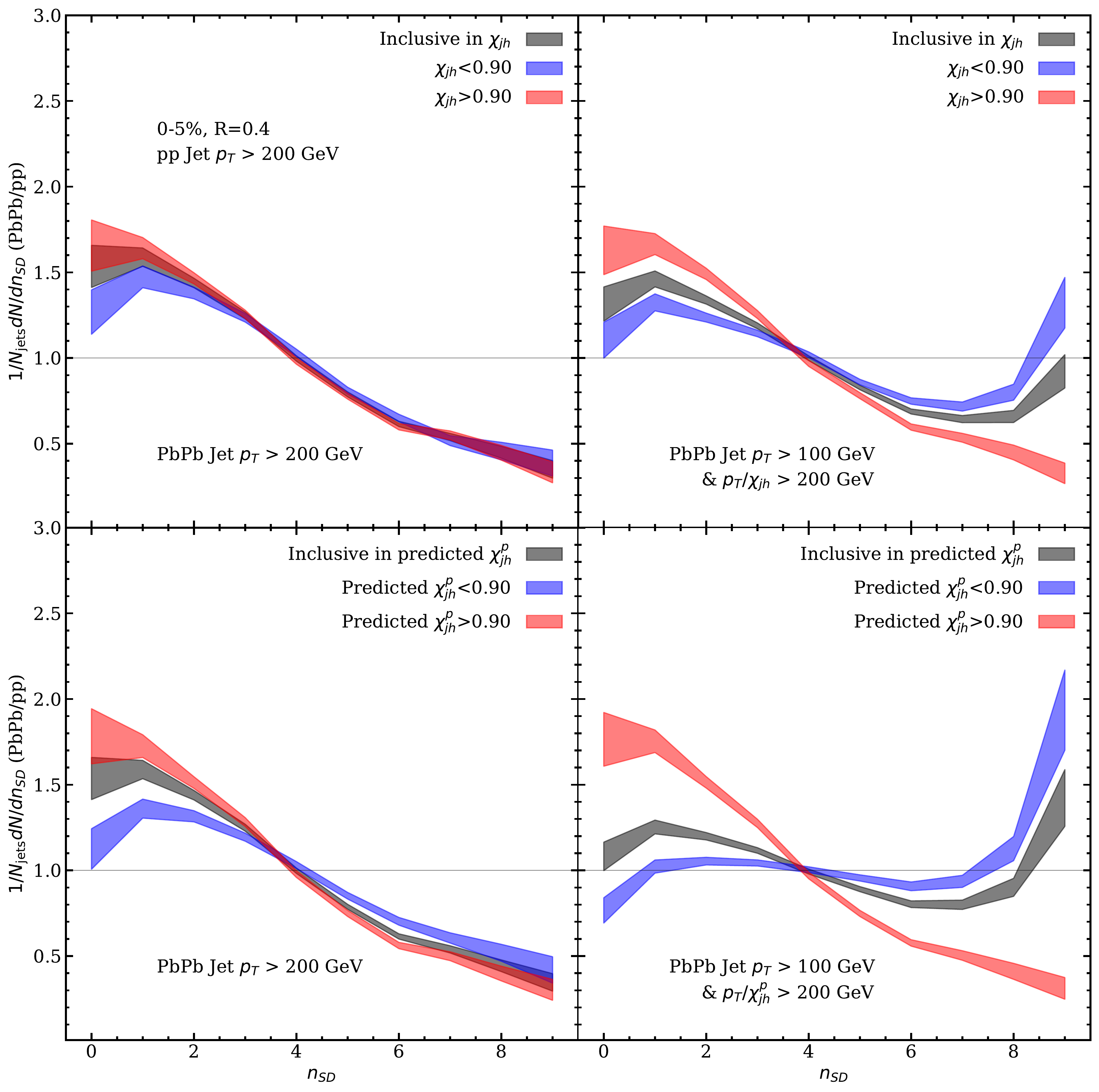}
\caption{Results comparing the $n_{SD}$ ratio between PbPb and pp for the true $\chi_{jh}$ (top row) and predicted $\chi_{jh}^p$ (bottom row), and the FES (left column) with the IES (right column).}
\label{nSD_pred}
\end{figure}

We start by discussing groomed observables within the Soft Drop procedure, with parameters $z_{\textrm{cut}}=0.1$ and $\beta=0$. In Fig.~\ref{nSD_pred}, we show the ratio between PbPb and pp for the number of SD splittings $n_{SD}$. This is a good proxy for the jet multiplicity (although in the iterative setup it only counts the primary emissions, this is, the ones along the hardest branch of the clustering tree) and has been measured in heavy-ion collisions~\cite{Acharya:2019djg,ALICE-PUBLIC-2020-006}. 
A reduction in average $n_{SD}$ has been observed in experiments. This feature has been attributed to energy loss and is well reproduced by jet quenching models~\cite{Caucal:2019uvr,Casalderrey-Solana:2019ubu}. With FES, we observe that these results are consistent between the ``quenched'' and the ``unquenched'' classes, as expected from the discussion above about the energy loss distributions from Fig.~\ref{chijh histogram}, see left column of plots in Fig.~\ref{nSD_pred}. Both classes contain jets that mainly lost little energy. Such jets tend to be of a special kind, and as we infer from the results in Fig.~\ref{nSD_pred}, they tend to have a smaller $n_{SD}$ than the average. Therefore, by looking only at the FES results, the actual effects of energy loss are obscured in this observable by the existence of the strong selection bias. 

We now turn to the observable constructed from the jet samples based on IES, which we expect to be more enriched in quenched jets. Now, the results inclusive in $\chi_{jh}$ no longer coincide with the ``unquenched'' class, revealing more of the genuine medium modification for this observable. 
Even though the ``quenched'' class still presents an overall shift towards fewer $n_{SD}$, characteristic of the jets sitting close to the $\chi_{jh}$ cut that lost relatively little energy, we observe that very quenched jets contribute to increase the number of large $n_{SD}\sim 8$ (see the more differential selection in $\chi_{jh}$ of Appendix~\ref{sec:observable-sliced-chijh}) due to the copious and spread number of soft particles originated from the excited wake. Even though there are not so many very quenched jets within the ``quenched'' class, their effect at large $n_{SD}$ is notable in the medium over vacuum ratio, since in vacuum there rarely are jets with such large \textcolor{black}{values} of $n_{SD}$.
Finally, it is worth emphasizing an extremely important feature of our analysis: we also observe that the ratio between the ``unquenched'' class and vacuum still \emph{deviates from unity}. This is a clear manifestation of the same selection bias we have discussed so far. The jet population that ultimately falls into the ``unquenched'' class tends to be narrower and less fragmented than the average jet population in vacuum.
In order to present a ratio around unity for the ``unquenched'' class, one would need to restrict the denominator to a specific vacuum jets population as well, i.e., a population of characteristically narrow jets from which the in-medium jets belonging to the ``unquenched'' class originated.
This kind of studies go beyond the regression task we have focused on in this work, and will be left for future studies.

\begin{figure}[t!]
\centering
\includegraphics[width=0.8\textwidth]{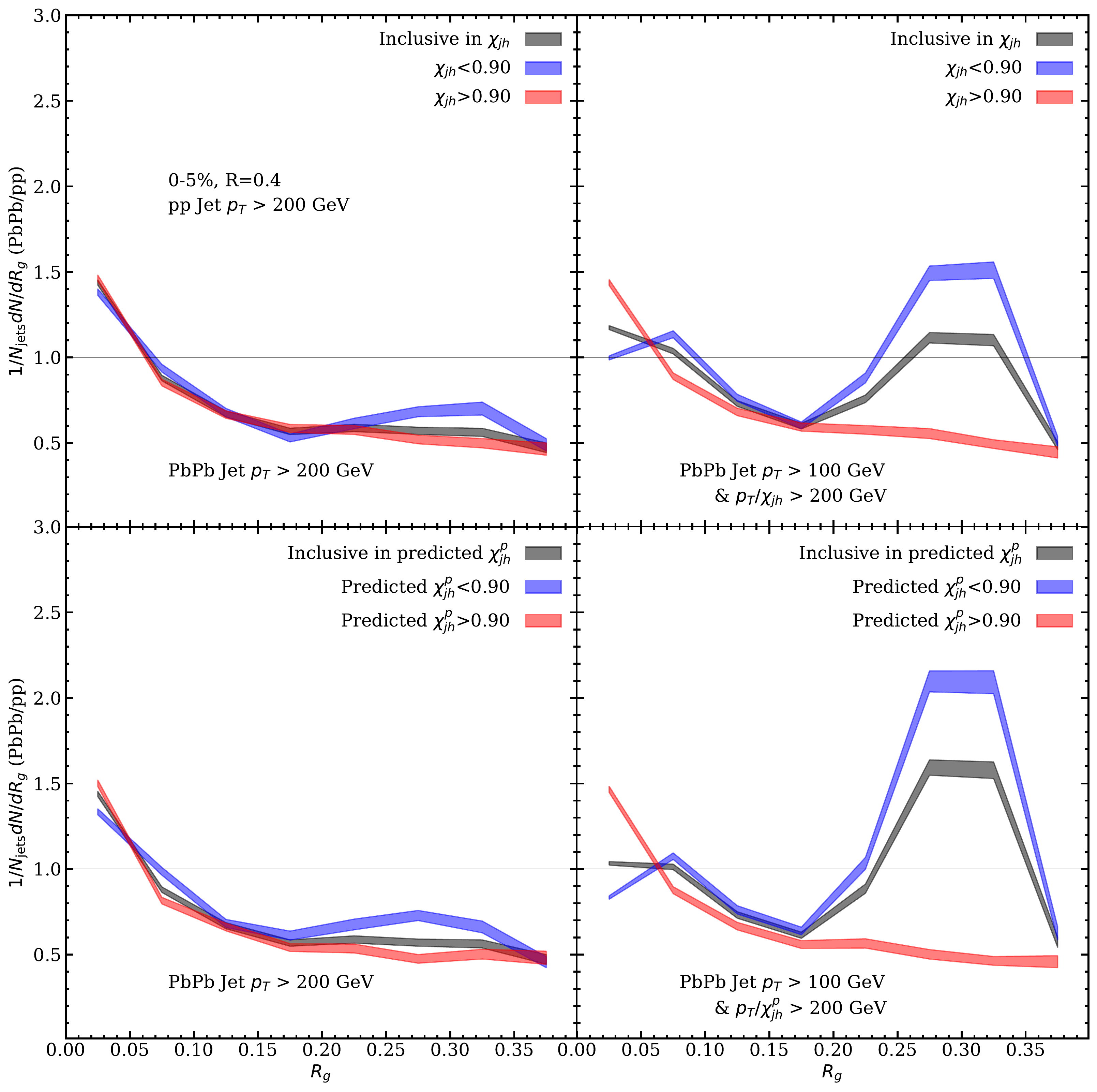}
\caption{Results comparing the $R_g$ ratio between PbPb and pp for the true $\chi_{jh}$ (top row) and predicted $\chi_{jh}^p$ (bottom row), and the FES (left column) with the IES (right column).}
\label{rgratio}
\end{figure}

An even more paradigmatic observable of the effect of the selection bias is the groomed angle $R_g$, shown in Fig.~\ref{rgratio}. In vacuum, jets with a larger $R_g$ are jets that typically explore a larger phase space of in-cone emissions, leading to higher multiplicity.
If the medium can resolve such internal structure, these wider jets will be more quenched than the average and they will fail to pass the cut~\cite{Casalderrey-Solana:2019ubu}. This results in a depletion of large $R_g$ jets  in PbPb compared to pp, as measured by experiments~\cite{Acharya:2019djg,ALICE-PUBLIC-2020-006} and well reproduced by several Monte Carlo simulations
~\cite{Caucal:2019uvr,Casalderrey-Solana:2019ubu,ALICE-PUBLIC-2020-006}. While it is true that, for the conventional FES setup, the $R_g$ distribution of the jets passing the cut is not significantly modified~\cite{Casalderrey-Solana:2020jbx}, see left column of Fig~\ref{rgratio}, this is the case only because significantly quenched jets escape such selection~\cite{Brewer:2020chg}. Indeed, as shown in the IES setup in Fig.~\ref{rgratio} (right column), energy loss effects do modify the $R_g$ of a jet through the creation of a new, semi-hard branch situated at large angles, composed mostly of the thermal particles originated in the medium response to the passage of the jet~\cite{Brewer:2020chg}.

\begin{figure}[t!]
\centering
\includegraphics[width=0.8\textwidth]{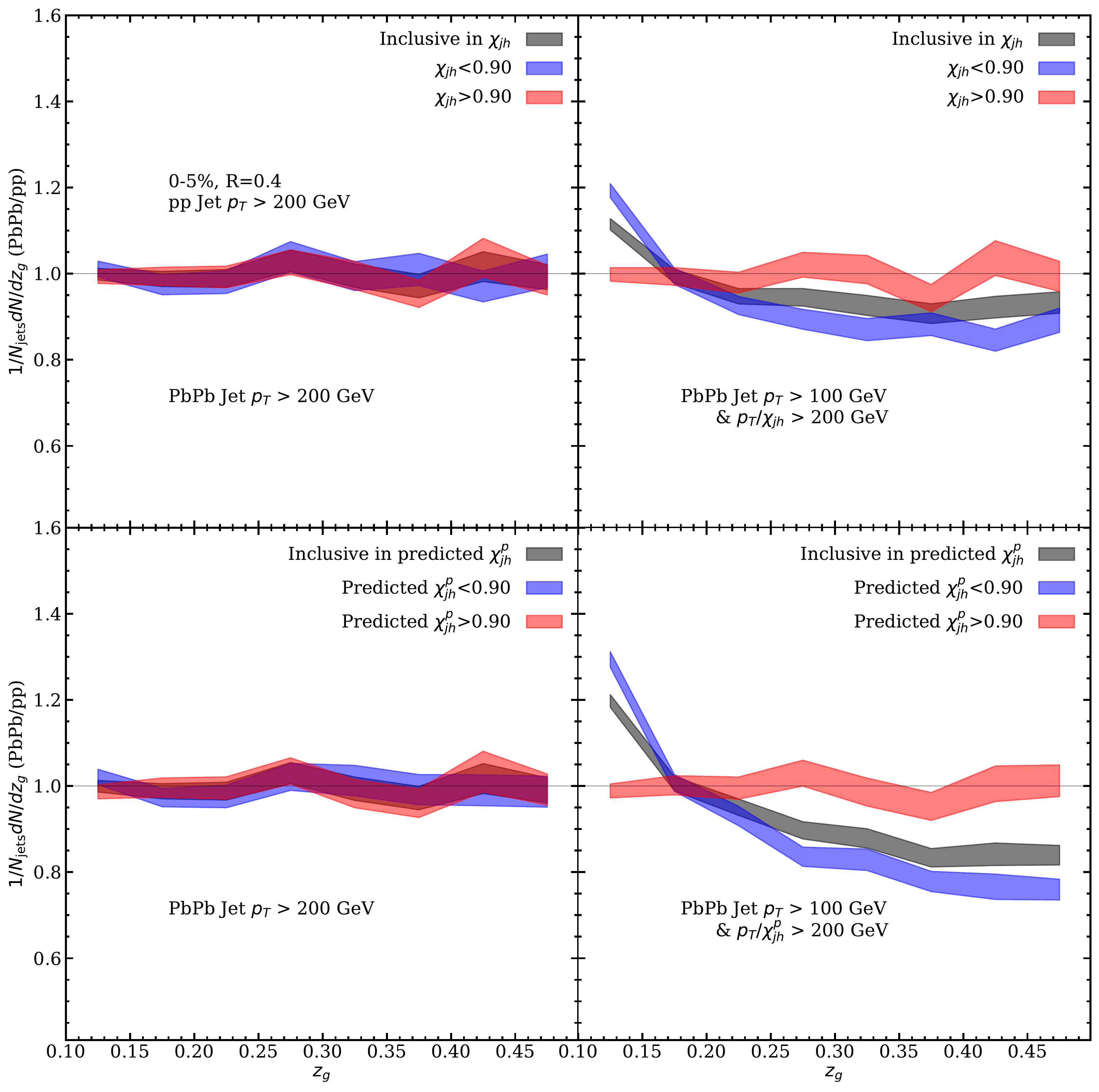}
\caption{Results comparing the $z_g$ ratio between PbPb and pp for the true $\chi_{jh}$ (top row) and predicted $\chi_{jh}^p$ (bottom row), and the FES (left column) with the IES (right column).}
\label{zgratio}
\end{figure}

The last of the groomed observables we analyze here is the groomed momentum sharing fraction $z_g$, shown in Fig.~\ref{zgratio}. It has been measured in heavy-ion collisions by CMS~\cite{Sirunyan:2017bsd} and ALICE~\cite{Acharya:2019djg,ALICE-PUBLIC-2020-006}, where an enhancement at small values of the $z_g $ distribution, when $R_g \gtrsim 0.1$, has been reported. Several theoretical calculations predicted an enhancement of the low $z_g$ part of the distribution due to a modification of the partonic splitting functions in the medium~\cite{Chien:2016led,Mehtar-Tani:2016aco,Chang:2017gkt,Caucal:2019uvr}, although other model studies have highlighted the importance of jet-medium recoil effects 
toward explaining the observed excess~\cite{Milhano:2017nzm}. 
Even so, the thermal hadrons from the wake of the hybrid model---that we employ in our studies---are typically too soft to produce any significant enhancement at low $z_g$ for the FES setup~\cite{Casalderrey-Solana:2019ubu}. Nevertheless, such results are consistent with measured data when one takes into account the contamination from the large fluctuating background, which precisely dominates the low $z_g$ part of the distribution at large angles~\cite{Acharya:2019djg,Casalderrey-Solana:2018xyb}. In contrast to the previously discussed observables, the fact that in the FES setup the ``unquenched'' class ratio between PbPb and pp sits at unity indicates that $z_g$ is \emph{not} significantly affected by the selection bias, as it does not determine the jet phase space in the way that $R_g$ does.\footnote{In principle, the splitting fraction $z$ of the first splitting inside the jet does affect its available phase space through $M^2/p_T^2\approx z(1-z) \theta^2$, although much more mildly than the opening angle~$\theta$, closely related to $R_g$.} Therefore, the ``unquenched'' class in the IES setup also presents a ratio around unity. However, sufficiently quenched jets now have a clear enhancement at low $z_g$ values, which are the soft branches at large angles dominated by the contribution from medium response particles.

\subsubsection{Ungroomed observables}
\label{sec:ungroomed-observables}
\begin{figure}[t!]
\centering
\includegraphics[width=0.8\textwidth]{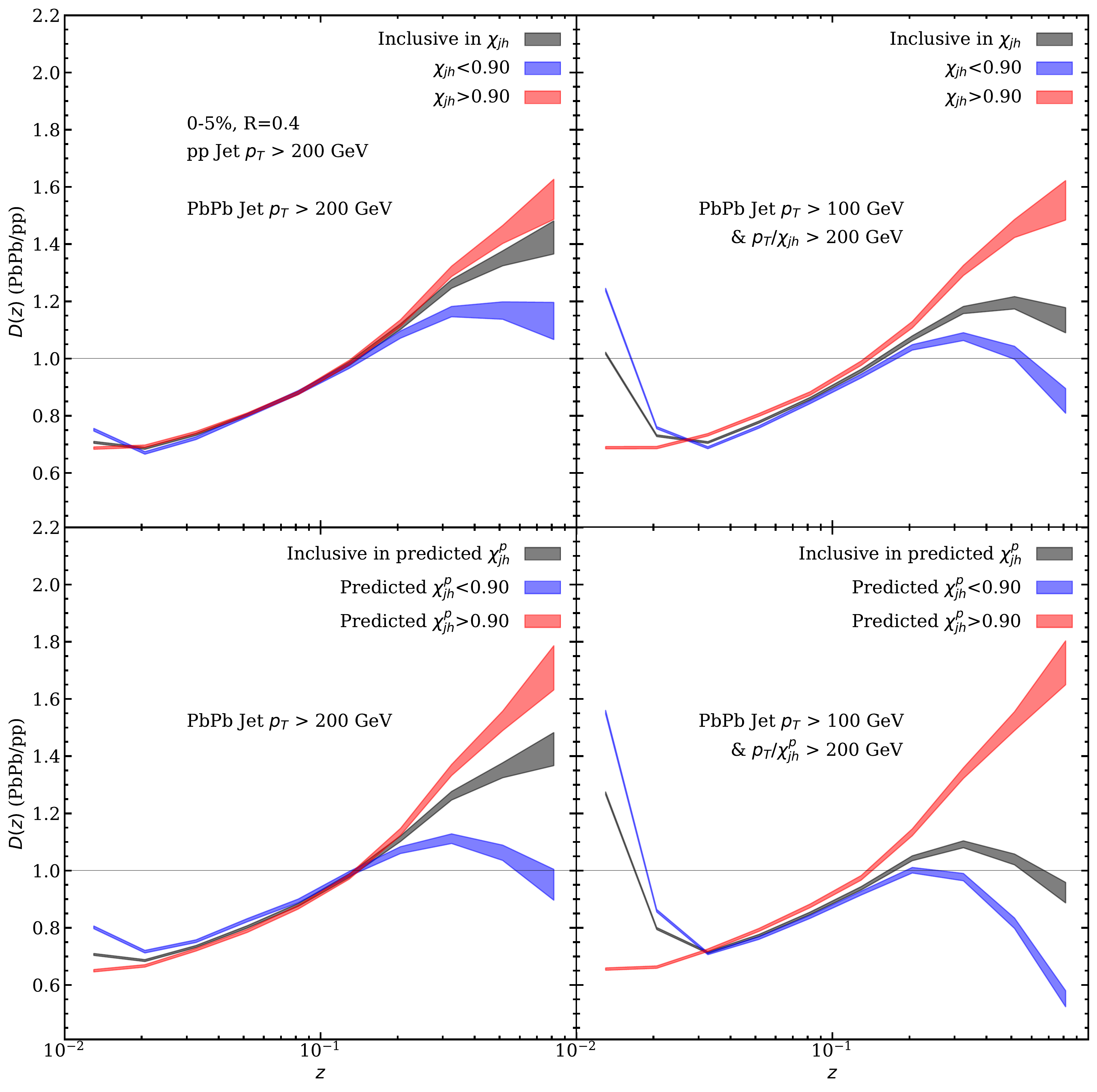}
\caption{Results comparing the JFF ratio between PbPb and pp for the true $\chi_{jh}$ (top row) and predicted $\chi_{jh}^p$ (bottom row), and the FES (left column) with the IES (right column).}
\label{ffratio}
\end{figure}

In this Section we analyze two of the most widely studied substructure observables in jet quenching, namely the jet fragmentation function (JFF) and the jet shape (JS). We apply the same methodology to unravel effects of jet quenching as in the previous Section.

From the left column of Fig.~\ref{ffratio} we can infer that the modification of the jet fragmentation function of the ``unquenched'' class is an indication of the effect of the selection bias in this observable. Indeed, we observe an enhancement of the large $z$ contribution, this is, a bias towards jets that are less fragmented than the average. Such hardening of the inclusive fragmentation function in the medium, also called jet collimation or core narrowing, has been measured by ATLAS~\cite{Aaboud:2017bzv,Aaboud:2018hpb} and is reproduced by several jet quenching models~\cite{Spousta:2015fca,KunnawalkamElayavalli:2017hxo,Casalderrey-Solana:2018wrw,Tachibana:2018yae,Caucal:2020xad}. Even though in the FES setup of Fig.~\ref{ffratio} we observe a non-negligible modification for the ``quenched class'', consisting in a moderate depletion of the energy of the hardest fragments in the jet, such modifications are greatly enhanced for the IES setup. In particular, in the latter selection we can see a clear enhancement of the contribution from the soft particles coming from medium response, which is both because there is a larger number of such particles in the jets belonging to the ``quenched'' class of IES, and also because their relative energy fraction $z$ with respect to the energy of the jet is now larger than for the higher $p_T$ jets from FES.

\begin{figure}[t!]
\centering
\includegraphics[width=0.8\textwidth]{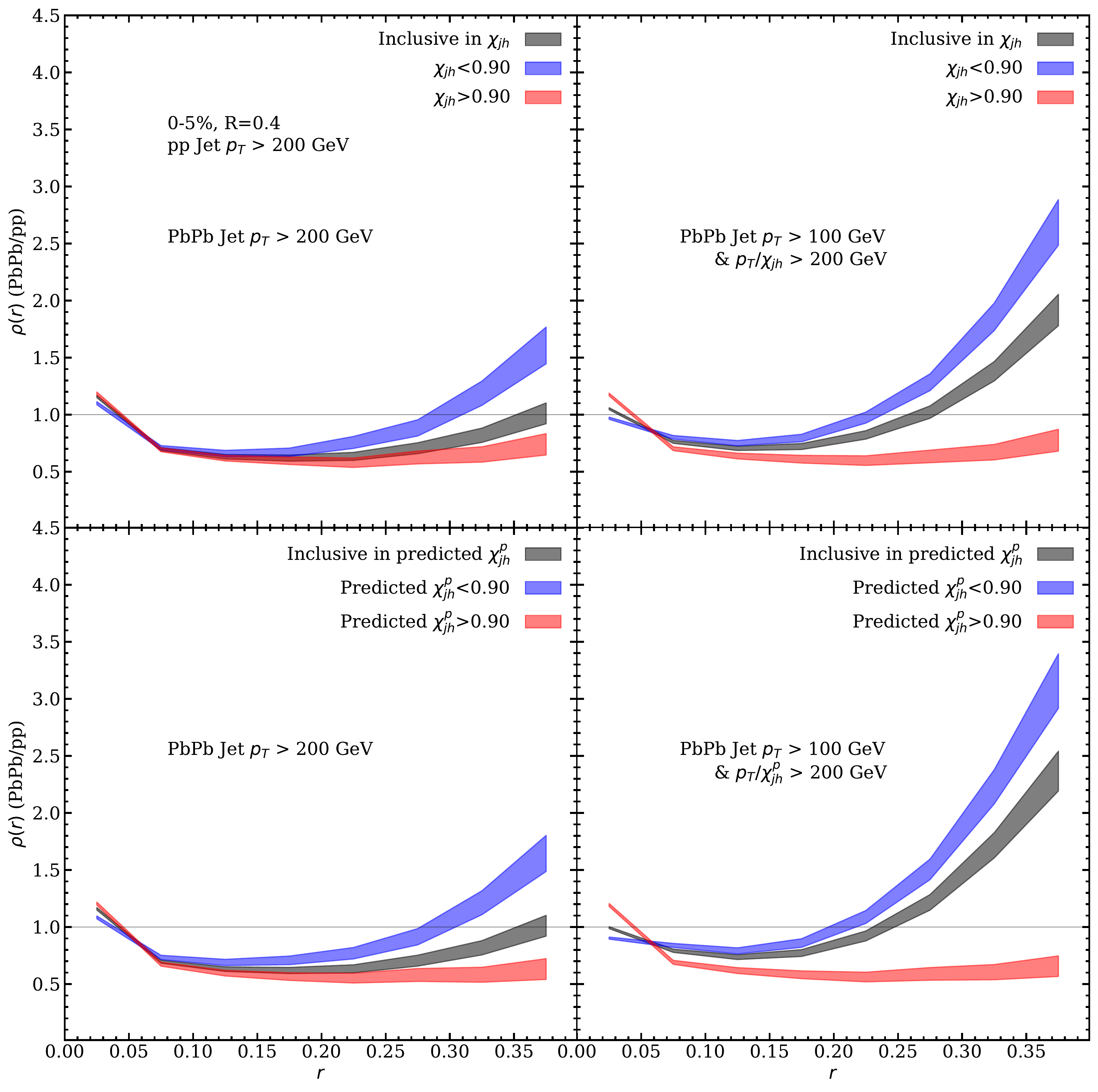}
\caption{Results comparing the JS ratio between PbPb and pp for the true $\chi_{jh}$ (top row) and predicted $\chi_{jh}^p$ (bottom row), and the FES (left column) with the IES (right column).}
\label{shapratio}
\end{figure}
Finally, very similar qualitative statements apply to the jet shape ratio shown in Fig.~\ref{shapratio}. Both the notable depletion of the energy from the partons at the core, accompanied by the sizeable enhancement of low $p_T$ particles at the periphery, is observed when going from the FES to the IES setup. However, for this observable, the separation between the ``quenched'' and ``unquenched'' class bands is wider throughout all bins in $r$ than for the fragmentation functions across the bins in $z$, as seen from Fig.~\ref{ffratio}. This is consistent with our findings in Section~\ref{sec:interpret} about the modification of JS containing more information on $\chi_{jh}$ than JFF. The inclusive jet shape has been measured in heavy-ion collisions by CMS~\cite{Chatrchyan:2013kwa,Khachatryan:2016tfj}, ATLAS~\cite{Aad:2019igg} and ALICE~\cite{Acharya:2019ssy}, and also for semi-inclusive boson-jet systems by CMS~\cite{Sirunyan:2018ncy}. Interestingly, the modifications observed in the IES setup resemble more those of the boson-jet systems, which is due to the fact that in such analysis the triggered boson $p_T$ serves as a proxy for the initial energy of the jet (this also applies to the results from JFF in Fig.~\ref{ffratio}, which have been measured in boson-jet systems by CMS~\cite{Sirunyan:2018qec} and ATLAS~\cite{Aaboud:2019oac}). These set of jet shape measurements are well reproduced by a variety of jet quenching models~\cite{KunnawalkamElayavalli:2017hxo,Casalderrey-Solana:2018wrw,Tachibana:2018yae,Caucal:2020xad,Chang:2019sae} and analytic computations~\cite{Chien:2015hda}.

\subsection{Tomography}
\label{sec:tomography}

The second application that we explore in this paper involves the relation between jet energy loss and traversed length $L$ across the QGP, as defined in Eqs.~\eqref{eq:lengtheq} and \eqref{eq:lengthjet}. By selecting jets that have suffered different amounts of energy loss using our newly introduced technique to extract $\chi_{jh}$, we are in fact selecting jets with different in-medium path lengths $L$. This new capability opens the path to the extraction of QGP properties through jet quenching analysis, or in other words, tomographic applications, in ways that were not possible before \footnote{See, however, a recent study based on gradient tomography~\cite{He:2020iow} which gives access to the creation point of an energetic jet using alternative methods.}.
\begin{figure}[t!]
\centering
\includegraphics[width=0.48\textwidth]{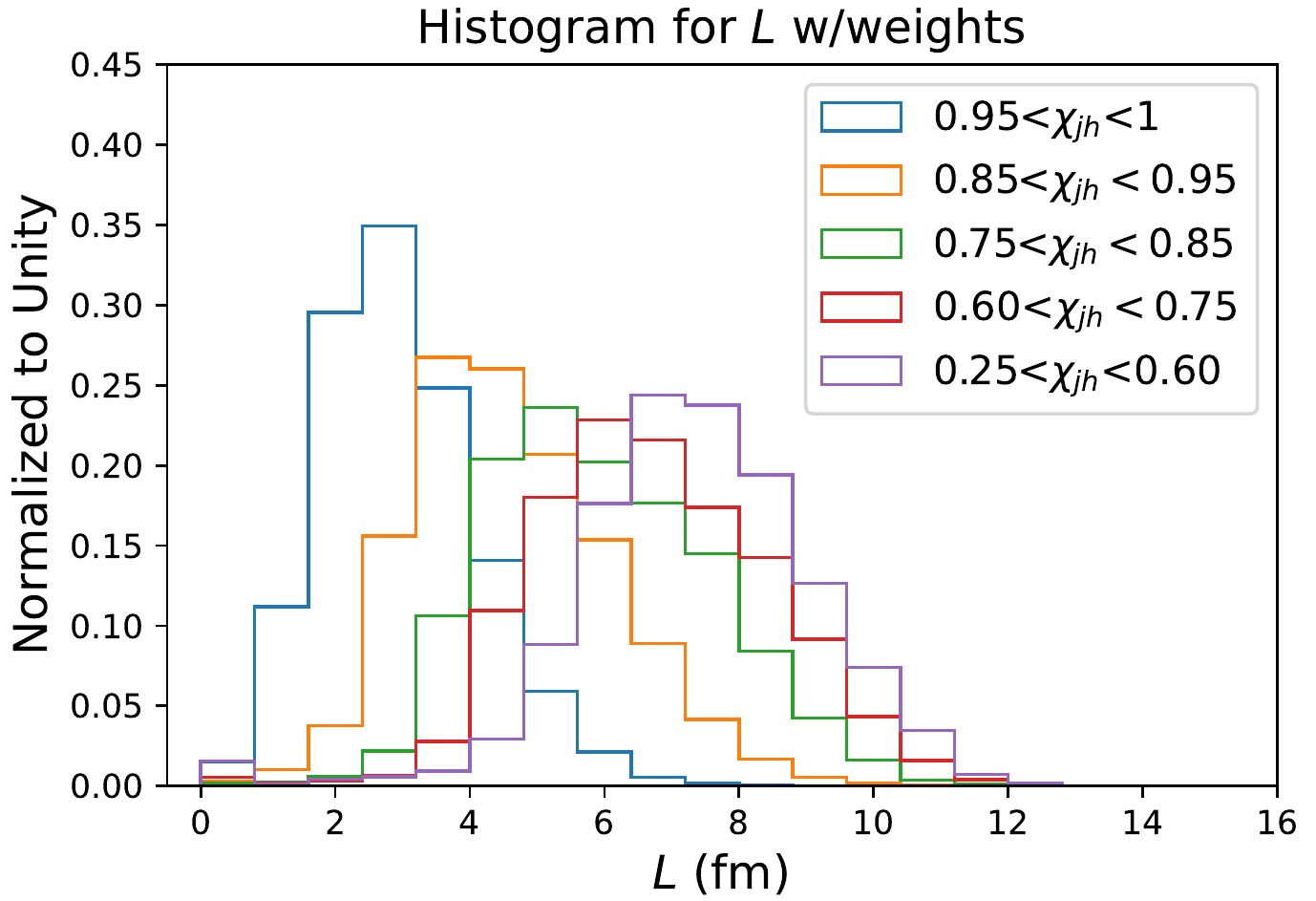}
\includegraphics[width=0.48\textwidth]{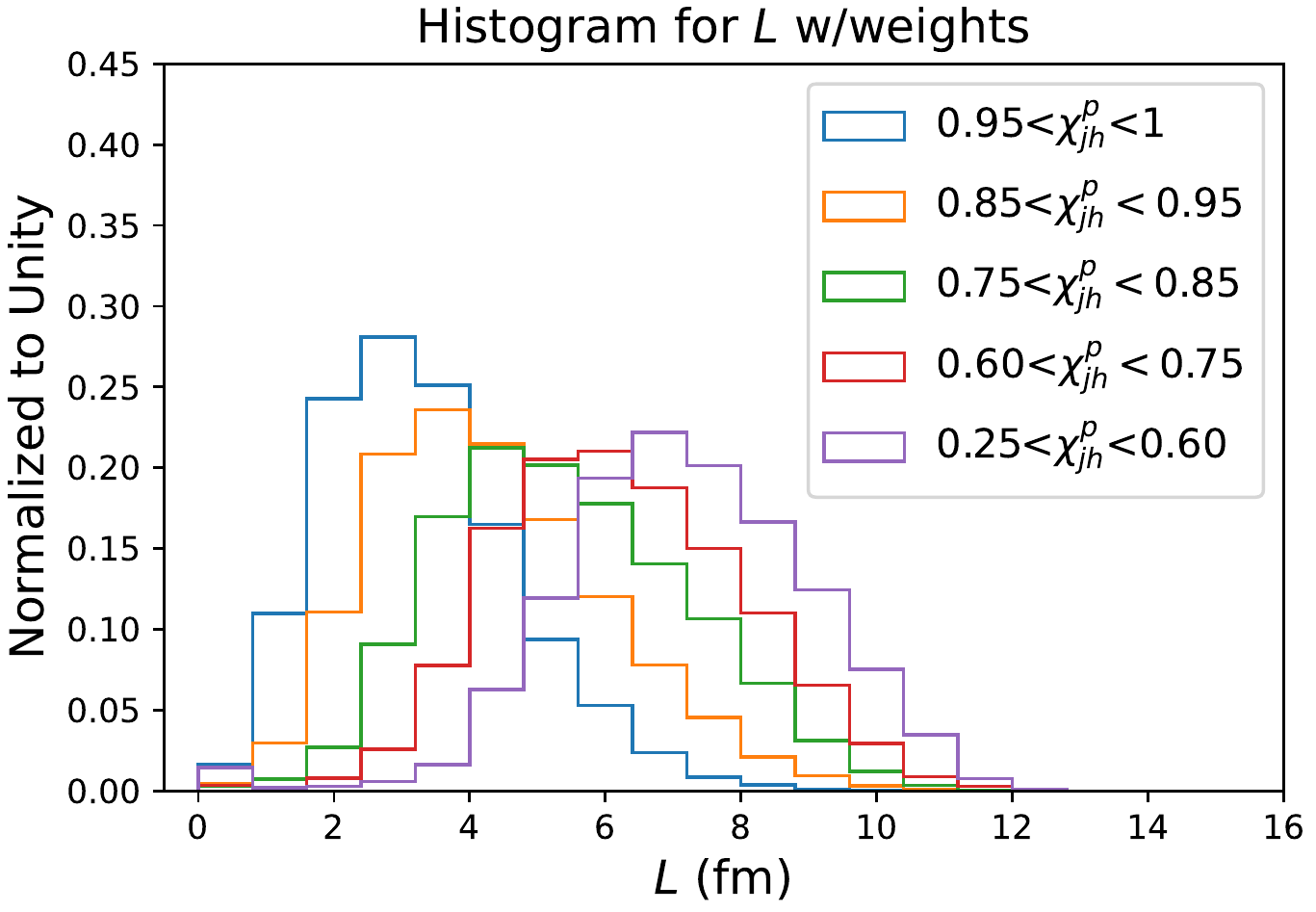}
\caption{Left: distributions for the length traversed in the QGP of jets within different ranges of the true value of $\chi_{jh}$. Right: same as left panel, but using the predicted value $\chi_{jh}^p$.}
\label{Ldists}
\end{figure}

\begin{figure}[t!]
\centering
\includegraphics[width=0.8\textwidth]{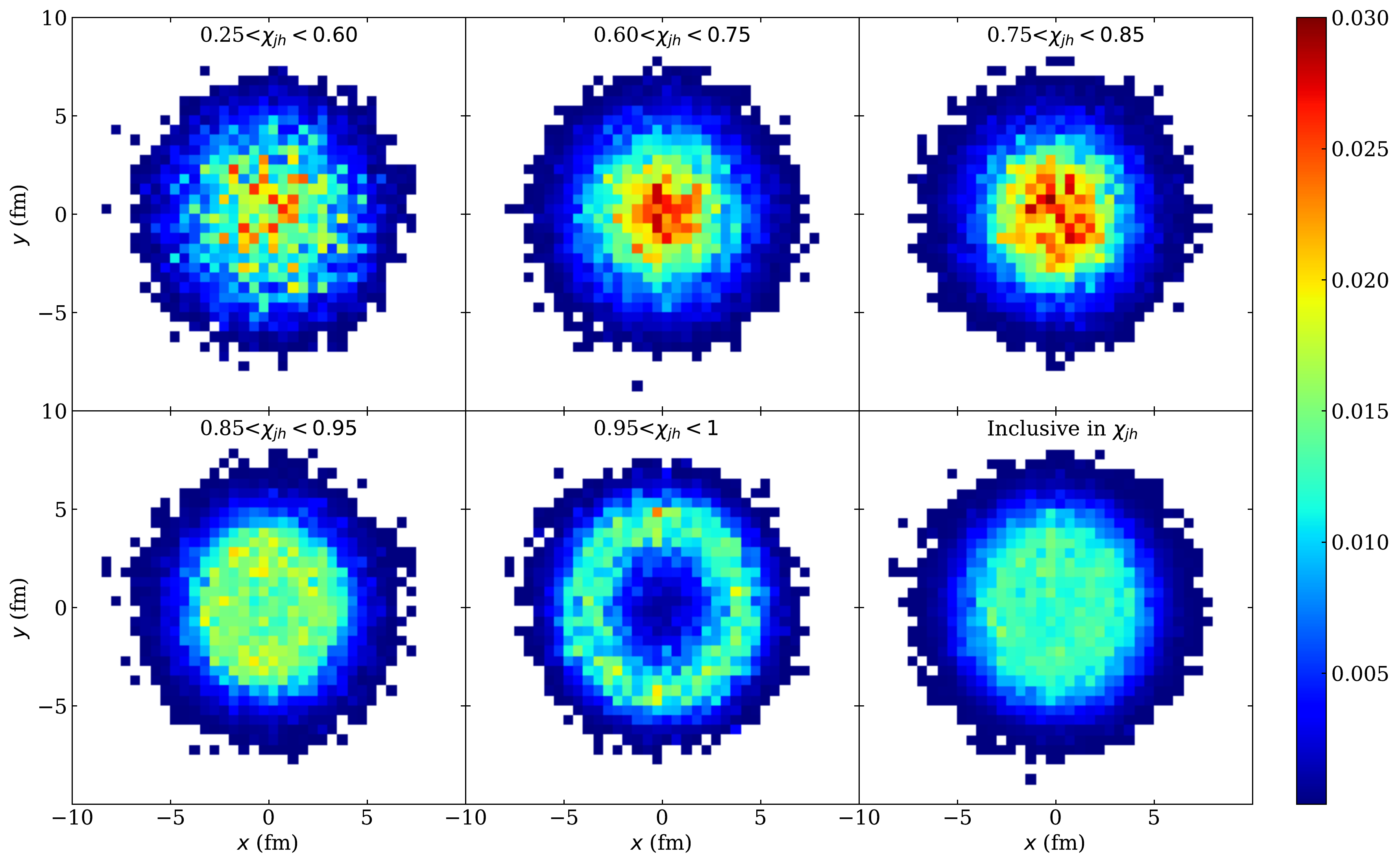}
\includegraphics[width=0.8\textwidth]{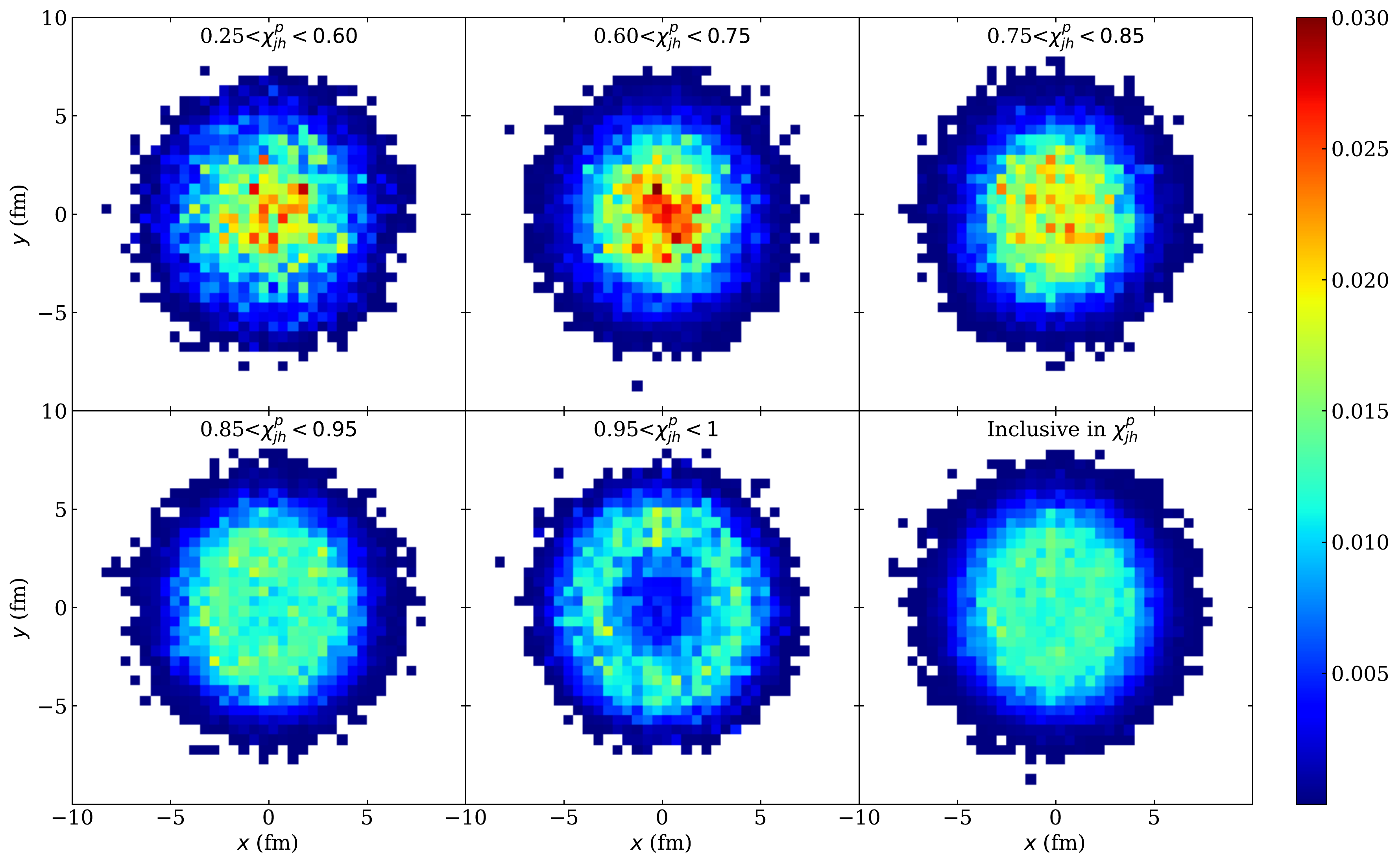}
\caption{Creation point distributions in the transverse plane for different ranges of the true value $\chi_{jh}$ (top panel) as well as the predicted value $\chi_{jh}^p$ (bottom panel).}
\label{creation}
\end{figure}

It is important to keep in mind that our machine learning setup does not learn to predict $L$ from the data.
It would indeed be desirable to extract the in-medium path length $L$ jet-by-jet directly, but carrying out such task with good enough precision has proven to be considerably more challenging than the extraction of $\chi_{jh}$ and will hopefully be tackled in future work.
The application in this Section is therefore ``hybrid'' in the sense that we supplement the information extracted by the machine learning setup, i.e., $\chi_{jh}$, with information provided by the model that produced the data.

The in-medium length distributions $L$ are shown in Fig.~\ref{Ldists}, both for the true value of $\chi_{jh}$ (left panel) and its predicted value $\chi_{jh}^p$ (right panel). We stress again that the value of $L$ for each jet is not extracted by our algorithm but it is taken directly from the Monte Carlo. One can clearly see that the average value of $L$ grows as the energy loss increases, as expected. This should mean that we are selecting jets that were created at different points in the transverse plane. Indeed, we can corroborate this statement by looking at the creation point distribution in the impact parameter plane $\lbrace x,y\rbrace$, presented in Fig.~\ref{creation}. Again, the pair of values $\lbrace x,y\rbrace$ are not extracted through our algorithm, but taken from the Monte Carlo instead. We observe that jets that are barely quenched were originated within a ring structure at the periphery of the collision. The production point shifts towards the centre of the initial geometry as energy loss is increased. Note that for the most quenched class, $0.25<\chi_{jh}<0.6$, the distribution becomes more spread than for the next less quenched class, $0.6<\chi_{jh}<0.75$. This is because the very quenched jets have had to go through the largest in-medium lengths, which can correspond to configurations in which the di-jet pair was created at the periphery and where one of the jets flew inwards, towards the center of the QGP. This observation points to a possible improvement of the current analysis by taking into account the orientation of a jet with respect to the event plane of the collision, which will be left for future work. 

\section{Conclusions and outlook}
\label{sec:Conclusions}

In this work we employed deep learning techniques to predict jet energy loss on a jet-by-jet basis using jet images and various physics-motivated features in a systematic and comparable way. Good performance has been achieved. We have also assessed and interpreted the success of the network using a multi-faceted approach.

We have found that the soft particles within the jet cone carry very relevant information about the magnitude of energy loss. In particular, the total number of activated pixels of the jet image, or jet multiplicity, is dominated by the production of thermal particles due to quenching and is found to be highly correlated with the energy loss ratio. This observation motivates us to construct the ``hard ratio'', which we define as the relative contribution of hard particles ($p_T>2$ GeV) to the total energy of the jet. Even though it exhibits a strong correlation with the energy loss ratio, its performance is notably worse than that obtained with deep learning techniques. Furthermore, jet shapes and jet fragmentation functions, which are the lower-dimensional projections of the normalized jet image, have also been used as inputs to the neural network.
The good results obtained with the jet shapes, comparable to those obtained via the normalized jet image, provide an alternative scheme and a candidate interpretation of the good performance given by the jet image. 
We corroborate these findings by constructing a universal 17-parameter fit that allows to extract the energy loss ratio with great precision merely by using the 8 bins of a given individual jet shape. 

The ability to extract an estimate of the energy lost by jets produced in heavy-ion collisions, achieved here through the extraction of an energy loss ratio, opens for a plethora of interesting applications and studies. As a proof of concept of the suitability of deep learning techniques to jet quenching phenomenology we have focused on two illustrative applications. First, we have used our knowledge about the relative energy loss that a given jet has suffered in order to study its observable modifications based on its initial energy.
This removes to a large extent the effect of the selection bias on jet observables and enhances the contribution of very quenched jets, allowing to study jets that have actually suffered considerable energy loss and modification. We note that it would be very interesting to extract not only the initial energy of the jet, but also some measure of its initial width, or some other equivalently relevant initial feature. This would allow us to construct nuclear modification ratios in which the vacuum distributions in the denominator would not simply correspond to the inclusive proton-proton sample, but rather to a restricted set of jets from which the medium jets belonging to a specific quenching class come. In this way, selection bias effects that are still present within each quenching class could also be removed. This will be pursued in a forthcoming study.

\begin{figure}[t!]
\centering
\includegraphics[width=0.8\textwidth]{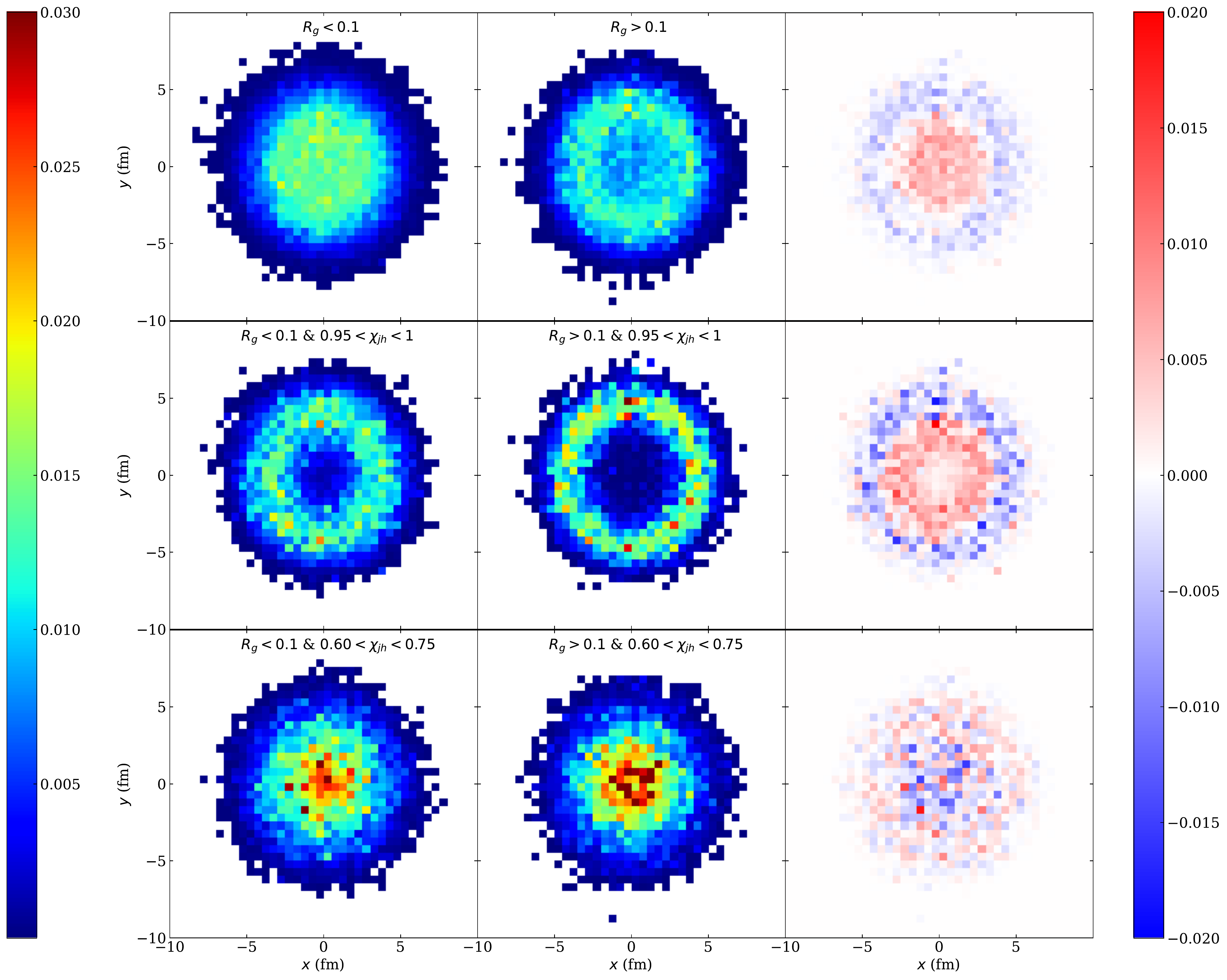}
\caption{The normalized distribution of the creation point in the transverse plane for jets with measured $R_g<0.1$ (left column) versus $R_g>0.1$ (middle column), and their difference (right column). Jets are required to have $p_T>100$ GeV and are inclusive in $\chi_{jh}$ (upper row), in unquenched class with $0.95<\chi_{jh}<1$ (middle row) and quenched class with $0.6<\chi_{jh}<0.75$ (lower row). Regression of $\chi_{jh}$ could be achieved by the neural network, while the production points are inputs from the hybrid model.}
\label{deltaRcreation}
\end{figure}

Second, we have shown that using the extracted energy loss ratio we have a handle on which jets have traversed longer distances within the QGP, which constitutes an important step towards the exploitation of the tomographic power of hard probes. Extensions to the limited scope of the presented analysis should include the orientation of the jet with respect to the event plane of the collision, as well as more differential selections based on measures of the jet width such as, for example, $R_g$. Indeed, given what we know about the dependence of energy loss on the width of a jet, and its relation to selection bias, as discussed in Section~\ref{sec:observables}, it is easy to anticipate that jets with different (initial) widths, and the same value of the energy loss ratio $\chi_{jh}$, will in general have traversed different amounts of length. In Fig.~\ref{deltaRcreation} we show some results supporting this picture (using the true value of $\chi_{jh}$). In this Figure we show the creation point distribution in the transverse plane, inclusive in $\chi_{jh}$, for jets with a measured (this is, final) $R_g$ smaller or larger than 0.1, with a measured $p_T>100$ GeV (here again, $\chi_{jh}$ could be estimated by the neural network while the production points are inputs from the hybrid model). We see that narrower jets tend to pass the momentum cut even if they were produced deep inside the medium, unlike wider jets which in comparison are pushed towards the surface. Similarly to what was discussed in Fig.~\ref{creation}, such ordering is reversed for the very quenched class shown in the bottom row of Fig.~\ref{deltaRcreation}, since narrow jets that get this quenched can tend to be produced towards the periphery and propagate inwards.

Finally, we bring up two potential shortcomings of our approach. Both are related to the sensitivity of soft radiation at the jet periphery. On the one hand, sensitivity to such modes leads to an expected drop of accuracy once jets are embedded into a realistic heavy-ion environment due to the presence of a large fluctuating background. Improving the performance under these conditions is highly desirable in order to extend our studies to the analysis of real experimental data. On the other hand, the modeling of such emissions is currently not under firm theoretical control, which could imply dependency on a particular model. Let us comment briefly on these two points below.


In order to check the robustness of our results in a realistic scenario, we embedded our jets into a thermal heavy-ion background and performed background subtraction using constituent subtraction. To this end, we use the JetToyHI tools available in \cite{JetToyHI}. The signal particles of the reconstructed jet conform what we here call the embedded jet image. As expected, performance is decreased due to the noise associated to the presence of the large, fluctuating background, consisting of particles whose $p_T$ is of the order of that of the soft particles originated through the quenching mechanisms. Improving the performance under the conditions in which jets are actually measured in experiments should be one of the main goals for the near future. However, the reduction of the performance (the obtained validation loss is 0.0072) still allows the analysis of many of the features addressed in this paper. We postpone a more thorough study in this direction to the future.

We also deem necessary to test the robustness of our procedure to extract the energy loss ratio by applying it to other jet quenching models. Those Monte Carlos that rely on a scale separation between vacuum and medium physics that results into a marked hierarchy on the formation time of the corresponding emissions are particularly well suited to this endeavour. Even though these models may present different modelling assumptions regarding the form of the medium induced radiation kernels, the inclusion of coherence effects,  the hydrodynamization rate or the values of certain medium parameters, to the extent that such models yield qualitatively similar predictions for jet observables one should expect our network to be capable of finding enough common generalizable features. Furthermore, given the generic, dominating effect of selection bias that affects the jet observables of all Monte Carlos, classifying jets into energy loss classes (preferably using the IES setup) will give us a clearer insight into the observable differences among the different models.

\acknowledgments
We thank Johannes Hamre Isaksen and Adam Takacs for their careful reading and helpful suggestions. This work is supported by the Trond Mohn Foundation under Grant No. BFS2018REK01 and the University of Bergen. Y. D. thanks the support from the Norwegian e-infrastructure UNINETT Sigma2 for the data storage and HPC resources with Project Nos. NS9753K and NN9753K.

\clearpage

\appendix

\section{Correlations between jet observables}
\label{app: variables correlations}
\begin{figure}[ht]
\centering
\includegraphics[width=0.8\textwidth]{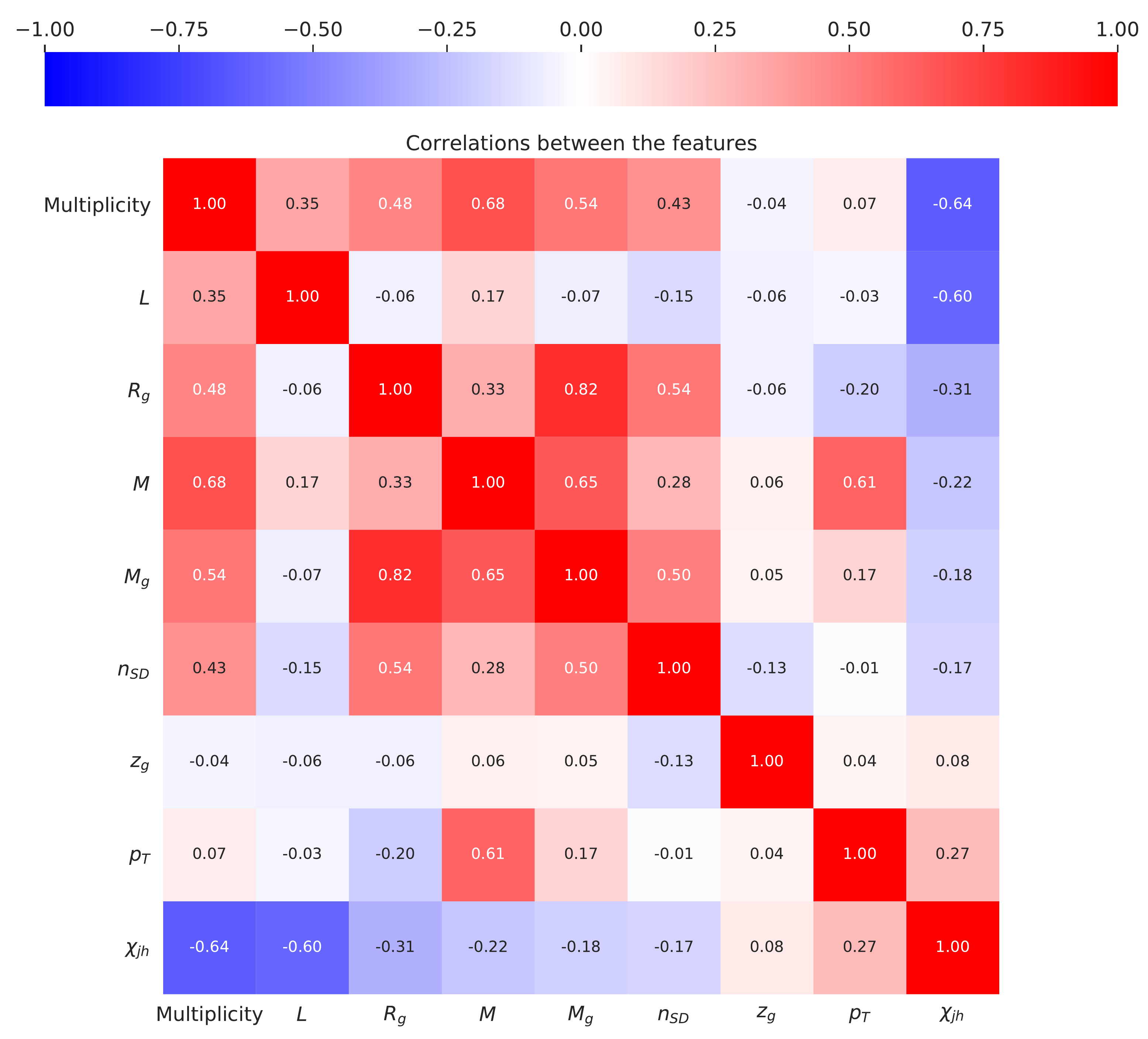}
\caption{Pearson correlation coefficient matrix between jet observables}
\label{fig:correlations_complete}
\end{figure}

Fig.~\ref{fig:correlations_complete} shows the Pearson correlation coefficient matrix between jet observables. We have several interesting observations. Firstly, the correlations between jet multiplicity, $R_g, M, M_g$ and $n_{SD}$ are all positive. In particular, the one between $R_g$ and $M_g$ is very strong. Secondly, the correlations between $z_g$ and other jet observables are generally very slight. Lastly, as we mentioned in the main text, the travel length $L$ is anti-correlated with $\chi_{jh}$, which means that longer $L$ will lead to more energy loss in general. However, its correlations with other jet observables are not always opposite, which implies that the travel length is not the unique factor governing the energy loss ratio.

\clearpage

\section{Prediction Performance versus jet observables}
\label{app:Prediction Performance}
\begin{figure}[H]
\centering
\includegraphics[width=0.46\textwidth]{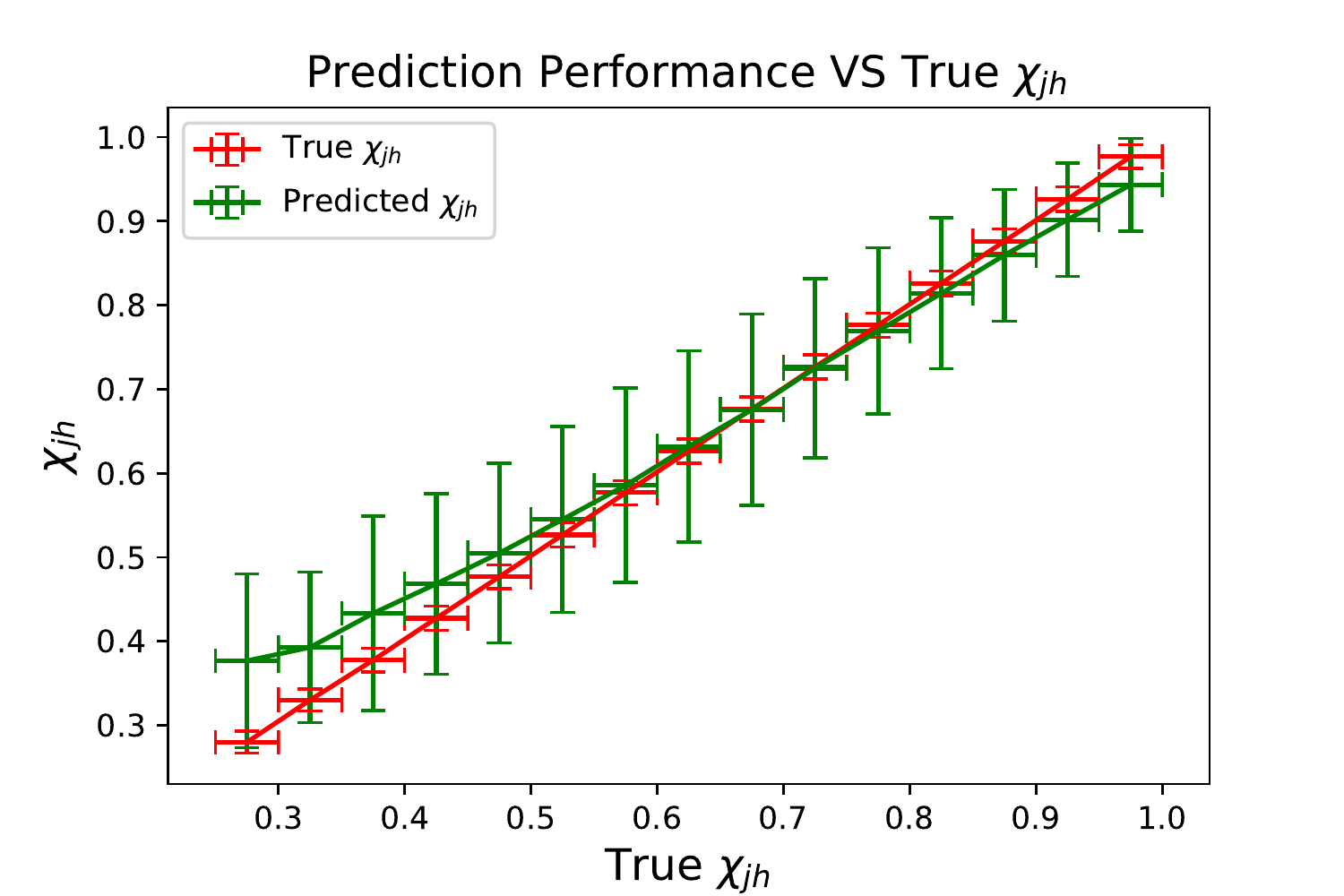}
\includegraphics[width=0.46\textwidth]{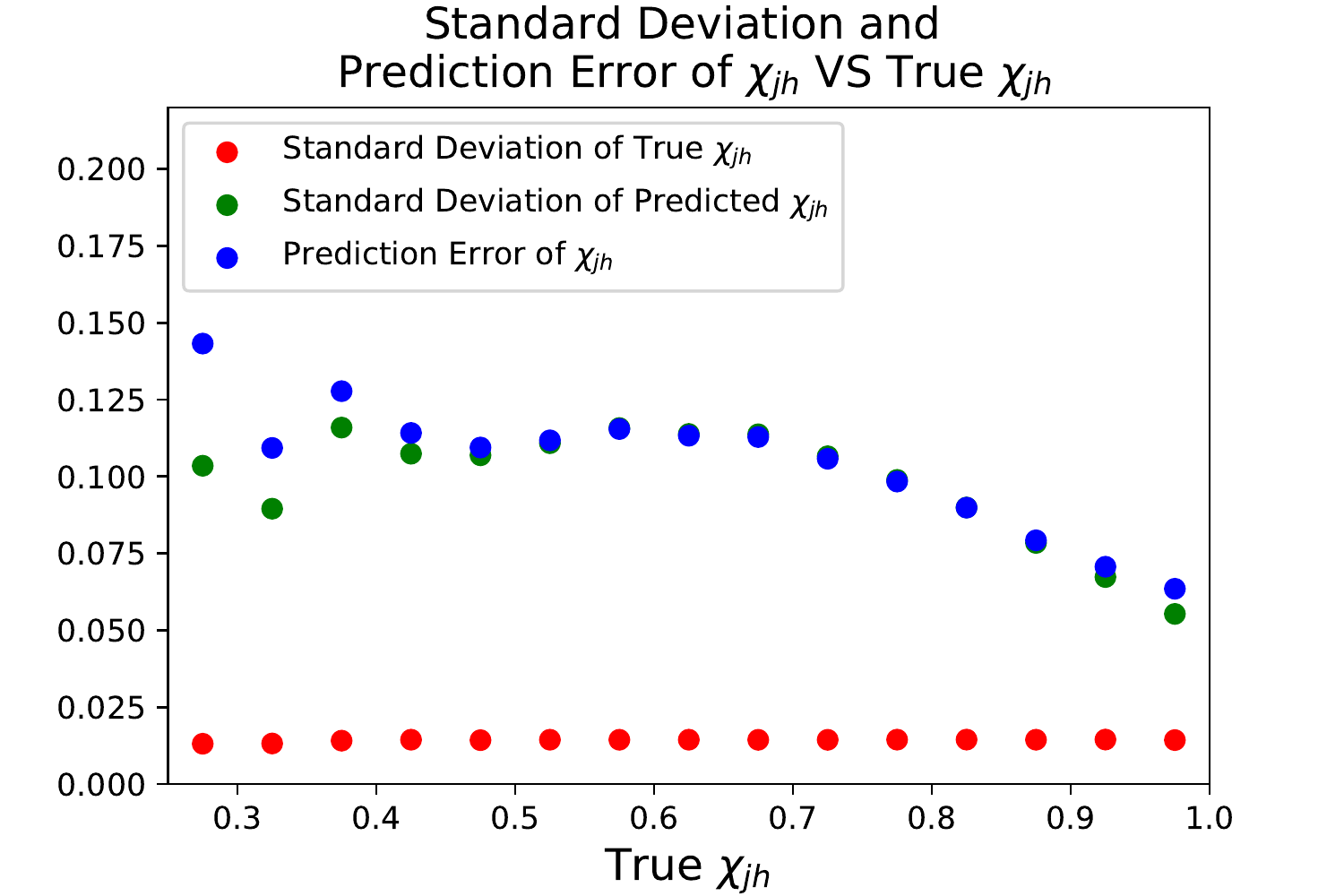}
\caption{Prediction Performance (left panel) and Standard Deviation, Prediction Error (right panel) VS $\chi_{jh}$}
\label{Performance chi}
\end{figure}

\begin{figure}[H]
\centering
\includegraphics[width=0.46\textwidth]{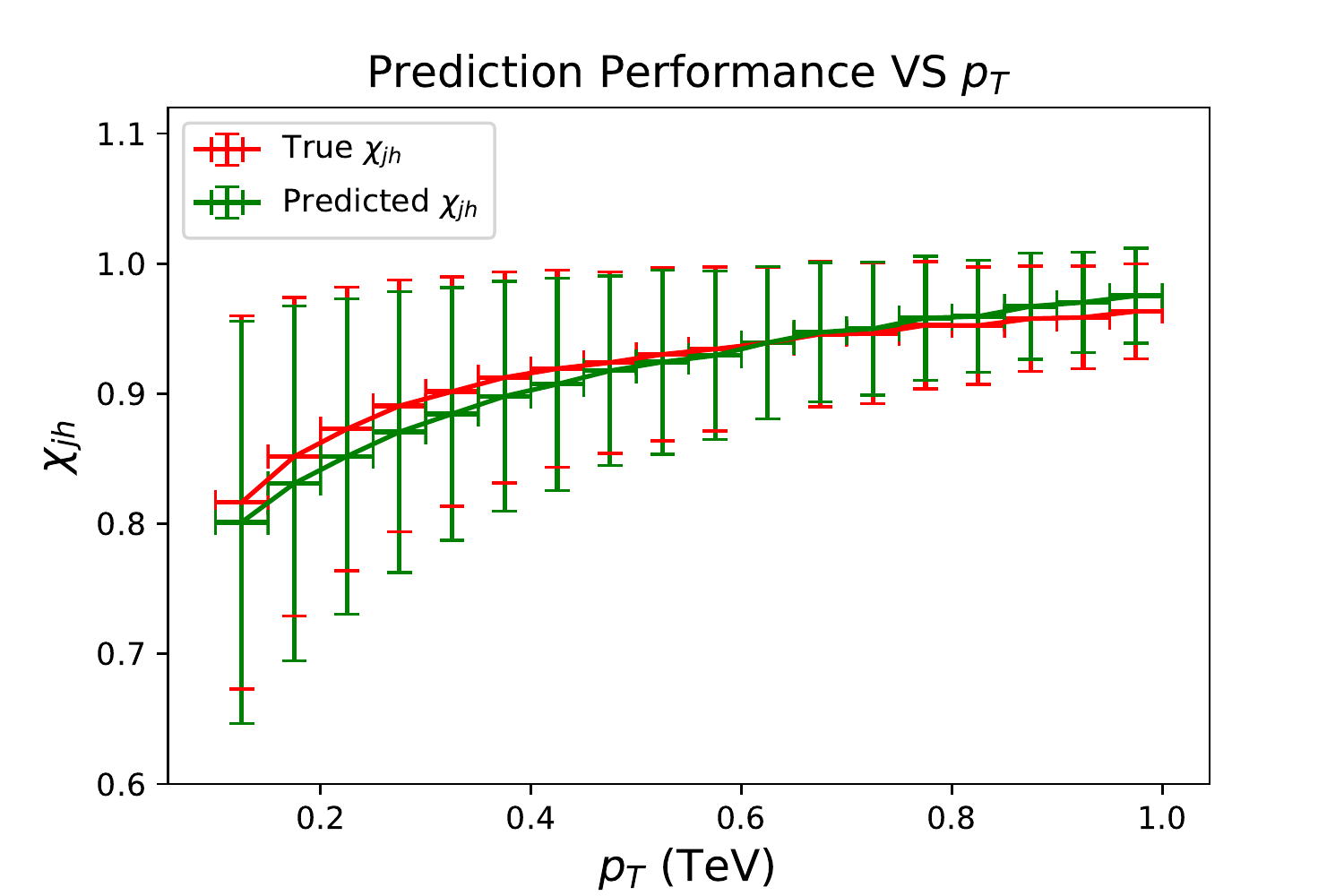}
\includegraphics[width=0.46\textwidth]{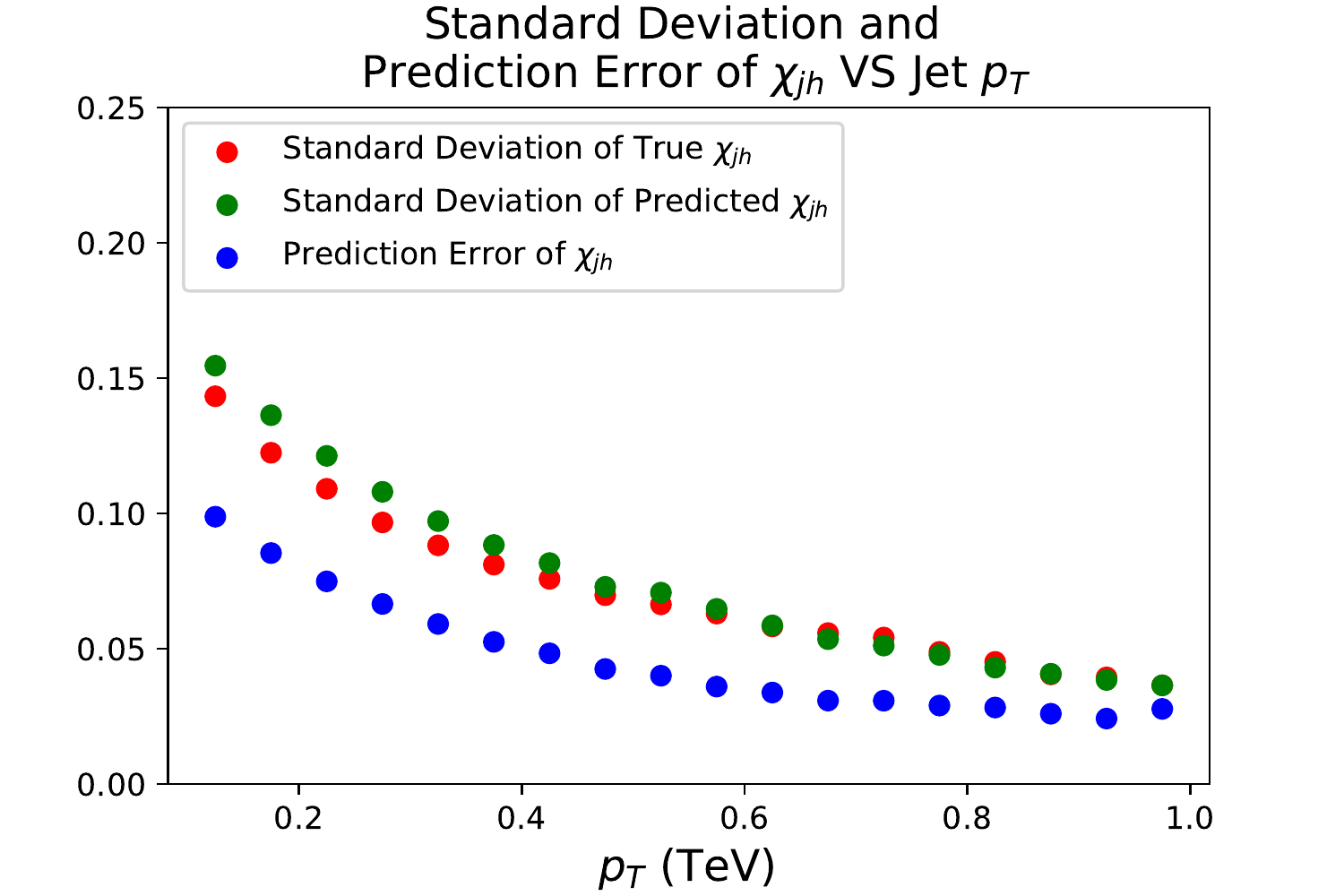}
\caption{Prediction Performance (left panel) and Standard Deviation, Prediction Error (right panel) VS Jet $p_T$.}
\label{Performance pT}
\end{figure}

\begin{figure}[H]
\centering
\includegraphics[width=0.46\textwidth]{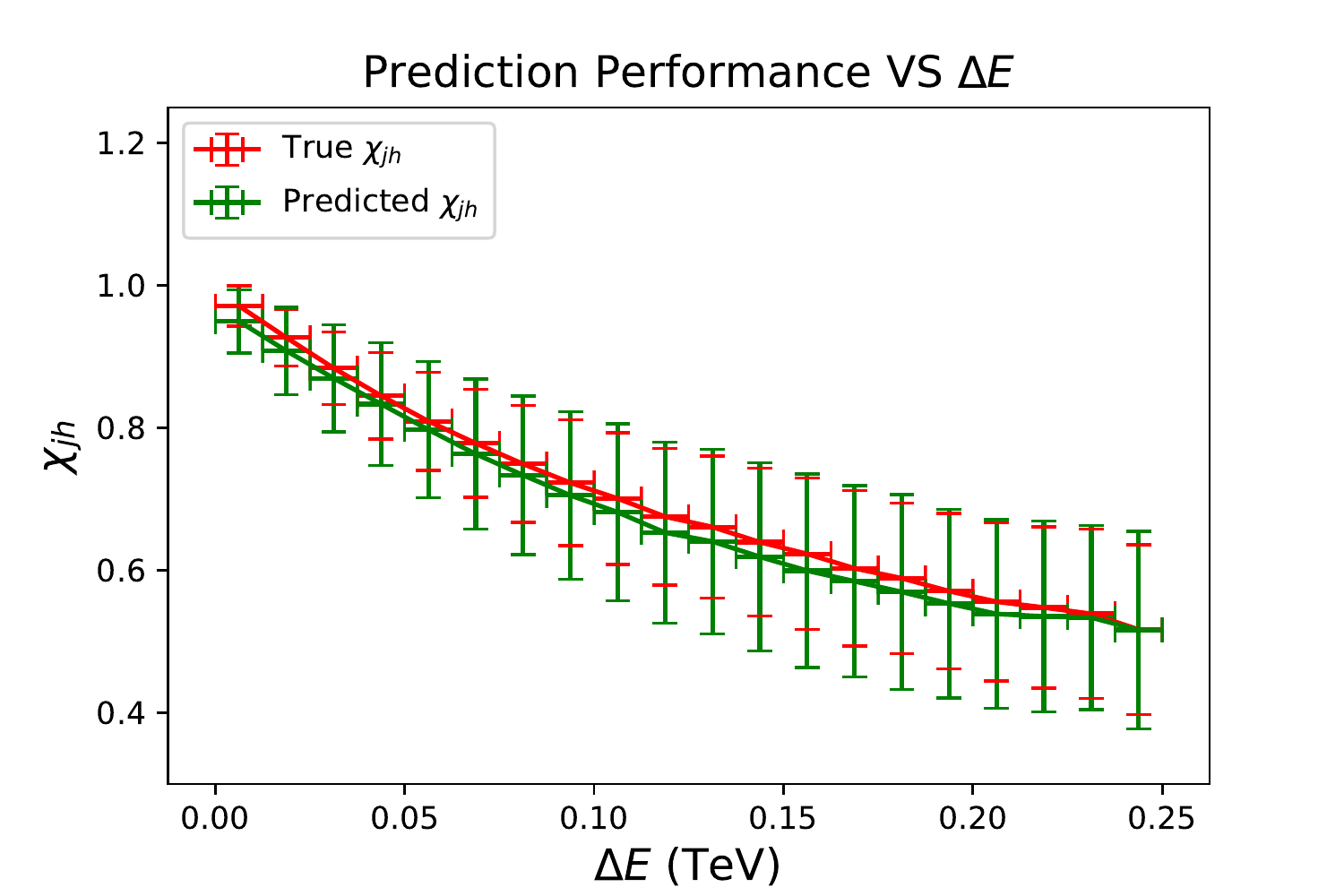}
\includegraphics[width=0.46\textwidth]{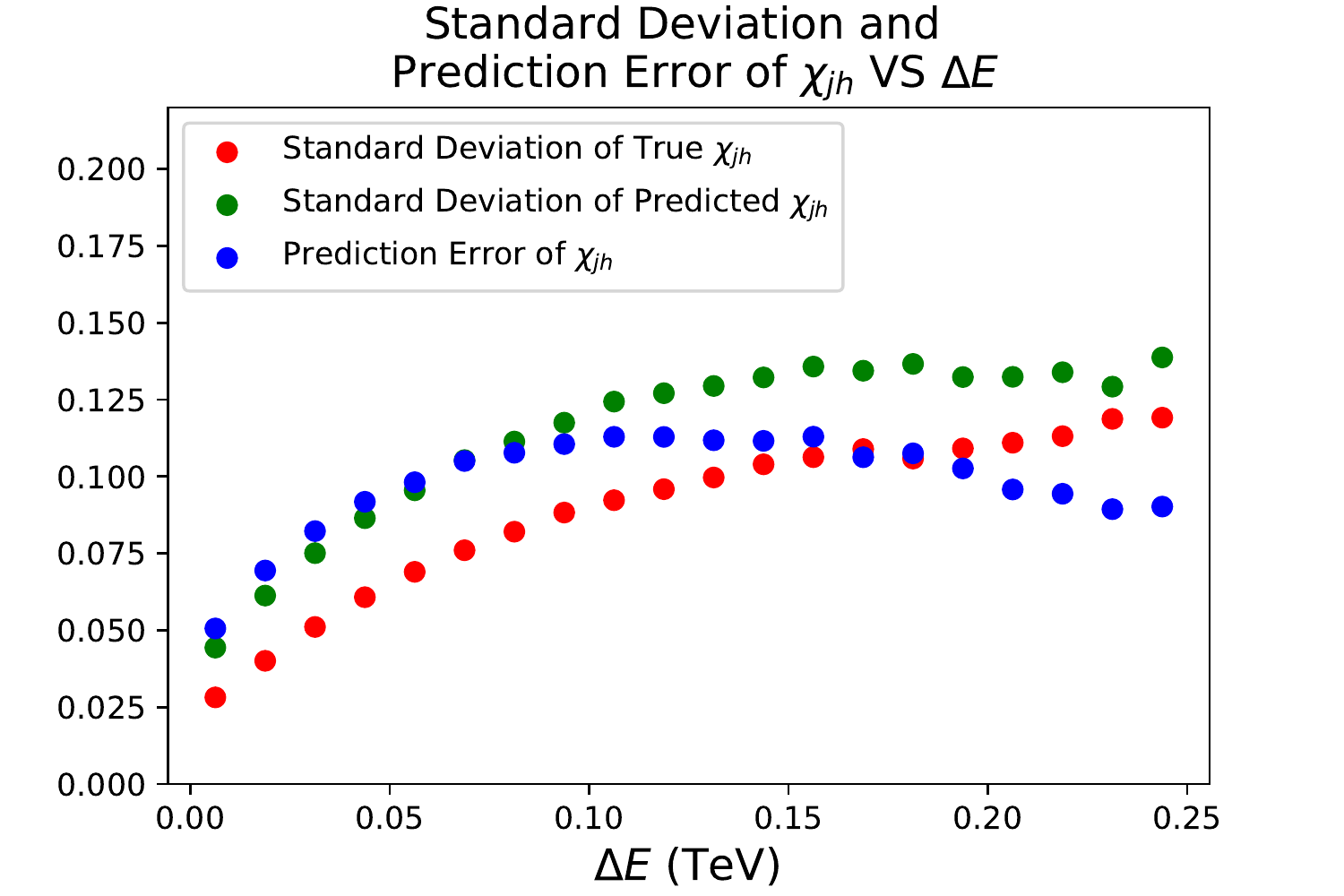}
\caption{Prediction Performance (left panel) and Standard Deviation, Prediction Error (right panel) VS Jet energy loss $\Delta E\equiv E_i^h-E_f^h$.}
\label{Performance DeltaE}
\end{figure}

\begin{figure}[H]
\centering
\includegraphics[width=0.46\textwidth]{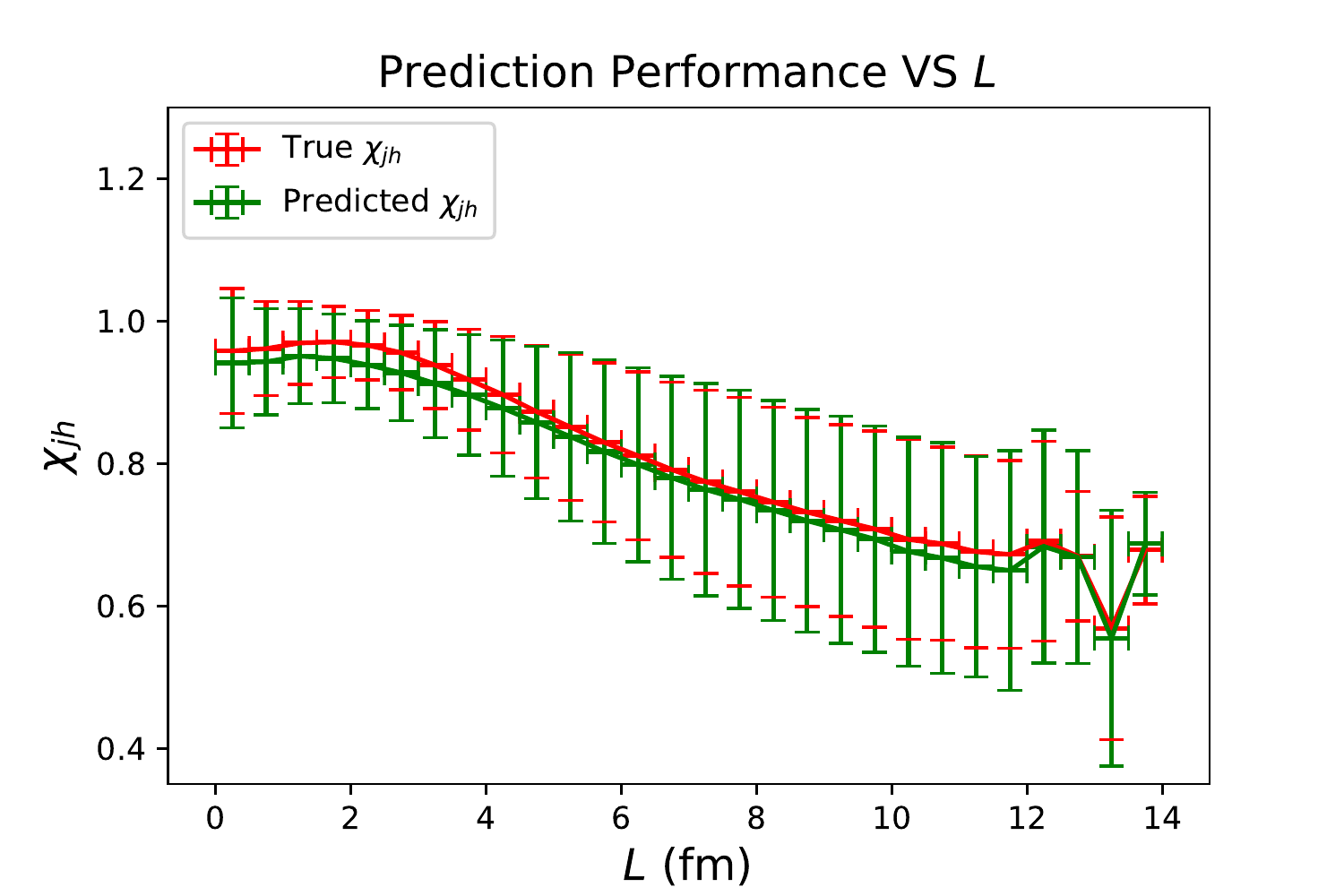}
\includegraphics[width=0.46\textwidth]{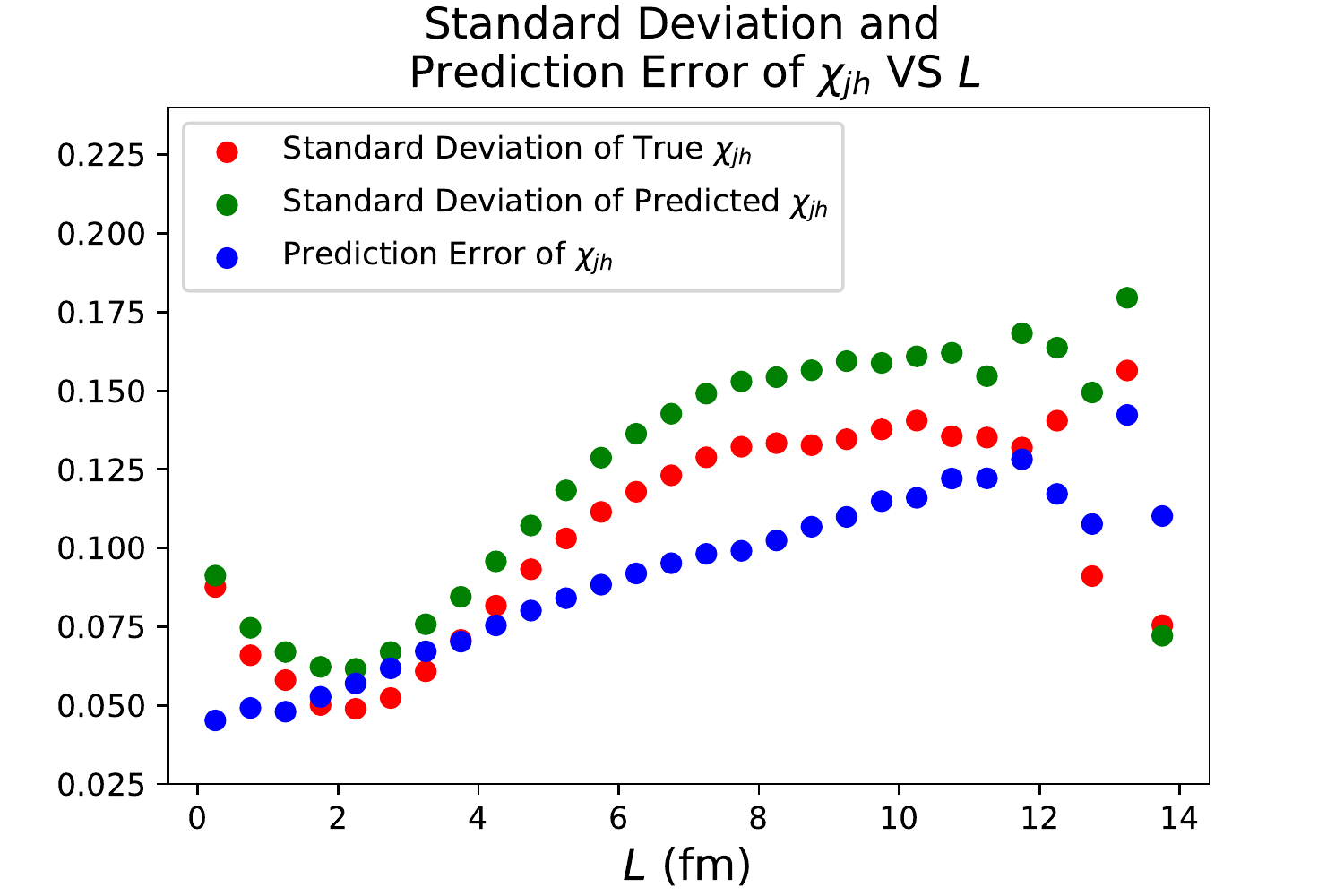}
\caption{Prediction Performance (left panel) and Standard Deviation, Prediction Error (right panel) VS $L$.}
\label{Performance L}
\end{figure}

\begin{figure}[H]
\centering
\includegraphics[width=0.46\textwidth]{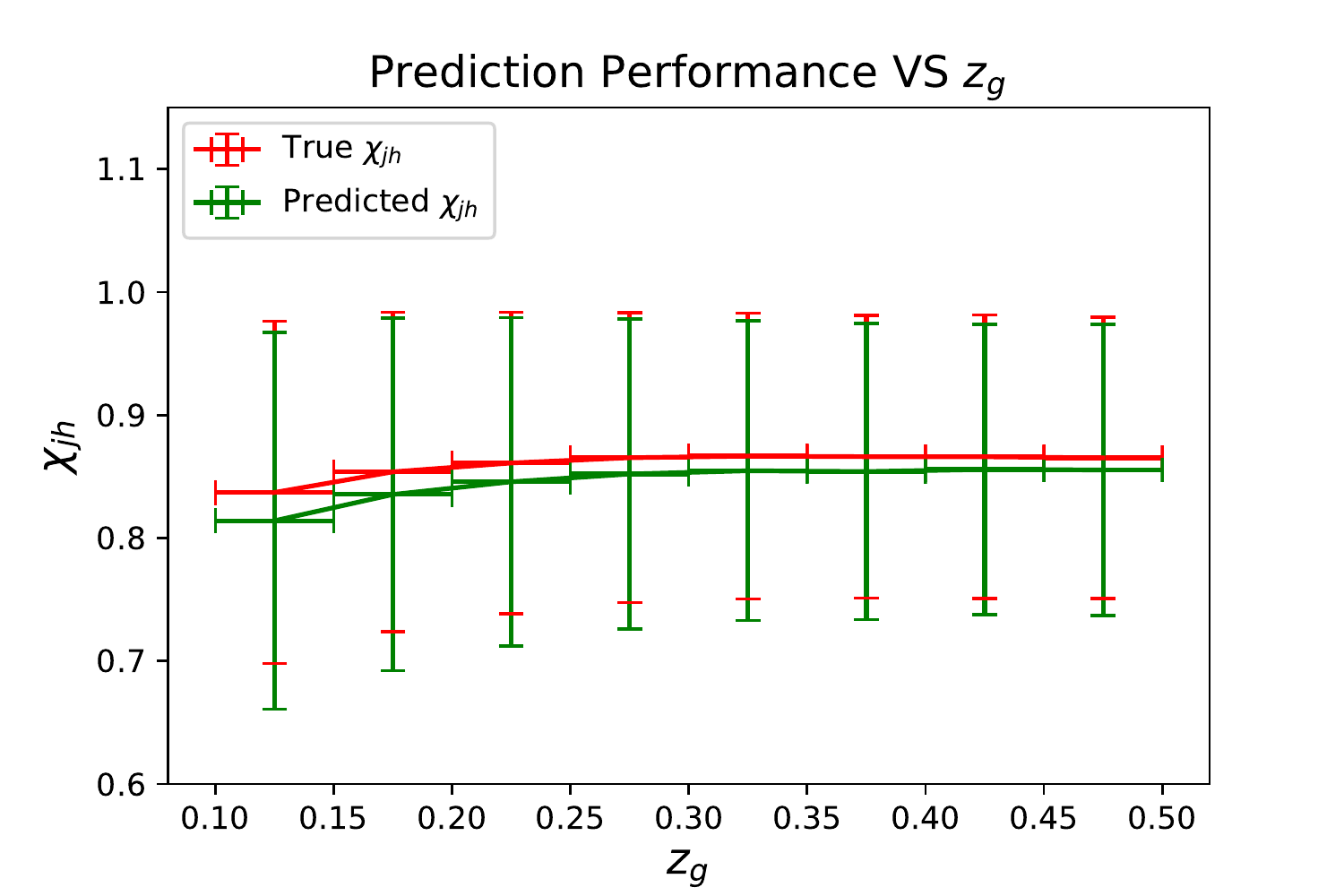}
\includegraphics[width=0.46\textwidth]{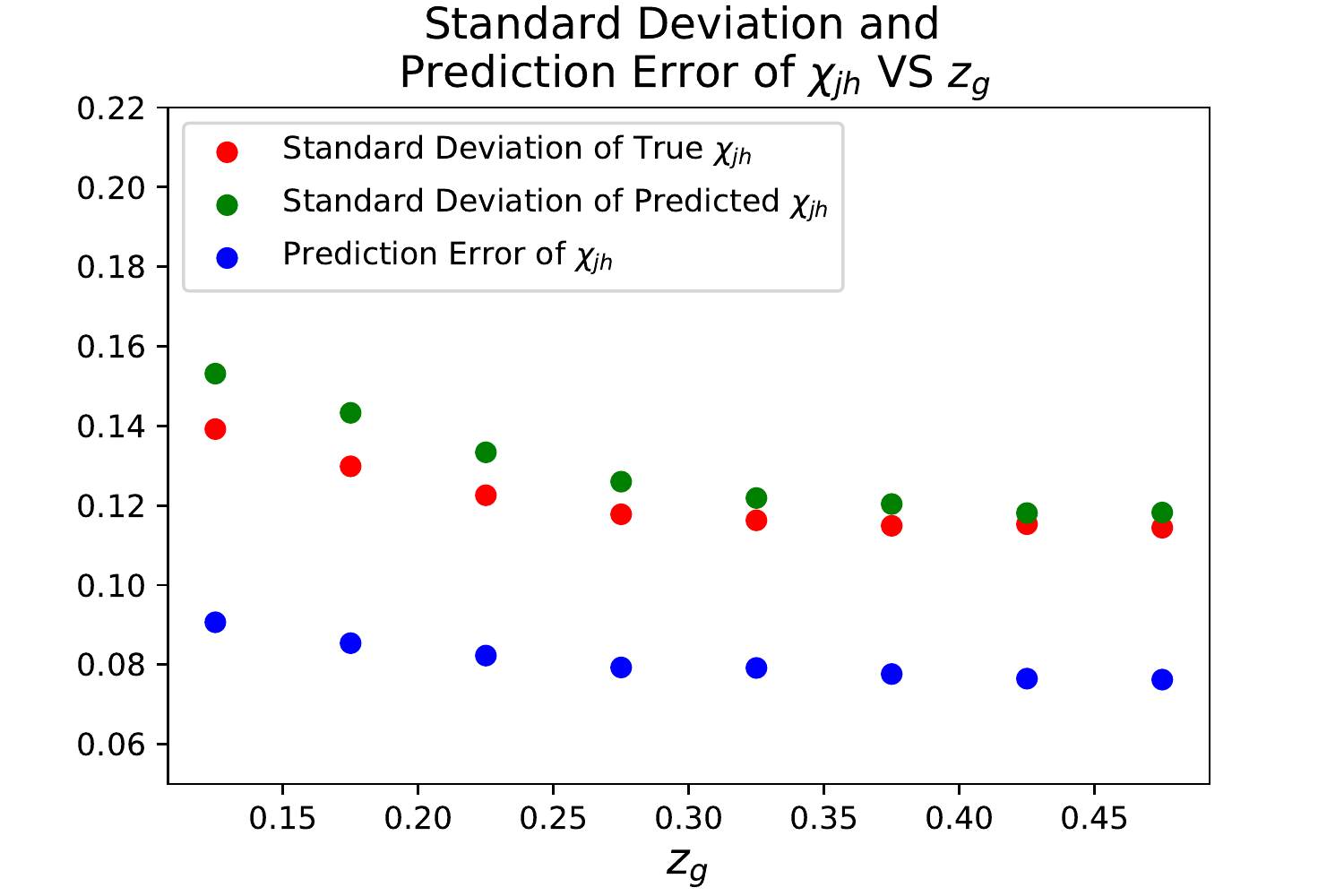}
\caption{Prediction Performance (left panel) and Standard Deviation, Prediction Error (right panel) VS $z_g$.}
\label{Performance zg}
\end{figure}

\begin{figure}[H]
\centering
\includegraphics[width=0.46\textwidth]{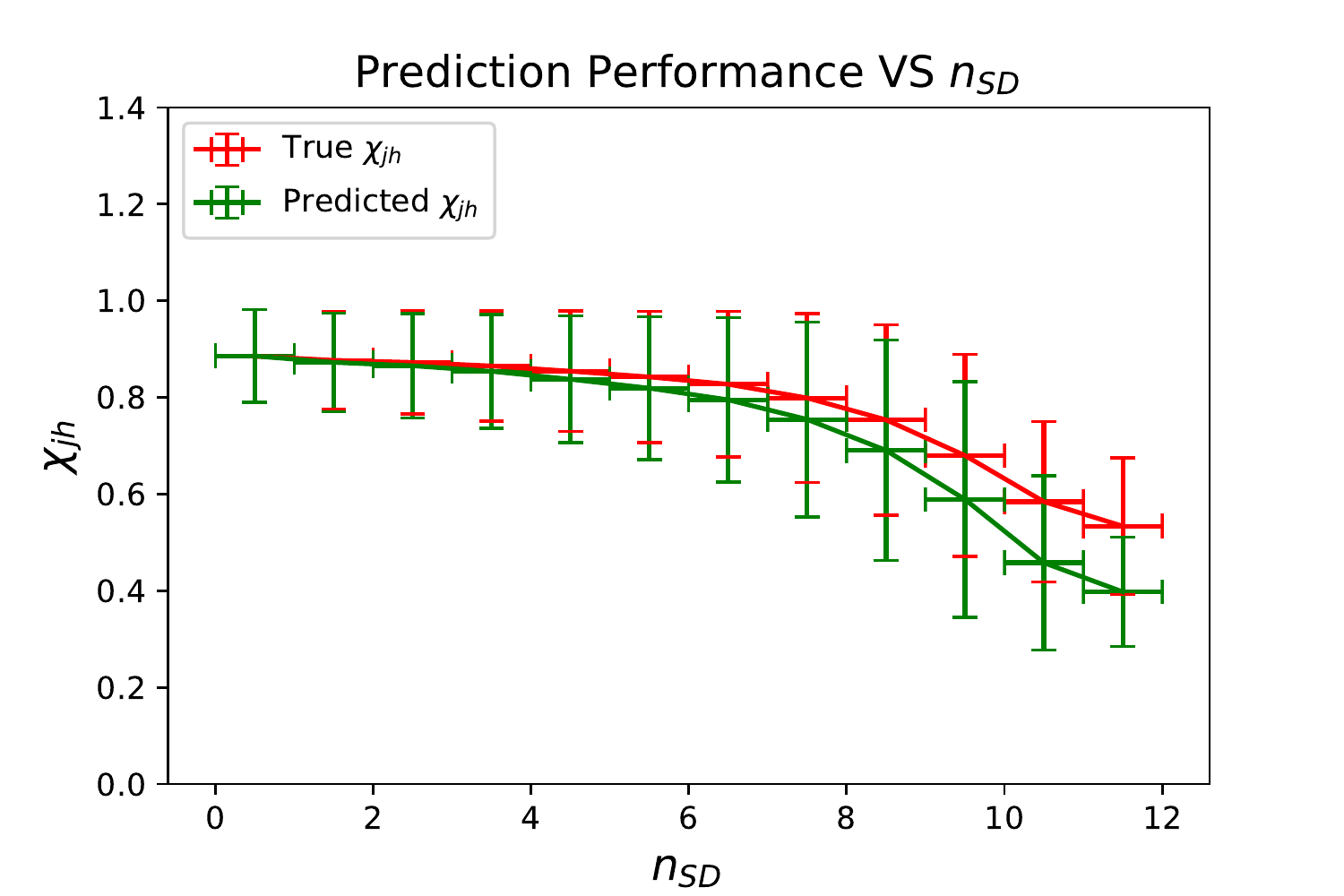}
\includegraphics[width=0.46\textwidth]{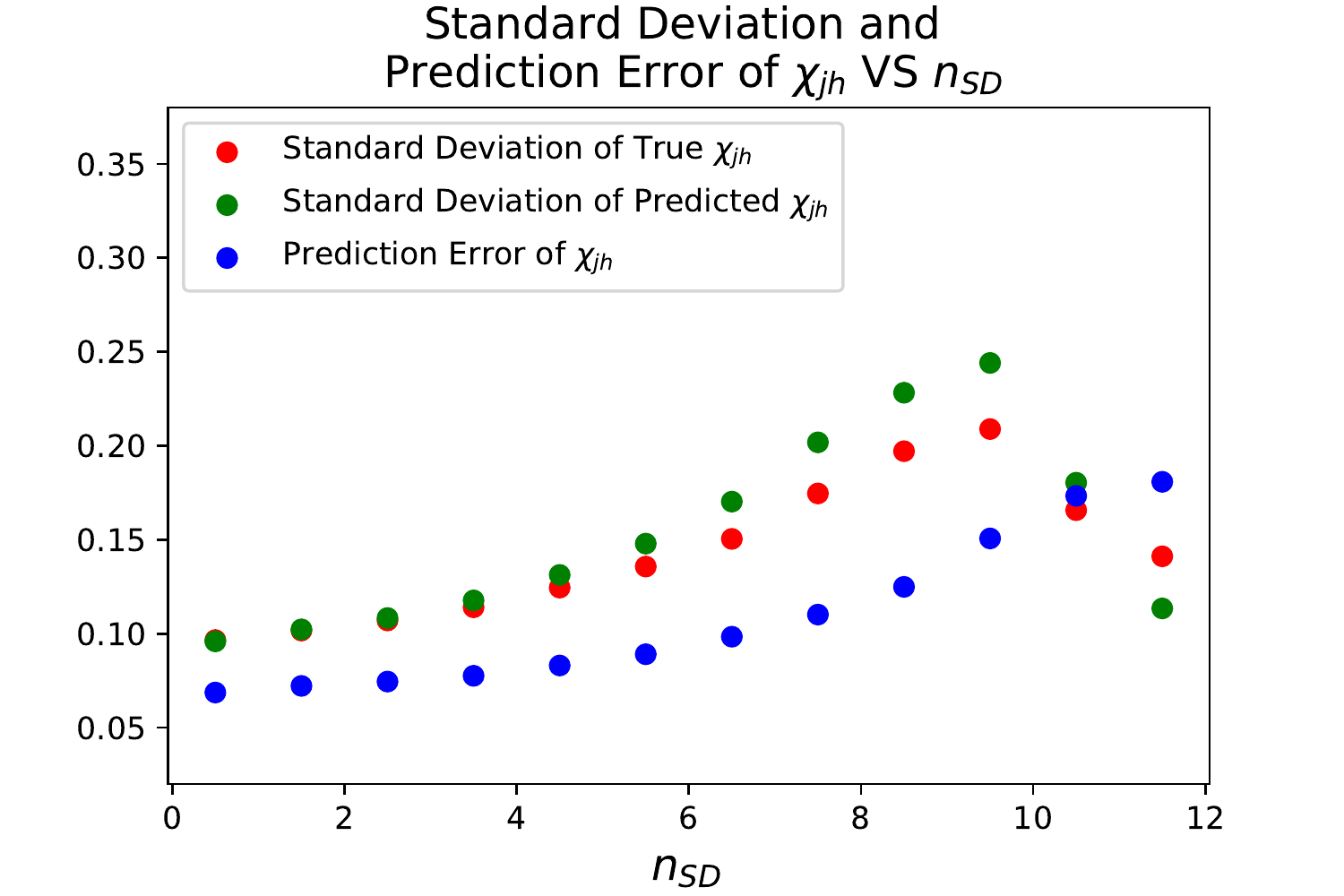}
\caption{Prediction Performance (left panel) and Standard Deviation, Prediction Error (right panel) VS $n_{SD}$.}
\label{Performance nSD}
\end{figure}

\begin{figure}[H]
\centering
\includegraphics[width=0.46\textwidth]{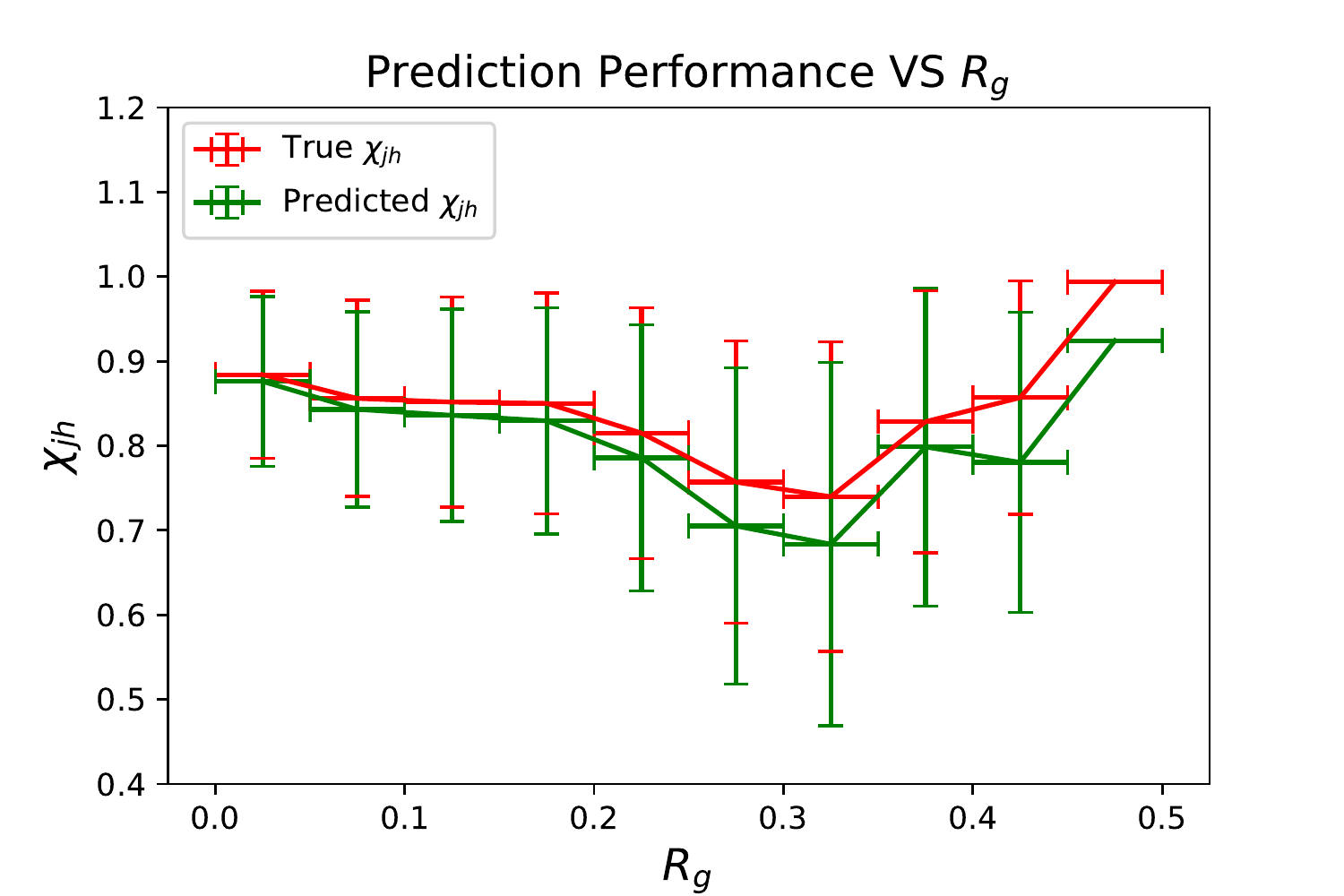}
\includegraphics[width=0.46\textwidth]{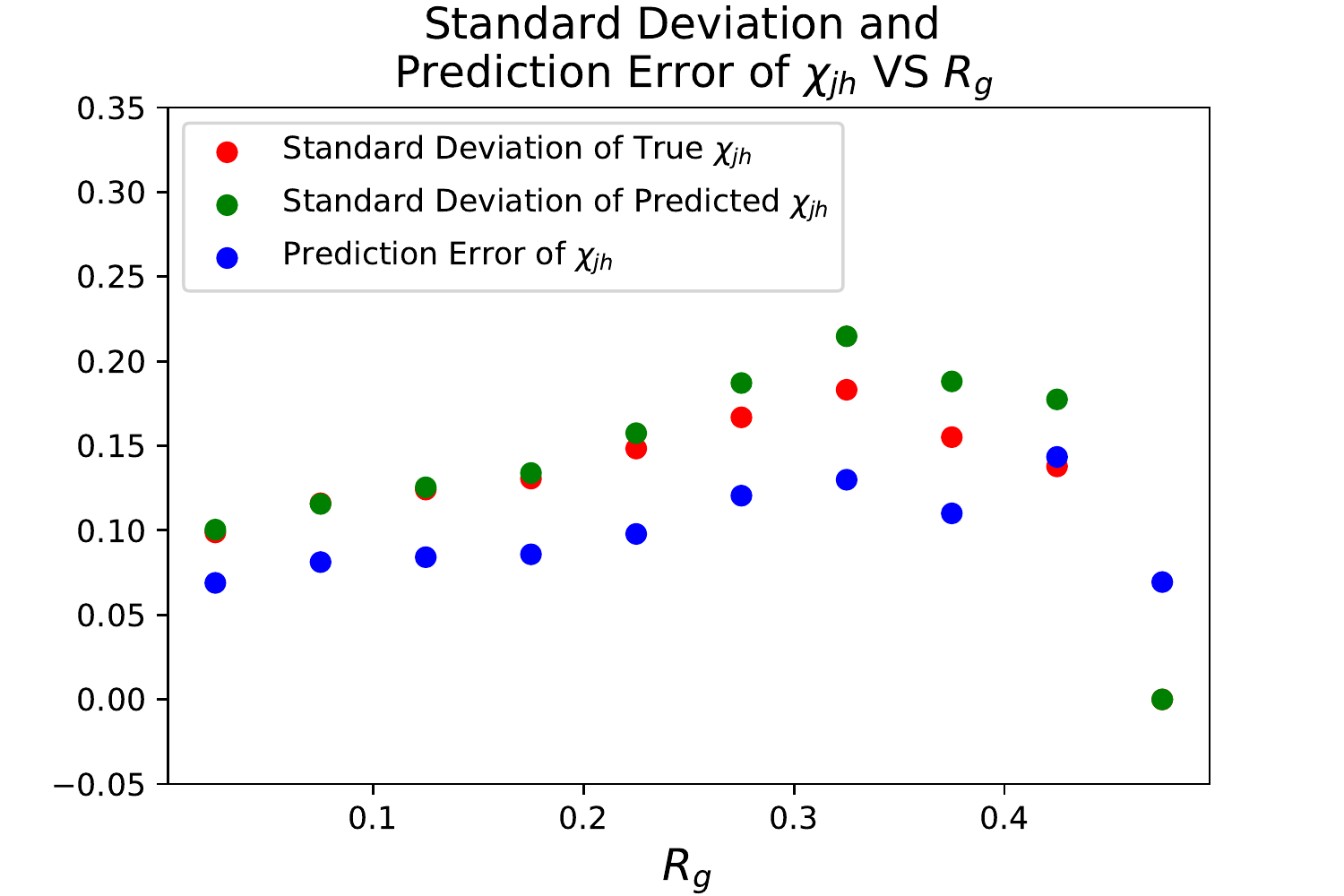}
\caption{Prediction Performance (left panel) and Standard Deviation, Prediction Error (right panel) VS $R_g$.}
\label{Performance DeltaR}
\end{figure}

\begin{figure}[H]
\centering
\includegraphics[width=0.46\textwidth]{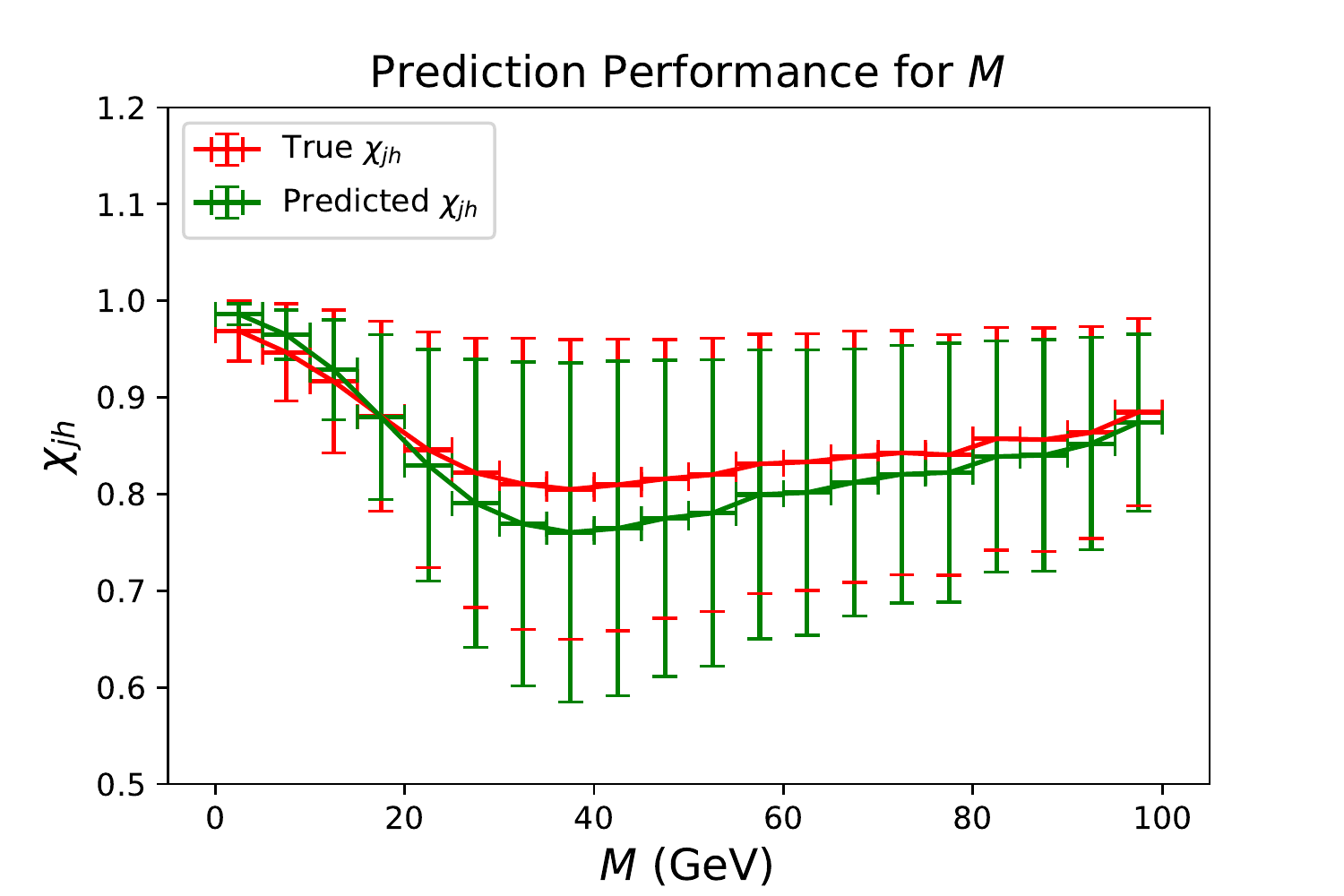}
\includegraphics[width=0.46\textwidth]{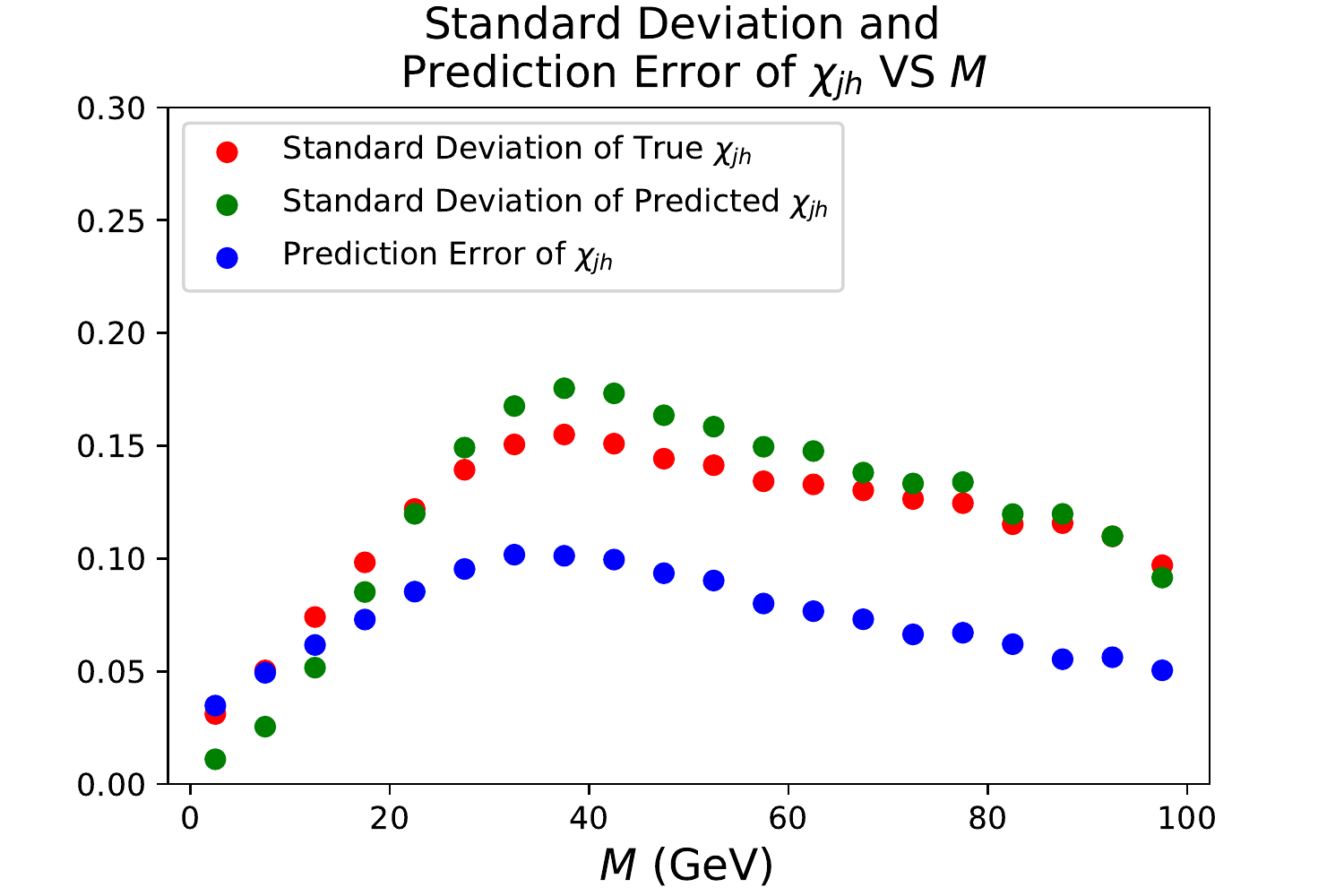}
\caption{Prediction Performance (left panel) and Standard Deviation, Prediction Error (right panel) VS $M$.}
\label{Performance M}
\end{figure}

\begin{figure}[H]
\centering
\includegraphics[width=0.46\textwidth]{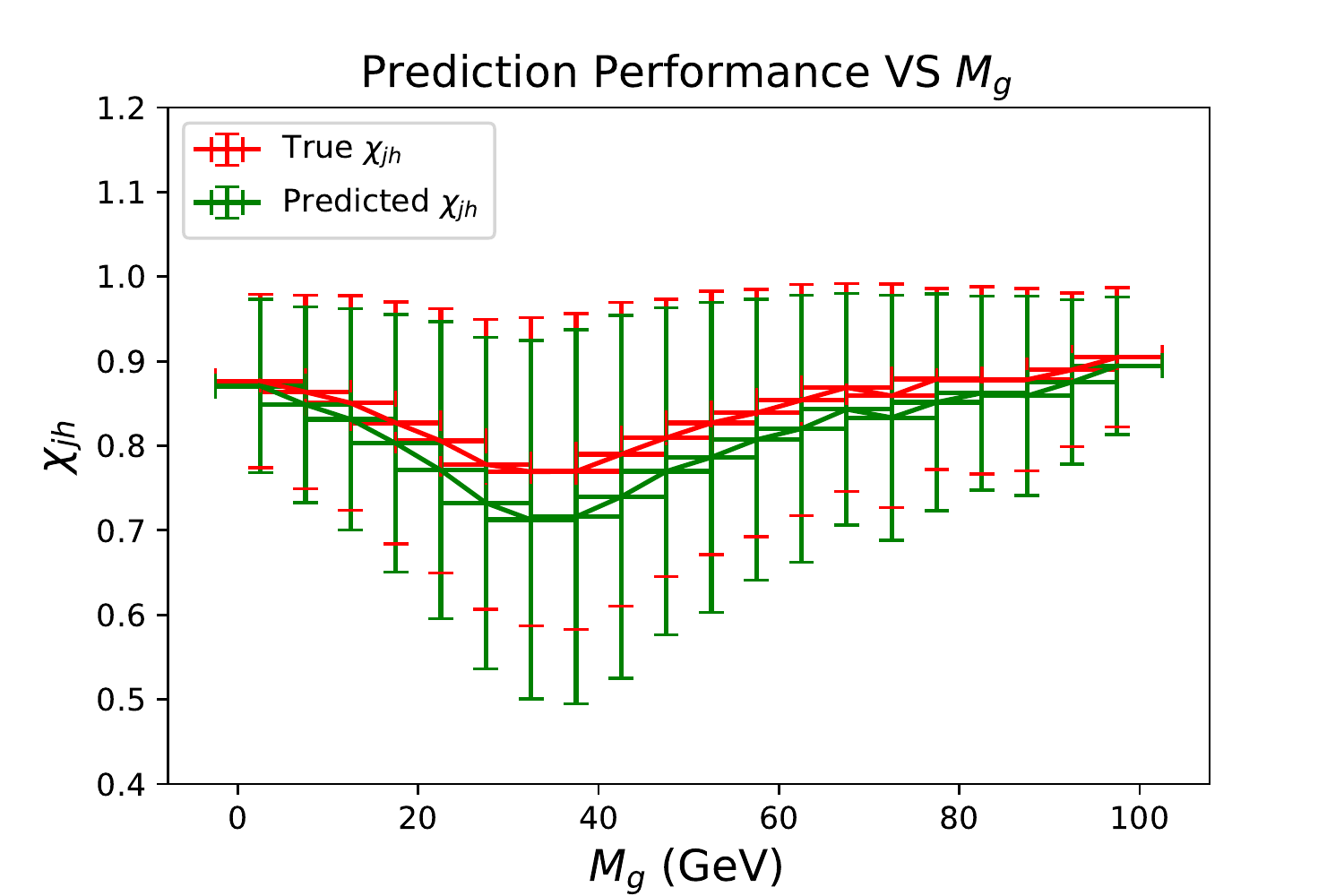}
\includegraphics[width=0.46\textwidth]{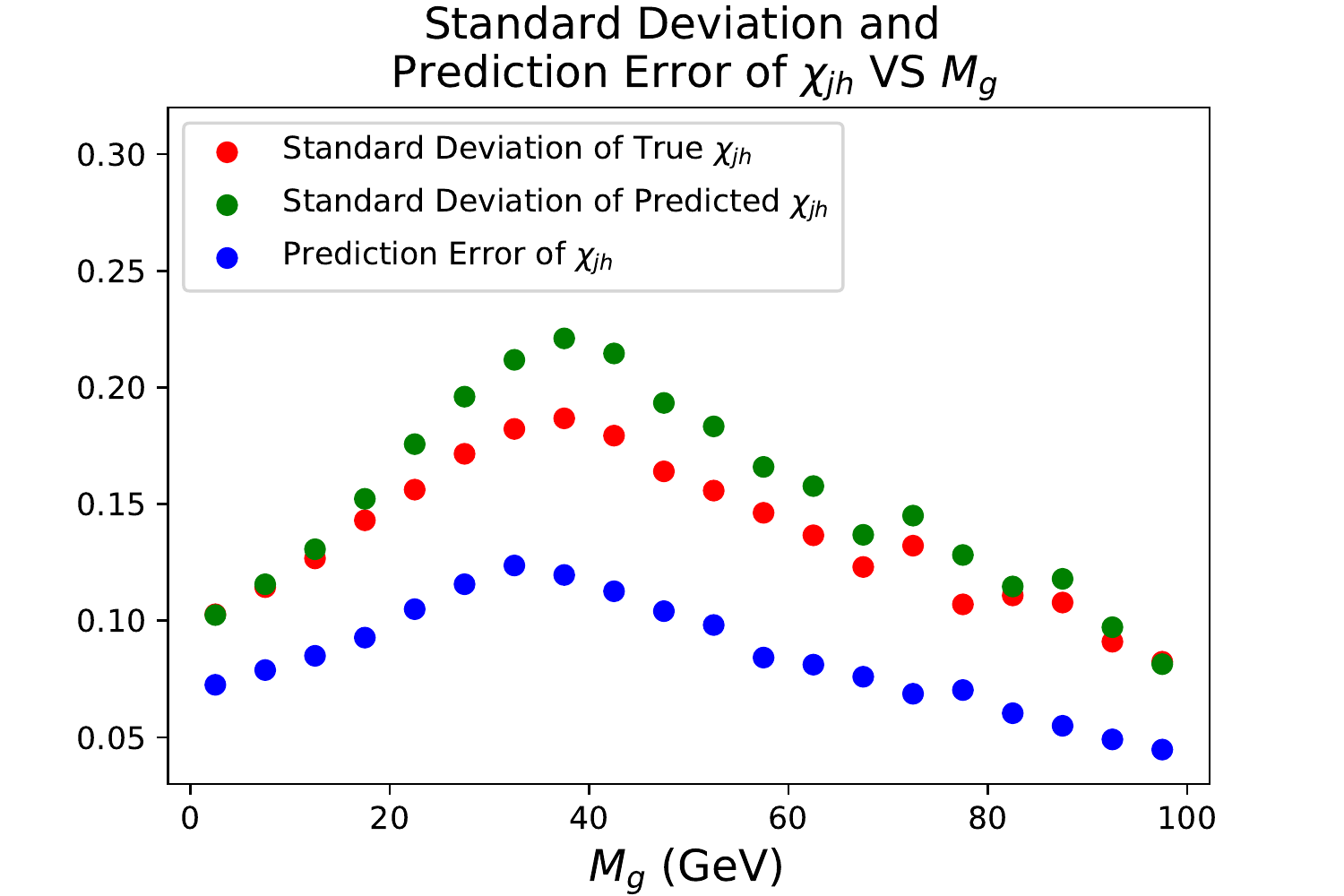}
\caption{Prediction Performance (left panel) and Standard Deviation, Prediction Error (right panel) VS $M_g$.}
\label{Performance Mg}
\end{figure}

In Figs.~\ref{Performance chi}-\ref{Performance Mg}, we show the prediction performance by comparing the average of true and predicted $\chi_{jh}$ with error bar and prediction error against various features. The error bars in the left panels are the standard deviation of the true and predicted $\chi_{jh}$ from the right panels, while the prediction error is the root mean square of the difference between the true and predicted $\chi_{jh}$ in the given bin of the selected features. In general, the average of the predicted $\chi_{jh}$ agrees well with the true one against various features, respectively. 

One can see that the prediction error of $\chi_{jh}$ has a relatively flat dependence on the true $\chi_{jh}$, which is attributed to the use of sample weights in the training. For the prediction of low $\chi_{jh}$ samples, the deviation is a bit larger because of the poorer statistics. The average of $\chi_{jh}$ increases with jet $p_T$, and the standard deviations and prediction error of $\chi_{jh}$ all decrease with jet $p_T$ even though the sample weights are applied in the training. This means that, on average, high $p_T$ jets are more likely to be less quenched and their $\chi_{jh}$ are more easily predicted due to the reasons discussed in the main text. 

The average of $\chi_{jh}$ decreases significantly with jet travel length $L$ and $n_{SD}$, which means that longer travel length $L$ will lead to more jet energy loss and number of Soft Drop splittings $n_{SD}$ is associated with larger energy loss. In contrast, the momentum sharing fraction between subjets, $z_g$, presents no clear impact on the average of $\chi_{jh}$. The physical interpretation of these observations are straightforward. Moreover, the average of $\chi_{jh}$ first decreases and then increases with $R_g$, jet mass $M$ and groomed jet mass $M_g$. The prediction error increases with length $L$ and $n_{SD}$ but has a very slight dependence on $z_g$ and $R_g$, while we observe that it first increases and then decreases with jet mass $M$ and groomed jet mass $M_g$.

\clearpage

\section{Jet observables sliced in $\chi_{jh}$}
\label{sec:observable-sliced-chijh}

\begin{figure}[H]
\centering
\includegraphics[width=0.33\textwidth]{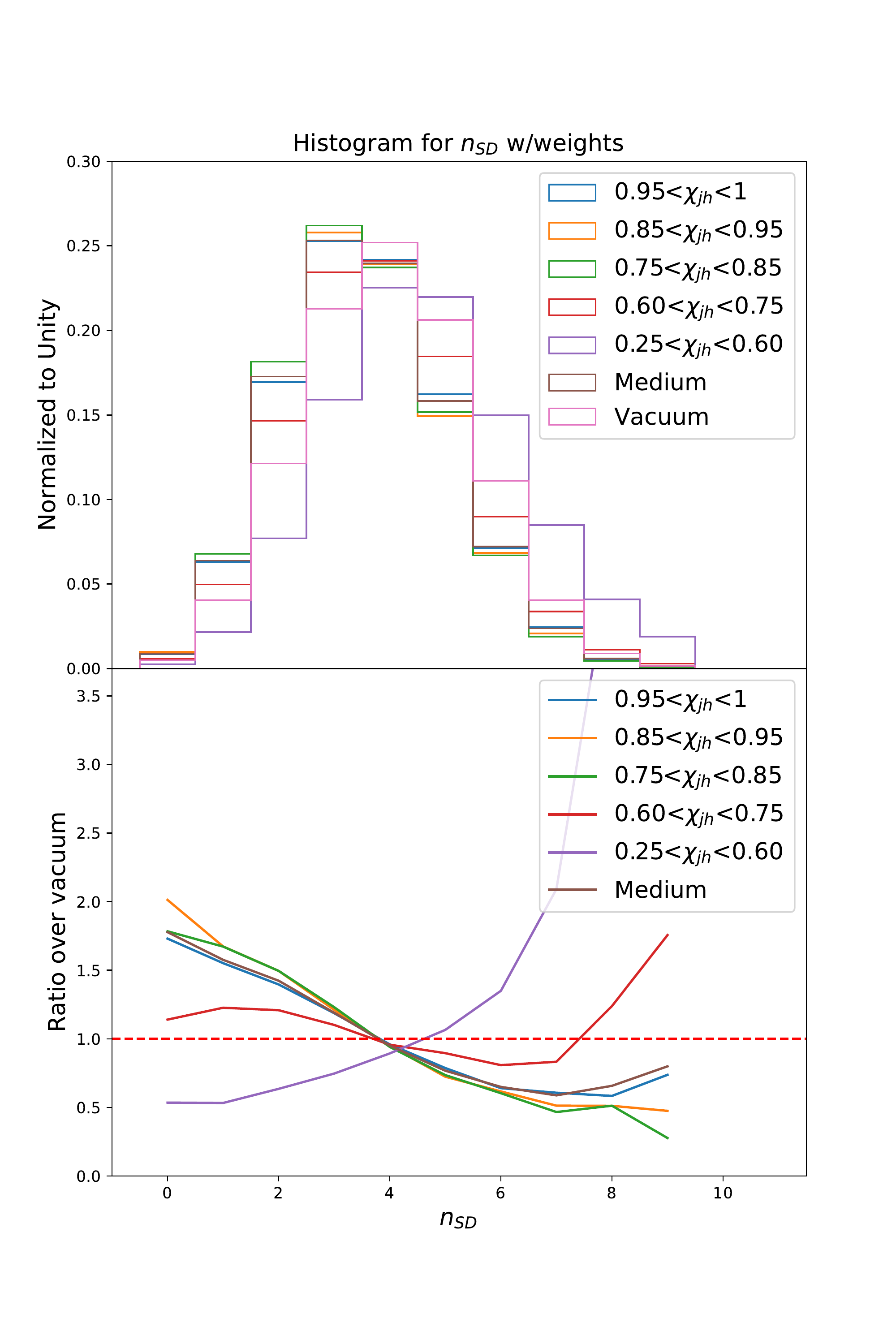}%
\includegraphics[width=0.33\textwidth]{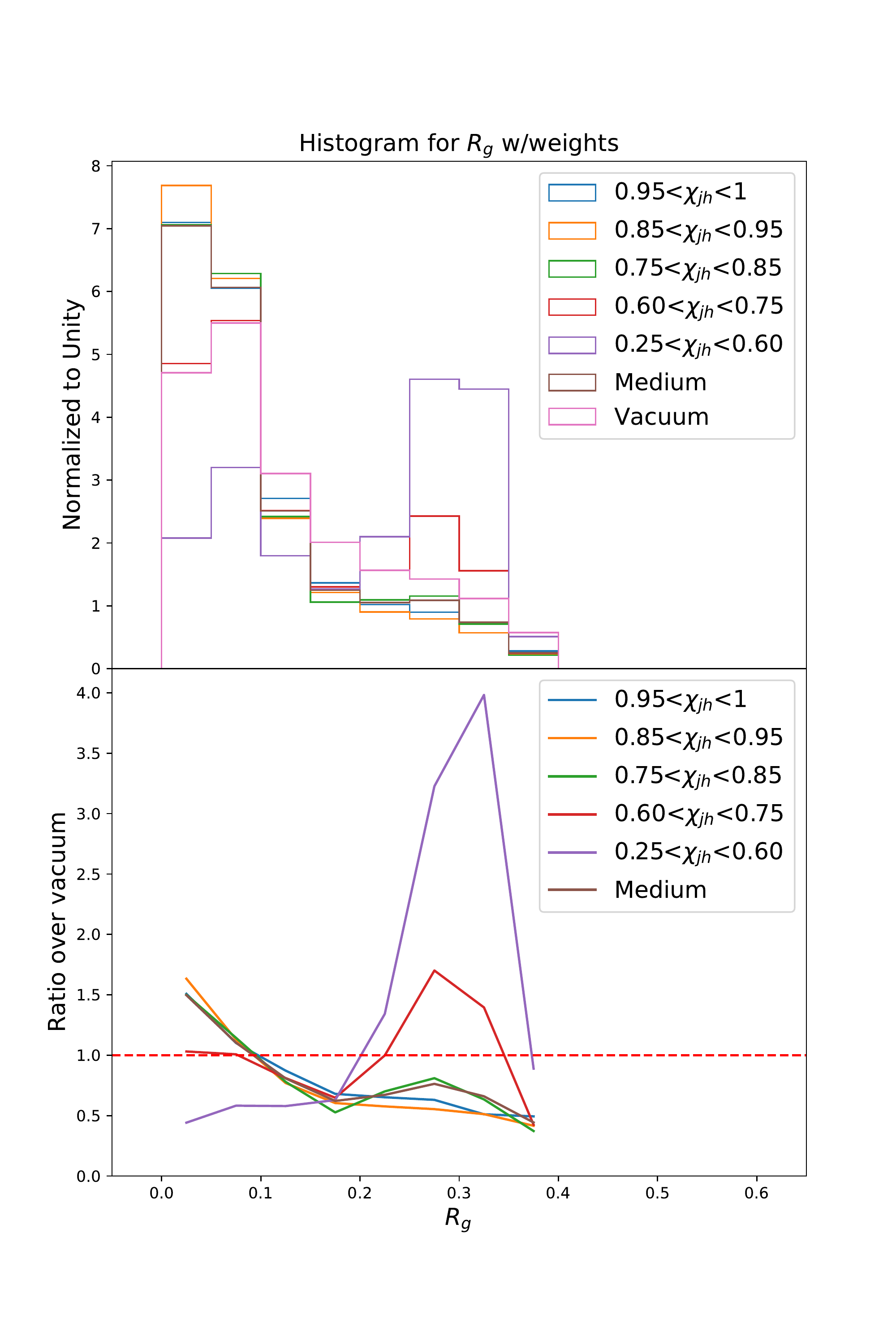}%
\includegraphics[width=0.33\textwidth]{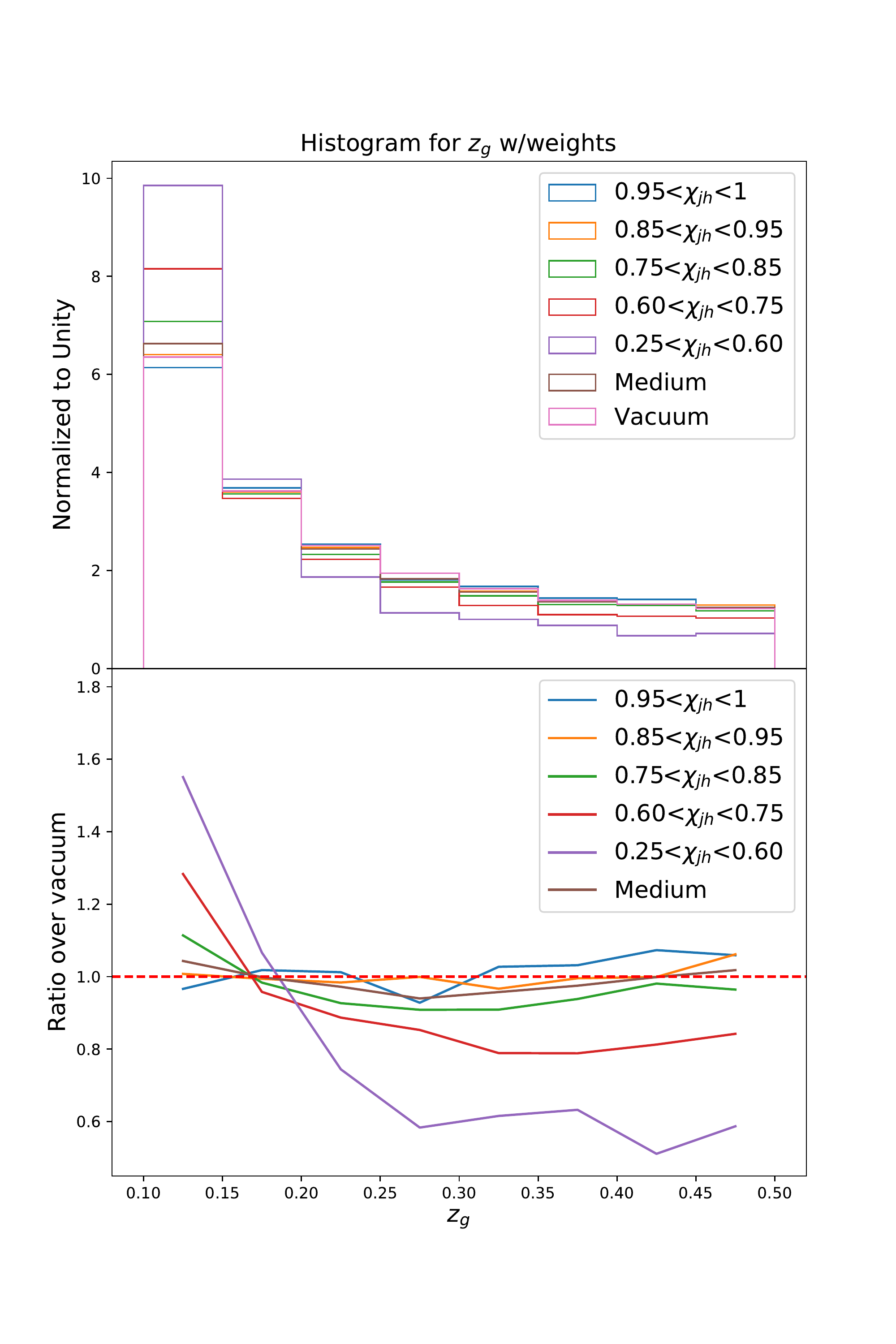}
\caption{Histograms of jet $n_{SD}$, $R_g$ and $z_g$ inclusive and sliced in $\chi_{jh}$ and their ratio over vacuum.}
\label{Rg nSD zg chi}
\end{figure}

\begin{figure}[H]
\centering
\includegraphics[width=0.46\textwidth]{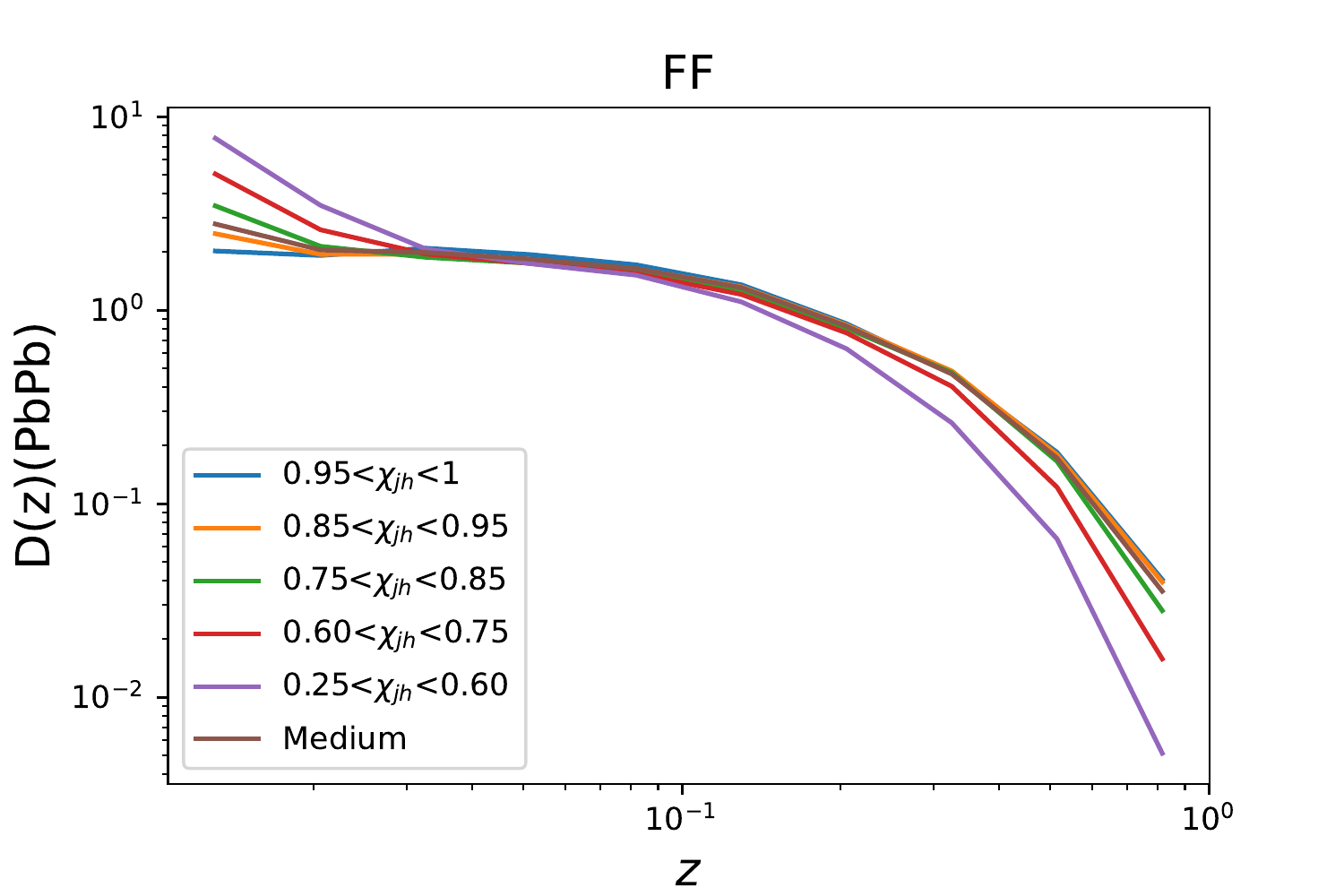}
\includegraphics[width=0.46\textwidth]{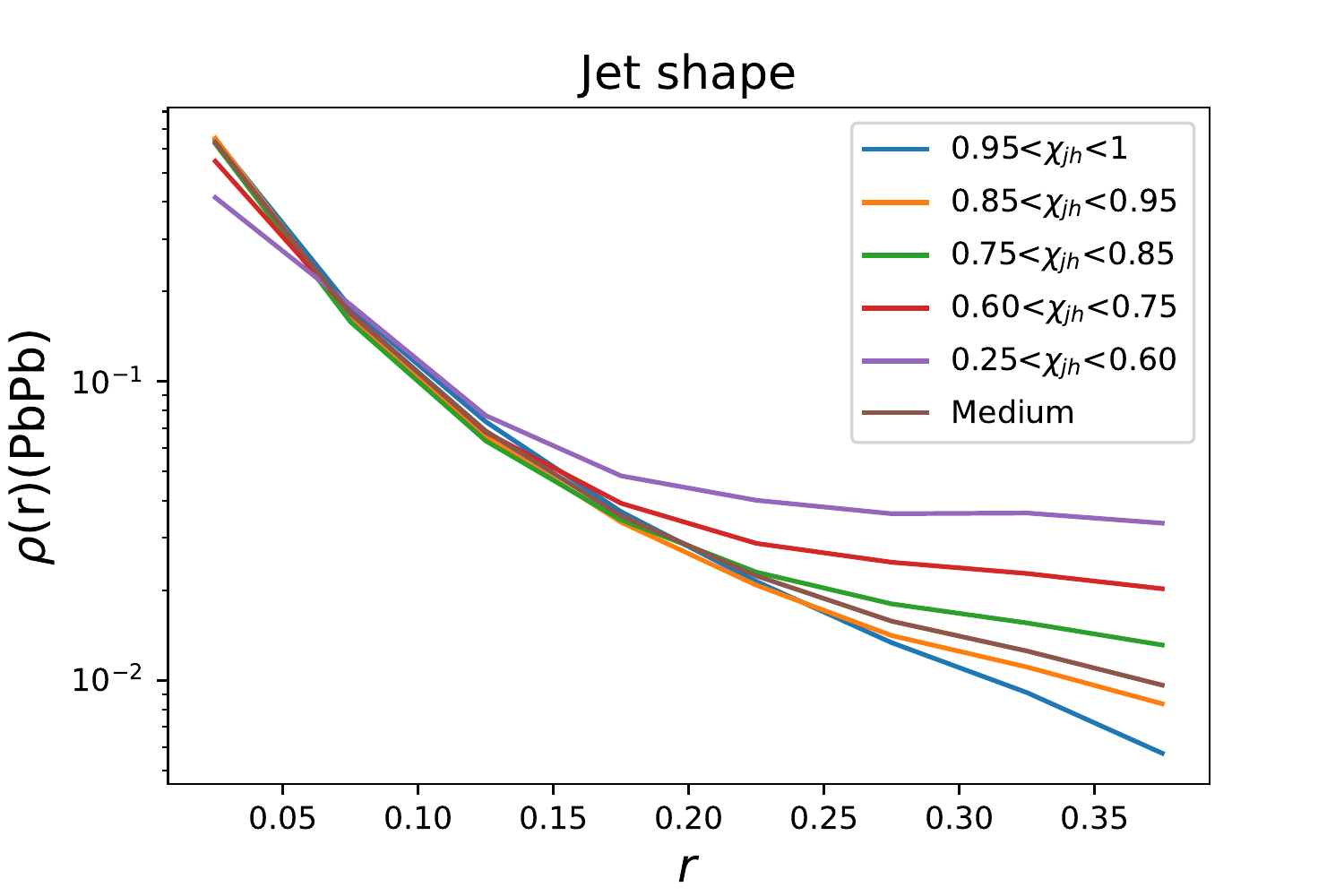}
\caption{JFF and jet shape inclusive and sliced in $\chi_{jh}$.}
\label{FFJS chi}
\end{figure}

Fig.~\ref{Rg nSD zg chi} shows normalized histograms of medium jet observables, including $n_{SD}$ (left), $R_g$ (middle) and $z_g$ (right) inclusive in $\chi_{jh}$, sliced in different $\chi_{jh}$ bins and the ratio over the vacuum one. Fig.~\ref{FFJS chi} shows JFF (left) as well as jet shape (right) inclusive in $\chi_{jh}$ and sliced in different $\chi_{jh}$ bins. All plots here are based on the conventional FES setup with $p_T^{\textrm{jet}}>100$ GeV for both vacuum and medium jets. In these Figures one can see with a greater resolution the extent to which these jet observables are sensitive to $\chi_{jh}$. In particular, both JFF and jet shape display a clear evolution with $\chi_{jh}$, which can be taken as an indication of their potential to predict $\chi_{jh}$.

\clearpage

\bibliographystyle{JHEP}
\bibliography{duyl}

\end{document}